\documentclass[preprint2]{aastex}
\newcommand{\ie}{{\em i.e.}}
\newcommand{\uHz}{\mbox{$\mu$Hz}}
\newcommand{\Frac}[2]{{#1}/{#2}}
\newcommand{\Deg}{^{\rm o}}
\shorttitle{MDI High-Degree Mode Frequencies}
\shortauthors{Korzennik et al.}      
\begin{document}
\title{On The Determination of MDI High-Degree Mode Frequencies}
\author{S.~G.~Korzennik}
\affil{Harvard-Smithsonian Center for Astrophysics, Cambridge, MA 02138, USA}
\author{M.~C.~Rabello-Soares}
\affil{Hansen Experimental Physics Laboratory, Stanford University, 
Stanford, CA 94305, USA}
\and
\author{J. Schou}
\affil{Hansen Experimental Physics Laboratory, Stanford University, 
Stanford, CA 94305, USA}
\begin{abstract}
The characteristic of the solar acoustic spectrum is such that mode lifetimes
get shorter and spatial leaks get closer in frequency as the degree of a mode
increases for a given order. A direct consequence of this property is that
individual $p$-modes are only resolved at low and intermediate degrees, and
that at high degrees, individual modes blend into ridges. Once modes have
blended into ridges, the power distribution of the ridge defines the ridge
central frequency and it will mask the true underlying mode frequency.  An
accurate model of the amplitude of the peaks that contribute to the ridge
power distribution is needed to recover the underlying mode frequency from
fitting the ridge.

 We present the results of fitting high degree power ridges (up to $\ell =
900$) computed from several two to three-month-long time-series of full-disk
observations taken with the Michelson Doppler Imager (MDI) on-board the Solar
and Heliospheric Observatory between 1996 and 1999.

 We also present a detailed discussion of the modeling of the ridge power
distribution, and the contribution of the various observational and
instrumental effects on the spatial leakage, in the context of the MDI
instrument. We have constructed a physically motivated model (rather than some
{\em ad hoc} correction scheme) resulting in a methodology that can produce an
unbiased determination of high-degree modes, once the instrumental
characteristics are well understood.

 Finally, we present changes in high degree mode parameters with epoch and
thus solar activity level and discuss their significance.
\end{abstract}
\section{Introduction}

\begin{figure}[!p]
\epsscale{1.}
\plotone{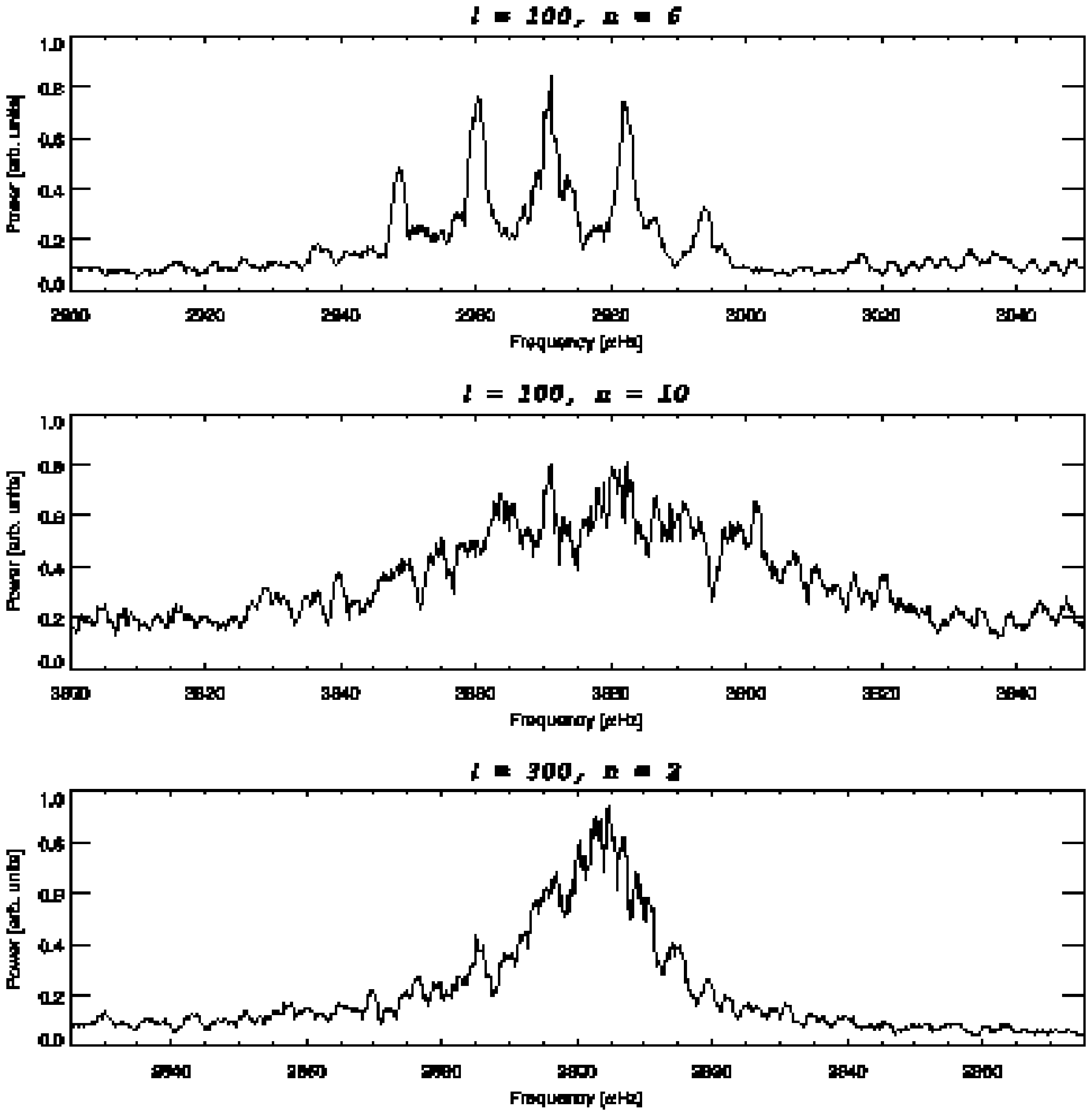}
\caption{MDI sectoral power  spectra, for {\em Dynamics} 1999 and for
         $(\ell,n) = (100, 6), (100, 10)$ and $(300, 2)$, computed using a 9th
         order sine multi-taper. At low degree and low order (top panel)
         individual modes are well resolved: the isolated peak at 2970 \uHz\
         is the target mode while the others are the spatial leaks. Otherwise
         the peaks are blended: the middle panel shows the spectrum for the
         same degree but at higher order where mode blending results from
         larger mode widths due to shorter life times. The bottom panel shows
         the spectrum at roughly the same frequency as in the top panel but at
         a higher degree, where mode blending results from a smaller mode
         spacing ($d\nu/d\ell$) as well as shorter life time.
         \label{fig:spectra}}
\end{figure}

  Since from a single vantage point we can only observe a bit less than half
of the solar surface, helioseismic power spectra computed for a specific
target mode with degree $\ell$ and azimuthal order $m$ also contains power
from modes with different -- and usually nearby -- $\ell$ and $m$ values. The
presence of these unwanted modes, or spatial leaks, complicates the fitting of
the resulting observed spectra and degrade the mode parameter estimates,
especially when the leaks have frequencies similar to that of the target mode.
As mode lifetimes get shorter and spatial leaks get closer in frequency (\ie,
$d\nu/d\ell$ becomes small), individual p-modes can no longer be
resolved. This mode blending occurs around $\ell=150$ for p-modes with
frequency near $3.3$ mHz and around $\ell=250$ for the f-modes. Once
individual modes blend into ridges, as illustrated in
Figure~\ref{fig:spectra}, the power distribution of the ridge masks the true
underlying mode frequency and the ridge central frequency is not a good
estimate of the target mode frequency. Moreover the amplitudes of the spatial
leaks have been shown to be asymmetric \cite[see for
example][]{korzennik99}. A direct consequence of this leakage asymmetry is to
offset the power distribution of the ridges. Such offset results in a
significant difference between the central frequency of the ridge and the
frequency of the targeted individual mode. To recover the underlying mode
frequency from fitting the ridge, an accurate model of the amplitude of the
peaks that contribute to the ridge power distribution (\ie, the leakage
matrix) is needed.

  The lack of unbiased determinations of mode frequencies and frequency
splittings at high degrees has so far limited the use of such high-degree data
in helioseismic inversions for constraining the near-surface structure and
dynamics of the sun. Since experiments like the Michelson Doppler Imager (MDI)
on board the Solar and Heliospheric Observatory (SOHO), with a
two-arcsec-per-pixel spatial resolution in full-disk mode, allows us to detect
oscillation modes up to $\ell \approx 1500$ \citep{scherrer95}, only a small
fraction of the observed modes are currently used.

  High-degree modes are trapped near the solar surface: for example, the lower
turning point of a mode of degree $\ell = 500$ and frequency around 3 mHz is
at 0.99 of the solar radius. This makes them exceptional diagnostic tools to
probe the near-surface region of the Sun, a region of great interest. Indeed,
it is there that the effects of the equation of state are felt most strongly,
and that dynamical effects of convection and processes that excite and damp
the solar oscillations are predominantly
concentrated. \citet{rabello-soares00} have shown that the inclusion of
high-degree modes (\ie, $\ell$ up to 1000) has the potential to improve
dramatically the inference of the sound speed in the outermost 2 to 3\% of the
solar radius.  Furthermore, inversion of artificial mode frequency differences
resulting from models computed with two different equations of state (\ie, MHD
and OPAL) recovered the intrinsic difference in $\Gamma_1$, the adiabatic
exponent, throughout the second helium ionization zone and well into the first
helium and hydrogen ionization zones, with error bars far smaller than the
differences resulting from using two different equations of state.  These
tests were carried out using the relatively large observational uncertainties
resulting from ridge fitting, but with the implicit assumption that systematic
errors were not present.  These tests show that we can probe subtle effects in
the thermodynamic properties of this region, but only when including such
high-degree modes.

  The first estimates of high-degree mode frequencies used $m$-averaged Big
Bear Solar Observatory data \citep{libbrecht88}. To recover the underlying
mode frequency from fitting a given ($n, \ell$) ridge, they used a simple
Gaussian profile as an approximation for the $m$-averaged leakage
matrix. Namely:
\begin{equation}
  C_r^2(\ell, \ell') = \exp({-(\frac{\Delta\ell - \epsilon\,\ell}{2s})^2})
  \label{eq:klCr}
\end{equation}
where $\Delta\ell = \ell' - \ell$ while the $\epsilon\,\ell$ term represents
the leakage asymmetry introduced by an image scale error of a fraction
$\epsilon$ of the image size.  The ridge centroid frequency was estimated
using a simple weighted average:
\begin{equation}
  \tilde{\nu}_{n,\ell} = 
     \frac{\sum_{\ell'} C_r^2(\ell, \ell') \, A_{n,\ell'} \, \nu_{n,\ell'}}
          {\sum_{\ell'} C_r^2(\ell, \ell') \, A_{n,\ell'}}
\end{equation}
where $A_{n,\ell'}$ is the individual mode power amplitude and $\nu_{n,\ell'}$
is the mode frequency. The frequency difference between the ridge and the mode
frequency, $\Delta \nu_{n,\ell} = \tilde{\nu}_{n,\ell}(\mbox{ridge}) -
\nu_{n,\ell}(\mbox{mode})$, can thus be estimated using a parametric
representation of the leakage coefficients.

  For intermediate degree modes ($50 < \ell < 150$) this frequency difference
can be directly measured by reducing the frequency resolution of the observed
power spectra as to force individual modes to blend into ridges. Using
observed $\Delta \nu$, the parameters $s$ and $\epsilon$ can be calibrated at
these intermediate degrees and the correction extrapolated to high-degree
modes.

  As \citet{libbrecht88} state in their paper, this is only a first step and
there is substantial room for improvement particularly at high $\ell$.
Following the same idea, but improving on the method, \citet{korzennik90} and
\citet{rhodes99} estimated high-degree mode frequencies using Mount Wilson and
MDI data respectively. In contrast, \citet{bachmann95}, using data from the
High-L Helioseismometer at Kitt Peak, calculated the $m$-averaged leakage
matrix, but neglecting the horizontal components, in order to estimate
high-degree mode frequencies.

  In this paper, we present the results of an extensive study of the various
elements that contribute to the precise value of the effective leakage matrix,
that in turn is key to the precise determination of high degree modes.  We
have attempted to construct a physically motivated model --- rather than an
{\em ad hoc} correction scheme --- in order to produce an unbiased
determination of the high-degree modes. Since \citet{korzennik99} has shown
that the inclusion of the horizontal component of the leakage matrix
calculation partially explains its observed asymmetry, we have included the
horizontal component in all our leakage matrix calculations.

  In Section 2, we describe the data we used and how we computed and fitted
the power spectra used in this work.  In Section 3 we present and discuss
ridge modeling for high degree modes while in Section 4 we address the issue
of estimating the ratio between the radial and horizontal components. In
Section 5 we discuss the instrumental effects specific to MDI that must be
included to properly model the ridge power distribution. Finally, in Section
6, we present the results of our analysis followed by our conclusions.

\section{Data Used and Fitting Methodologies}

  The data used in this work consists of four separate time series of full-disk
Dopplergrams. These Dopplergrams were obtained by the MDI instrument while
operating in its full disk $2''$ per pixel resolution mode and during the {\em
Dynamics Program}. The instrument is operated in this mode some 3 months every
year, when the available telemetry bandwidth is large enough to bring down
full-disk images. Table~\ref{tab:observationPeriods} lists the starting time
and duration of each of the four {\em Dynamics} data time series.

  The spherical harmonics decomposition up to $\ell = 1000$ was carried out by
the MDI Science Team \citep{schou99}.  For the same periods, mode frequencies
and rotational splitting coefficients of individual p-modes were computed for
low and intermediate degrees as part of the {\em Structure Program}
\citep{schou99}.

\begin{deluxetable}{llc}
\tablewidth{0pt}
\tablecaption{{\em Dynamics Program} Epochs
             \label{tab:observationPeriods}}
\tablehead{\colhead{Data Set} & \colhead{Starting Date} & \colhead{Duration} \\
           \colhead{[year]}   & \colhead{month/day}     & \colhead{[days]}}
\startdata
 {1996} & 05/23 & 63  \\
 {1997} & 04/14 & 93  \\
 {1998} & 01/09 & 92  \\
 {1999} & 03/13 & 77  \\
\enddata
\end{deluxetable}

  For each {\em Dynamics Program} period we computed and fitted power spectra
using two distinct approaches. The purpose of the first one was to get some
observational estimate of the leakage matrix asymmetry, using an optimum
estimate of the sectoral limit spectra computed with a high frequency
resolution where individual p-modes can be clearly resolved.

 We used only sectoral spectra because the leakage matrix for the sectoral
modes is diagonal at a fixed radial order, $n$ (namely non-zero only when
$\delta m = \delta\ell$), resulting in a simple leakage pattern, with modes
separated by at least the quantity $\Frac{\partial\nu}{\partial\ell}$. For
zonal and tesseral modes the leakage pattern is more complex resulting in
leaks very close to each other, some separated only by even multiples of
$\Frac{\partial\nu}{\partial m}$.  For these modes, only for very low
frequencies (where the lifetime is long enough), and if the time series is
long enough, can the main peak be resolved from its closest spatial leaks.

  The second approach is aimed at computing optimum power spectra for
high-degree modes ridge fitting. Namely spectra with a lower frequency
resolution --- since individual modes are not resolved -- while in the process
averaging out the realization noise that a high frequency resolution spectrum
would reveal. Moreover, at low and intermediate degrees individual modes blend
into ridges as a result of the low frequency resolution. This approach creates
an overlap between mode fitting and ridge fitting where the frequency offset
introduced by ridge fitting can be directly measured. By using this overlap,
we can check the model we have developed to infer mode parameters from the
fitted ridge parameters.

\subsection{Limit Spectra for Resolved Modes at Intermediate Degrees}
\label{sec:limitSpectra}

  Estimates of limit spectra were computed for intermediate degree sectoral
modes, when individual modes could be resolved, using the following procedure.
First, each time series of spherical harmonic coefficients was detrended using
a 21-minute-long running mean. Then, the 9th order sine multi-tapered power
spectrum was computed, using the two- to three-month-long time series and thus
corresponding to a frequency resolution between 0.18 \uHz\ and 0.24
\uHz. Finally, for a range of degrees (15 consecutive $\ell$) at a fixed
order, $n$, small fractions of power spectra (about 300 $\mu$Hz), centered
around each mode frequency, were averaged. The sectoral mode frequencies were
computed using a table of frequencies and frequency splittings resulting from
mode fitting of MDI data.

  A seven-component profile was fitted in the least-squares sense to these
limit spectra using the following parameterization:
\begin{eqnarray}
  P_{n, \ell} (\nu) & = & \sum_{\ell'=\ell-3}^{\ell+3} \: 
     \left( A_{n, \ell'} \; \frac{1 + \alpha_{n, \ell} 
        (x - \alpha_{n, \ell}/2)}{1 + x^2} \right)  \nonumber \\
& ~ & + \, b_{n, \ell} + s_{n, \ell} \, (\nu-\nu_{n, \ell})
\label{eq:7comp}
\end{eqnarray}
where
\begin{equation}
   x = x_{n, \ell, \ell'}(\nu) = \frac{\nu - \nu_{n, \ell'}}{\gamma_{n, \ell}}
\label{eq:eqx}
\end{equation}
and where $A_{n, \ell'}$ and $\nu_{n, \ell'}$ are the individual mode
amplitude and frequency respectively; $\gamma_{n, \ell}$ and $\alpha_{n,
\ell}$ are the mode width and asymmetry parameter respectively -- assumed
constant for all seven components; and $b_{n, \ell}$ and $s_{n, \ell}$
represent the background power.

  Note that the shape of the profiles used in this formulation is equivalent
to the one defined by Equation~4 of \citet{Nigam+Kosovichev:1998}. After some
rudimentary algebra, one can identify their asymmetry coefficient, $B$, to the
one we use, $\alpha$, namely $\frac{\alpha}{2} = B\,(1-\frac{\alpha^2}{2})$.

\subsection{Power Spectra at High Degrees and Ridge Fitting}
\label{sec:ridgeFitting}

  For each period listed in Table~\ref{tab:observationPeriods}, the time
series of spherical harmonic coefficients were sub-divided in 4096-minute-long
intervals.  For each spherical harmonic degree, $\ell$, and each azimuthal
order, $m$, each interval was first detrended using a 17-minute-long running
mean, and then Fourier transformed with an oversampling of 2 to estimate the
power spectrum. The power spectra corresponding to each 4096-minute-long
interval in a given period were then averaged together to produce high
signal-to-noise ratio spectra with a frequency resolution of 4.1 \uHz.  As a
result of this low frequency resolution, the individual modes down to $\ell
\approx 100$ blended into ridges, while the ridge width itself is still very
much oversampled.

  The following asymmetric profile plus a background term was fitted in the
least squares sense to each low frequency resolution spectrum, for $100 \le
\ell \le 900$:
\begin{eqnarray}
  P_{\ell,m}(\nu) & = &
  \sum_n \left(\tilde{A}_{n, \ell, m} \: \frac{1 + \tilde{\alpha}_{n, \ell, m} 
    (\tilde{x}_{n, \ell, m} - \tilde{\alpha}_{n, \ell, m}/2)}
  {1 + \tilde{x}_{n, \ell, m}^2} \right)  \nonumber \\
& ~ & + \, B_{\ell, m}(\nu) 
\label{eq:asymprof}
\end{eqnarray}
where
\begin{equation}
  \tilde{x}_{n, \ell, m} = \frac{\nu - \tilde{\nu}_{n, \ell, m}}
                                {\tilde{\gamma}_{n, \ell, m}}
\end{equation}
and
\begin{equation}
 \log \, B_{\ell, m} (\nu) = \sum_{j=0}^2 \, b_j (\ell, m) \: \nu^j 
\end{equation}
such parameterization of the background power is better suited for the
fit performed here, namely over a broad frequency range.

  In this case, $\tilde{A}_{n, \ell, m}$, $\tilde{\gamma}_{n, \ell, m}$,
$\tilde{\nu}_{n, \ell, m}$ and $\tilde{\alpha}_{n, \ell, m}$ are the {\em
ridge} amplitude, width, frequency and asymmetry parameters for a given ($n,
\ell, m$) ridge fitting. For practical reasons, this fitting was carried out
only every 5th $\ell$ for $100 \le \ell \le 250$ and every 10th $\ell$ for
$250 < \ell \le 900$, and only for some 50 equally spaced $m$ values at each
$\ell$. From the resulting {\em ridge} frequencies, the {\em ridge} frequency
splittings were parameterized in terms of Clebsh-Gordon coefficients
\citep[$\tilde{a}_i$, for $i=1,6$, see][]{Ritzwoller+Lavely:91}.

  A single asymmetric function was used to to fit each ridge power density
profile (Eq.~\ref{eq:asymprof}). Since the ridge power density profile results
from the overlap of the target mode and the spatial leaks, it is
mathematically not a similar single asymmetric function. In the specific case
of high degree modes and at the frequency resolution we have used we believe
that our approach is both justified and practical, based on the following
considerations. 
First, the profile resulting from the overlap of nearby profiles is reasonably
well modeled by a single profile when the ratio of the width of the individual
profiles ($\Gamma_{\rm eff}$) to the separation ($\partial\nu/\partial\ell$)
become large. Indeed, for a ratio of 2 -- a ratio that roughly corresponds to
the mean ratio around $\ell=300$ -- the root-mean-squares (RMS) of the
difference between a superposition of profiles and a single profile is only
4\% (when normalizing the resulting fitted profile maximum to unity). For a
ratio of 4 ($\ell \approx 500$) the RMS of this difference drop to 2\%, while
for a ratio of 10 ($\ell \approx 700$) this RMS drop to 0.6\%. These numbers
are to be compared to the RMS of the residuals to the fit that are around 5 to
7\% (when using the same normalization) at all degrees and are dominated by
the realization noise. The second reason to use a single profile is practical:
fitting to the resulting power density profile the sum of individual
overlapping profiles is underconstrained, unless some assumptions on the
separations and/or the relative amplitudes are made. Since neither of these
quantities are known {\em a priory} with enough precision such approach is
impractical.

\section{Ridge Modeling}

  We describe here our physically motived ridge model, from which corrections
to the ridge parameters are estimated.  Once the offsets between ridge and
mode parameters are determined, the observed ridge parameters can be corrected
and the mode parameters for high-$\ell$ obtained, making it possible to study
the near-surface region of the Sun.

\subsection{Leakage Matrix: Radial and Horizontal Components}

  Since the spatial dependence of solar oscillations can be described in terms
of spherical harmonics $Y^m_{\ell}(\theta,\phi)$ of co-latitude $\theta$ and
longitude $\phi$, the velocity signal resulting from these oscillations can be
written as:
\begin{equation}
  \vec{v}_{n, \ell, m} = (V_r \, Y^m_{\ell}, \,\,
                          V_h \, \partial_{\theta} Y^m_{\ell}, \,\,
                          V_h \, \frac{1}{\sin\theta} \, \partial_{\phi} Y^m_{\ell}).
\end{equation}
and the leakage matrix components, resulting from the spherical harmonic
decomposition of the observed line-of-sight velocity, are given by:
\begin{eqnarray}
   C^r_{\ell, m; \ell', m'}\hspace*{-1em} & = & 
   \oint W  Y^{m*}_{\ell}  Y^{m'}_{\ell'} 
        \sin\theta \cos\phi \, d\Omega
   \label{eq:theoleakr} \\
   C^{\theta}_{\ell, m; \ell', m'}\hspace*{-1em} & = &
  \!\!\!\! -\frac{1}{L} \oint W  Y^{m*}_{\ell} 
  \partial_{\theta}  Y^{m'}_{\ell'} \cos\theta \cos\phi \, d\Omega
   \label{eq:theoleakt} \\
   C^{\phi}_{\ell, m; \ell', m'}\hspace*{-1em} & = & 
     \frac{1}{L} \oint W  Y^{m*}_{\ell} 
   \partial_{\phi}  Y^{m'}_{\ell'} \sin\phi \, d\Omega
   \label{eq:theoleakp}
\end{eqnarray}
where $^*$ represents the complex conjugate operator and $W(\theta,\phi)$ is
the spatial window function of the observations -- \ie, a function that
delimits the angular span of the observations and that includes additional
spatial attenuations like spatial apodization.

  The complete leakage matrix is the sum of the radial component $C^r$ and the
horizontal components $C^{\theta} + C^{\phi}$.  Thus a mode with a given
($\ell', m'$) will leak in the ($\ell, m$) power spectrum with an amplitude
attenuated by a factor $C^2_{\ell,m;\ell',m'}$ where:
\begin{eqnarray}
C_{\ell, m; \ell', m'} &=& C^r_{\ell, m; \ell', m'}  + \nonumber \\
  &~& \beta_{n, \ell} \, (C^{\theta}_{\ell, m; \ell', m'} +
C^{\phi}_{\ell, m; \ell', m'}),
\label{eq:complete}
\end{eqnarray}
and where $\beta_{n,\ell}$ is the horizontal-to-vertical displacement
ratio. Using a simple outer boundary condition, \ie\ that the Lagrangian
pressure perturbation vanish ($\delta p = 0$), the small amplitude
oscillations equations for the adiabatic and non-magnetic case lead to an
estimate of the ratio $\beta$, given by \citep{cd97}:
\begin{equation}
  \beta_{n, \ell} = \frac{G\;M_{\odot}\;L}{R_{\odot}^3\; \omega^2_{n, \ell}} =
\frac{\nu_{0, \ell}^2}{\nu_{n, \ell}^2}
\label{eq:beta}
\end{equation}
where $G$ is the gravitational constant, $M_{\odot}$ is the solar mass,
$R_{\odot}$ is the solar radius, $\omega$ is the cyclic frequency
($\omega=2\pi\nu$) and $L^2 = {\ell(\ell+1)}$. It can be reduced to the
square of the ratio of the frequency of the fundamental ($n=0$) for the given
degree to the mode frequency, since the frequency of the fundamental is given
by \citep{GoughEtal:80}
\begin{equation}
  \nu_{0, \ell}^2 = \frac{G\;M_\odot\;L}{4 \pi^2 R^3_\odot}
\end{equation}
Since for a given $\ell$, the f-mode frequency is smaller than any 
p-mode frequency, Equation~\ref{eq:beta} implies that $0 < \beta \leq 1$.

  Estimates of $\beta$ resulting from computing the adiabatic and non-magnetic
eigenfunctions using a standard solar model \citep{cd96} at the appropriate
atmospheric height for the MDI instrument and using a different upper boundary
condition than the one leading to Equation~\ref{eq:beta} gives a very similar
result. These calculations produce values that, above 2 mHz, decrease with
frequency slightly faster than predicted by Equation~\ref{eq:beta} (\ie,
15$\%$ smaller at 3 mHz).  However, the non-adiabatic nature of the
oscillations in the solar atmosphere are not taken into account in these
calculations nor is it taken into account in the derivation of
Equation~\ref{eq:beta}. Since the solar interior is non-adiabatic near the
surface and since the effect of the magnetic field, if any, is likely to be
mostly near the surface, the {\em real} value of $\beta$ might be somewhat
different.

  The estimate of the coefficients of the leakage matrix given by
equations~\ref{eq:theoleakr} to \ref{eq:theoleakp} is a useful guide but
remains oversimplified. Indeed, these equations do not take into account the
pixel size of the detector or the instrumental modulation transfer function
(MTF), effects that are substantial at high degrees. Therefore, rather than
integrating these equations, we opted to compute the leakage matrix
coefficients by constructing simulated images corresponding to the
line-of-sight contribution of each component of a single spherical harmonic
mode and decompose that image into spherical harmonic coefficients using the
numerical decomposition used to process the observations. Such an
approach allows us to easily include effects like the finite pixel size of the
detector as well as other instrumental effects like the instrumental MTF, a
plate scale error or some image distortion.

  Since this calculation is computationally expensive and since the leakage
coefficients vary smoothly with $\ell$ and $m$, it was carried out only for a
subset of modes, from which all the required values were interpolated.

\begin{figure}[!p]
\epsscale{1.}
\plotone{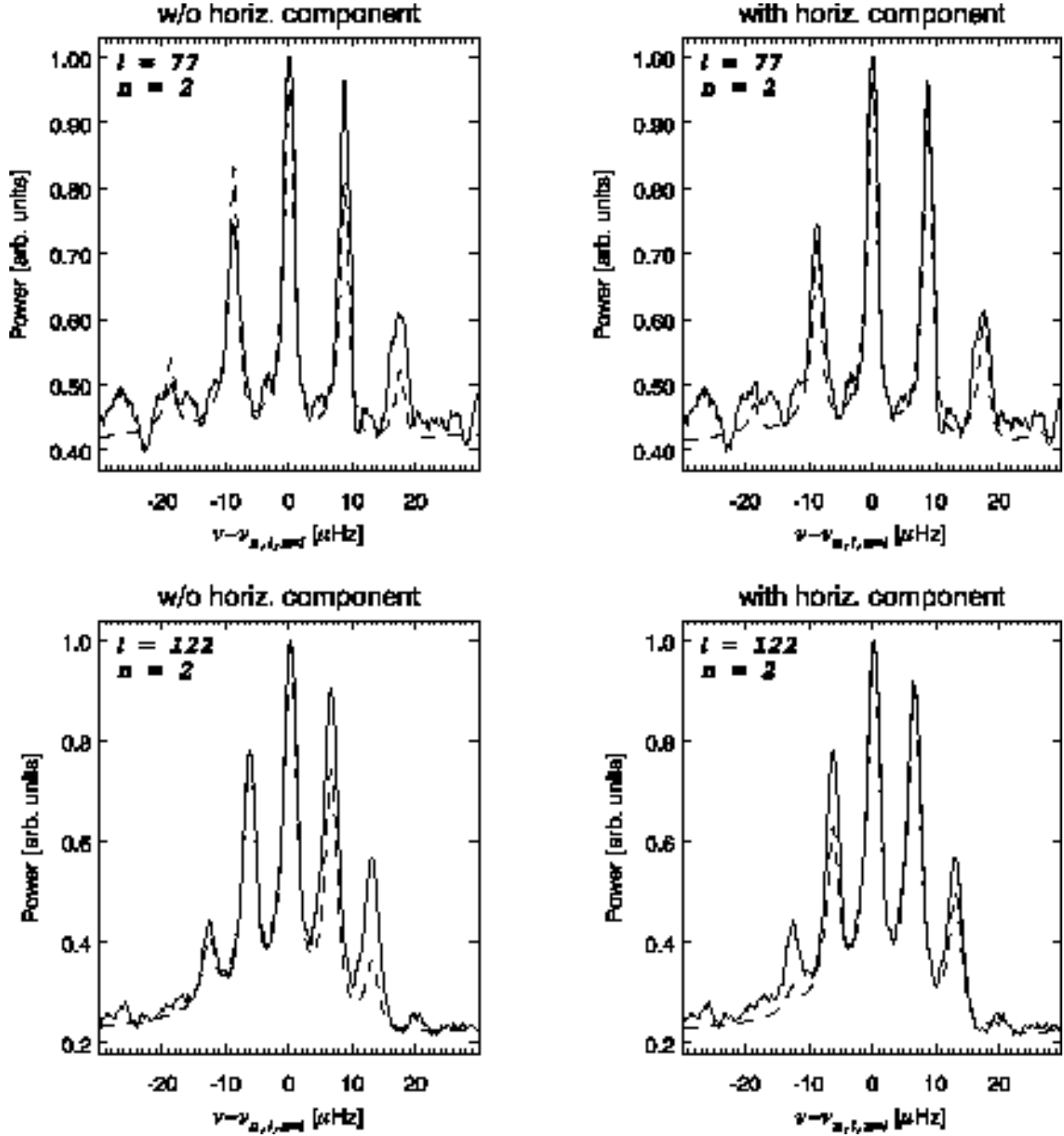}
\caption{Synthetic (dashed line) and observed (continuous line) sectoral limit
         spectra, {\em Dynamics} 1996, for $n=2$ and $\ell=77$ and $\ell=122$
         in the top and bottom rows respectively.  Synthetic spectra computed
         using leakage matrices without and with the inclusion of the
         horizontal component are plotted in the left and right columns
         respectively.  \label{fig:obspowbeta}}
\end{figure}

   Figure~\ref{fig:obspowbeta} shows some typical examples of observed and
synthetic spectra where the amplitude of the leaks corresponds to leakage
matrices calculated with and without including the horizontal component (right
and left panels respectively). This figure clearly shows not only that the
observed amplitudes of the spatial leaks are asymmetric, but the importance of
the contribution of the horizontal component in the leakage matrix
calculation, since the radial component of the leakage matrix is nearly
symmetric.  It should also be noticed that, although the complete leakage
matrix agrees in broad terms with the observed spectra, it does not agree in
detail, as already pointed out by \citet{korzennik99}. This was the motivation
for the work we present here, where we discuss in details how to best compute
the correct {\em effective} leakage matrix, \ie\ the one resulting from the
{\em actual} data analysis.

\subsection{Perturbation of the Leakage Matrix by the Solar Differential
            Rotation}
\label{sec:PertRot}

  \citet{woodard89} has computed the distortion of the eigenfunctions by a
slow, axisymmetric differential rotation. He showed that the distorted
eigenfunctions can be expressed as a superposition of the undistorted
ones. Under the assumption of axial symmetry, each of the superposed functions
has the same azimuthal order $m$, while symmetry with respect to the solar
equatorial plane implies that the values of the spherical harmonic degree
included in the superposition be either all odd or even. The Coriolis
acceleration is also assumed to be small which is valid for high-degree
p-modes ($\ell > 100$). This additional simplification reduces the perturbed
eigenfunctions to be approximately a superposition of unperturbed
eigenfunctions of the same radial order $n$.

  Also, the rotation rate $\Omega$ for high degree modes ($\ell > 100$
corresponds roughly to the radius range $0.9R_{\odot} < r < R_{\odot}$) is
primarily a function of latitude, and can be parameterized as:
\begin{equation}
  \Omega(\theta) = 2\pi(B_0 + B_2 \cos^2 \theta + B_4 \cos^4 \theta)
\label{eq:omega}
\end{equation}
where $B_0 = 473$ nHz, $B_2 = -77$ nHz and $B_4 = -57$ nHz
\citep{snodgrass90}, and $\theta$ represents the co-latitude.

The perturbed leakage matrix can be expanded in terms of the unperturbed one
as: 
\begin{equation}
  \tilde{C}_{\ell, m; \ell', m'} = 
  \sum_{\ell''} G_{\ell', \ell''} \, C_{\ell, m; \ell'', m'} 
\label{eq:rotdist}
\end{equation}
where:
\begin{equation}
G_{\ell', \ell''} = \frac{(-1)^p}{2\pi}
      \int_{-\pi}^{\pi} \cos\left[ p\theta + 
                                   \delta\Phi(\theta) \right] \, d\theta
\label{eq:Gll}
\end{equation}
\begin{eqnarray}
  \delta\Phi(\theta) & = & \frac{1}{2} \frac{m}{\partial\nu/\partial\ell}
                       \left[
                       \frac{1}{2} (   B_2\, y_{\ell'}^2
                                     + B_4\, y_{\ell'}^4) \sin\theta
\right. \nonumber \\ 
& & \left.-\frac{1}{16}   B_4\, y_{\ell'}^4  \sin(2\theta)
                       \right]
\label{eq:delPhi}
\end{eqnarray}
and where $G_{\ell', \ell''}$ is zero unless $\ell' - \ell''$ is even,
$2p=\ell'-\ell''$, $y_{\ell}^2 = 1 - (m/L)^2$.

\begin{figure}[!p]
\epsscale{.9}
\plotone{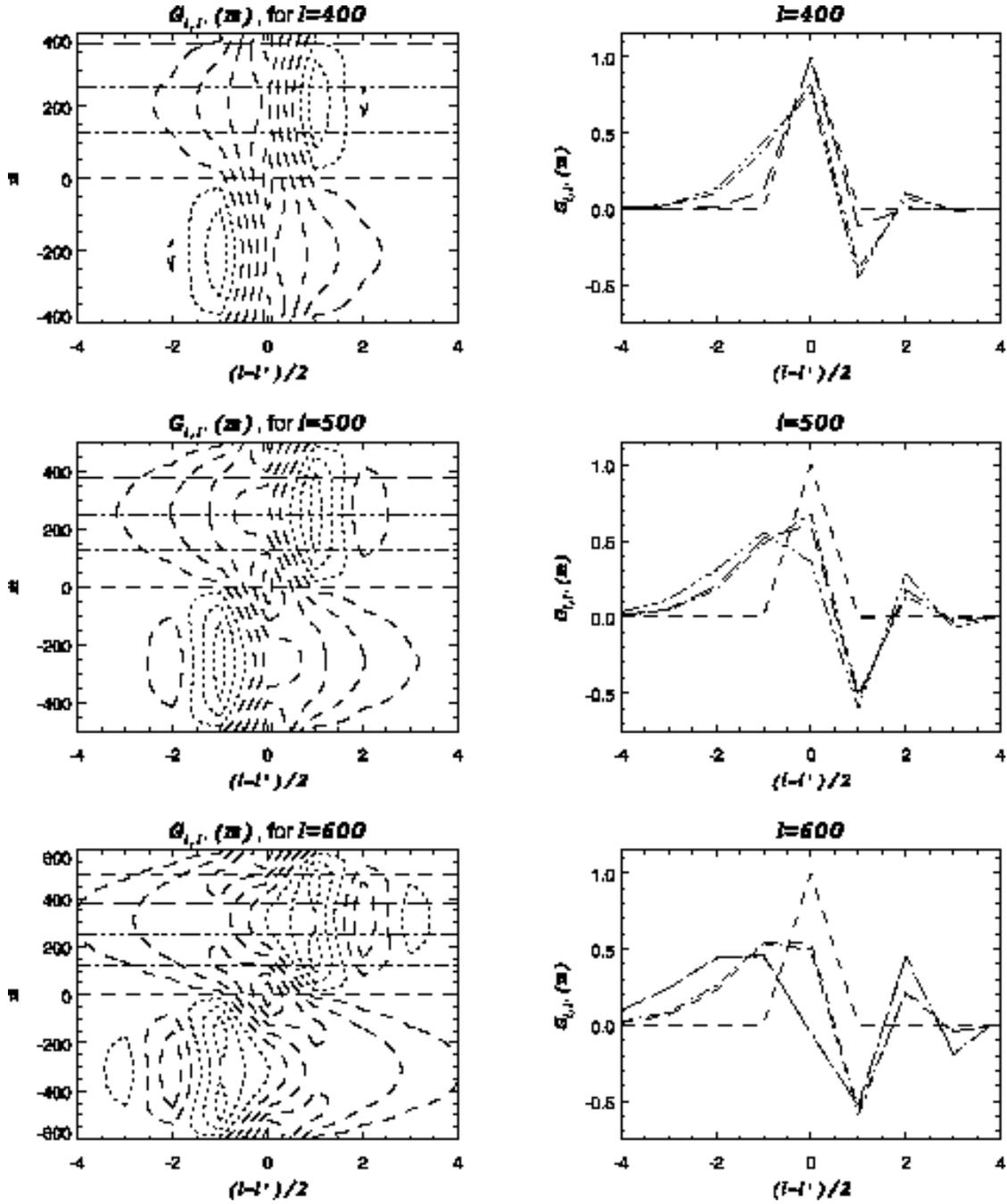}
\caption{Perturbation of the leakage matrix by the solar differential
         rotation: $G_{\ell,\ell'}(m)$, for $\ell=400, 500$ and $600$, at
         $n=3$, see definition in the text. The panels on the left show the
         $G_{\ell,\ell'}$ coefficients as a function of $p=(\ell-\ell')/2$ and
         $m$ (dash lines and dotted lines correspond to positive and negative
         contour levels respectively), while the panels on the right show
         profiles at fixed $m$, for the $m$ values indicated by the horizontal
         lines on the corresponding left panel. \label{fig:gll}}
\end{figure}

\begin{figure}[!p]
\epsscale{1.}
\plotone{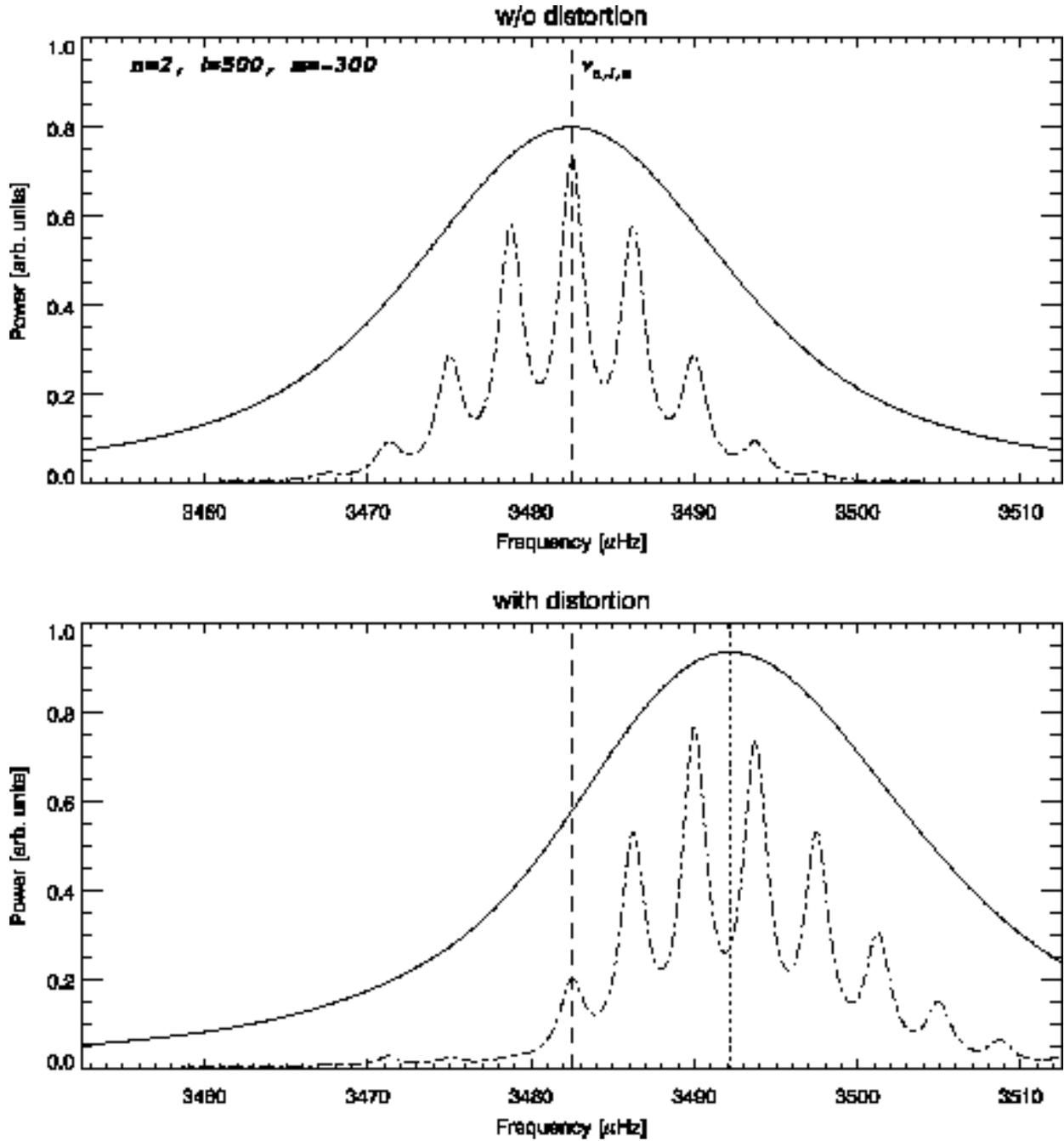}
\caption{Synthetic power spectrum computed for the $n=2$, $\ell=500$, $m=-300$
         mode without and with the inclusion of the effect of the distortion
         of the eigenfunctions by the differential rotation (top and bottom
         panel respectively), and neglecting the horizontal component of the
         leakage matrix. The dot-dashed lines show the power spectra that
         would be observed if the mode linewidth was one tenth of the observed
         value.  The vertical dash and dot lines indicate the mode frequency
         and the peak of the ridge respectively --- displaced in this case by
         some 9.7 \uHz.  \label{fig:woodard}}
\end{figure}

  This distortion is illustrated in Figures~\ref{fig:gll} and
\ref{fig:woodard}, where we show the variation of $G_{\ell', \ell''}$ with
respect to $m$ and $\ell-\ell'$ for selected values of $\ell$, and a synthetic
spectrum with and without distortion by differential rotation for $\ell =
500$. We should also add that \citet{Woodard:00} has shown that including
meridional circulation affects these calculations at the 10\% level.

\subsection{Theoretical Ridge Parameters}

  Synthetic power spectra were computed by overlapping individual modes as
they leak in the simulated spectrum according to:
\begin{eqnarray}
 \hat{P}_{n, \ell, m} (\nu) & = &
   \sum_{\ell'=\ell-3}^{\ell+3} \sum_{m'=m-3}^{m+3}
    \tilde{C}^2_{\ell, m; \ell', m'}
   (\beta_{n, \ell})\, {\times} \nonumber \\
 & &
 \frac{1 + \hat{\alpha}_{n, \ell} 
     (\hat{x}_{n, \ell', m'} - \hat{\alpha}_{n, \ell}/2)}{1 + \hat{x}^2_{n, \ell', m'}}
\label{eq:ridgemodel} \\
\hat{x}_{n, \ell', m'} & = &
  \frac{\nu - \hat{\nu}_{n, \ell', m'}}{\hat{\gamma}_{n, \ell}}
\end{eqnarray}
where $\hat{\nu}$, $\hat{\gamma}$ and $\hat{\alpha}$ are the mode frequency,
width and asymmetry parameter used to generate the model.
We used for ($\hat{\nu}$, $\hat{\gamma}$, $\hat{\alpha}$) values based on
observed parameters and used Equation~\ref{eq:beta} for the values of $\beta$.
We generated these synthetic spectra for values of $\ell$ between 100 and 900
-- but only every 5th $\ell$ up to 250 and every 10th $\ell$ above 250,
replicating the sampling we used in the fitting of the observed MDI spectra as
described in Section \ref{sec:ridgeFitting}.

  We generated two sets of synthetic power spectra, one where we did not
include the effect of the distortion of the eigenfunction due the differential
rotation (\ie, $G_{\ell',\ell''} = \delta_{\ell',\ell''}$), while the other
did include this effect.  We then used on these two sets of simulated ridges
the same fitting procedure we used on the actual observations, as described in
Section~\ref{sec:ridgeFitting} and Equation~\ref{eq:asymprof}, to compute {\em
ridge} frequencies and frequency splitting coefficients.  If our model is
correct and complete, our simulations should produce frequency and splitting
coefficient offsets smilar to the one resulting from MDI observations.

\begin{figure}[!p]
\epsscale{1.}
\plotone{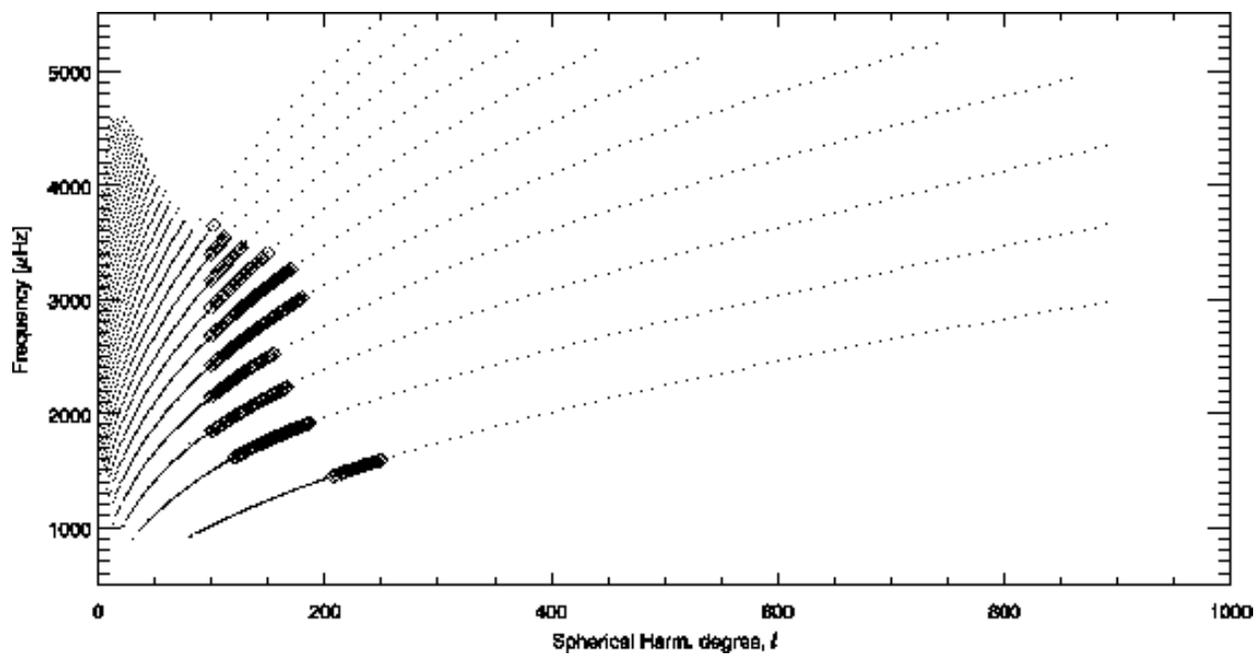}
\caption{This $\ell$--$\nu$ diagram illustrates the region of overlap between
         mode fitting and ridge fitting. The overlapping modes are marked as
         diamonds. \label{fig:overlap}}
\end{figure}

\begin{figure}[!p]
\epsscale{1.}
\plotone{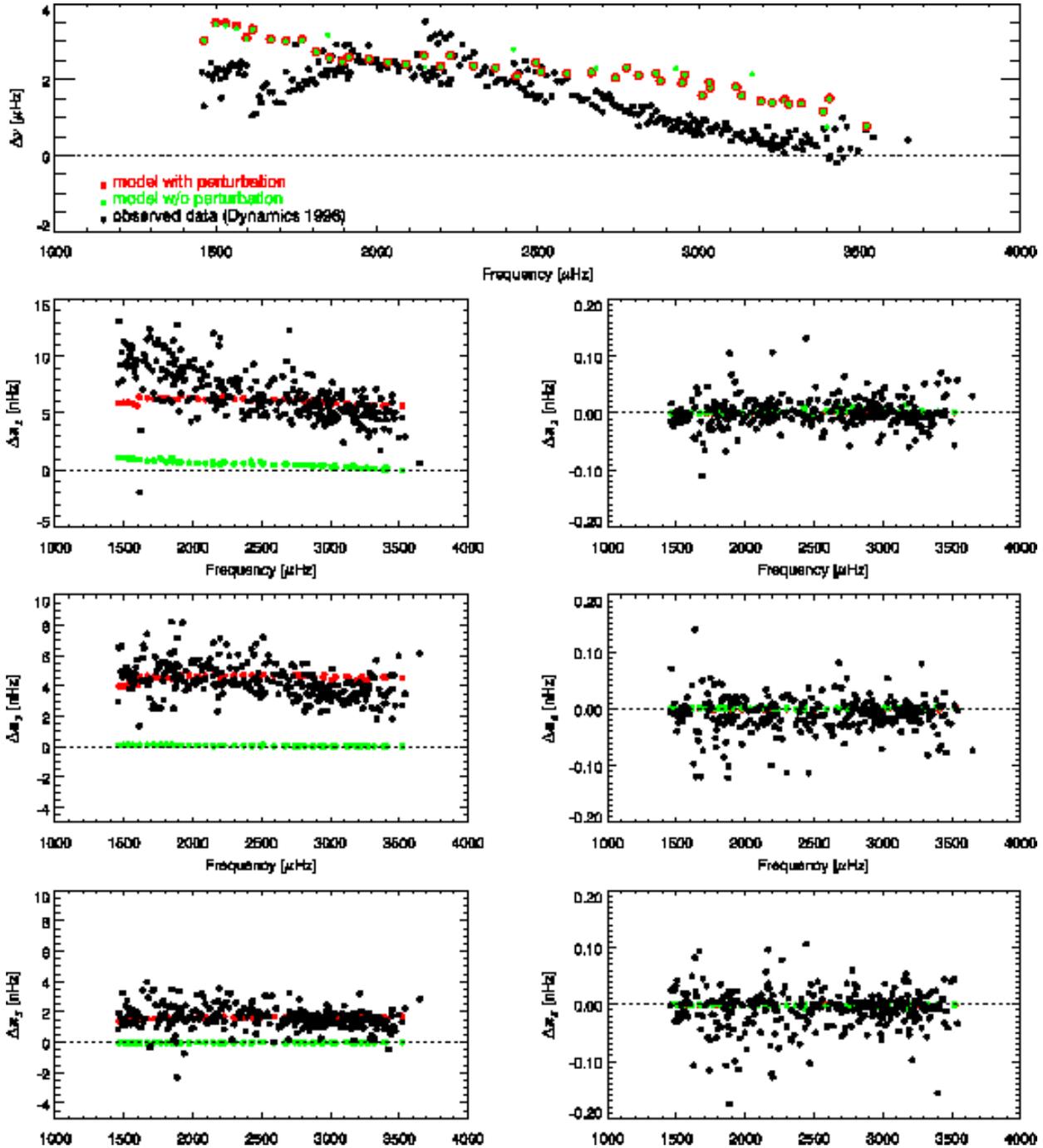}
\caption{
         Differences between ridge and mode frequencies, and ridge and mode
         splitting coefficients, as a function of frequency, for the {\em
         Dynamics} 1996 data (black points). These differences are compared to
         theoretical differences (for the same set of modes) computed from
         simulated power spectra without and with the inclusion of the effect
         of the distortion of the eigenfunctions by the differential rotation
         (green and red points respectively). 
\label{fig:del96a}}
\end{figure}

\begin{figure}[!p]
\epsscale{1.}
\plotone{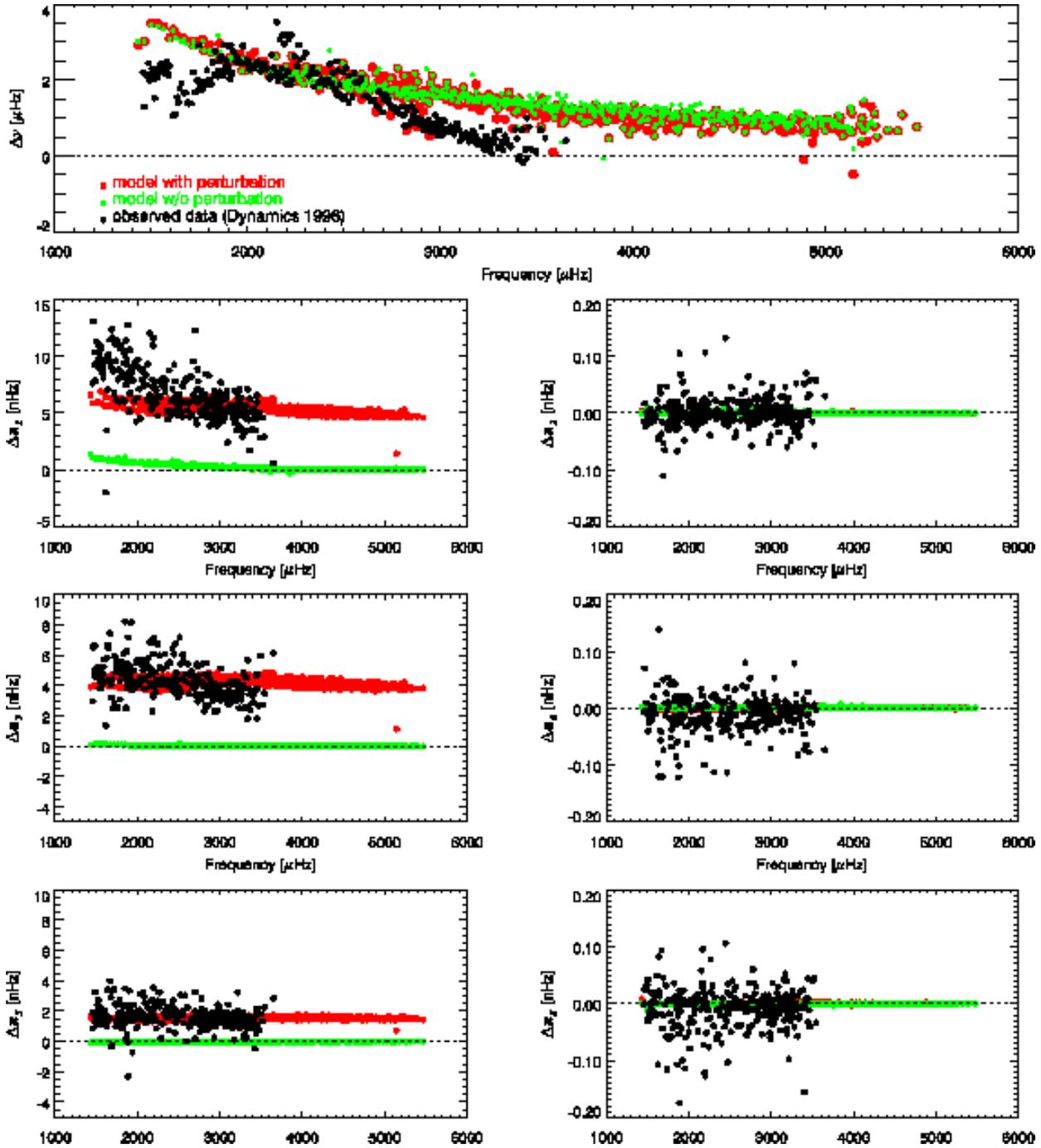}
\caption{Plot similar to Fig.~\ref{fig:del96a}: differences between ridge and
         mode frequencies, but with theoretical differences extending to
         spherical harmonic degrees up to $\ell=900$. \label{fig:del96b}}
\end{figure}

\begin{figure}[!p]
\epsscale{1.}
\plotone{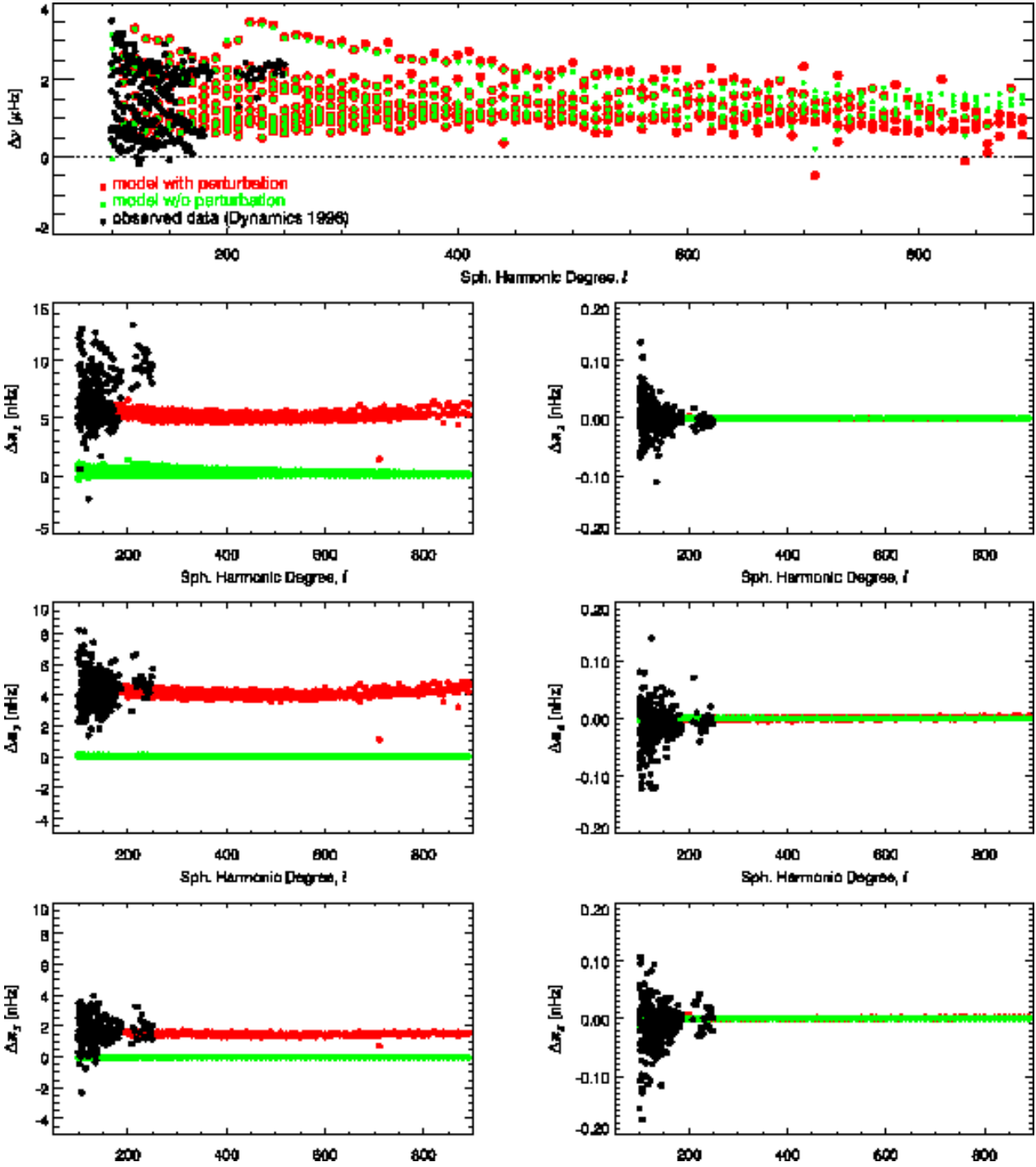}
\caption{Plot similar to Fig.~\ref{fig:del96b}: differences between ridge and
         mode frequencies, but shown as a function of spherical harmonic
         degree, $\ell$. \label{fig:del96bl}}
\end{figure}

   Figure~\ref{fig:overlap} illustrates, in an $\ell$--$\nu$ diagram format,
the region of overlap between mode fitting and ridge fitting. This range of
frequencies and degrees is useful to valide our methodology but too
restrictive to dare to carry out any extrapolation.
  In figures~\ref{fig:del96a} to \ref{fig:del96bl} we show frequency and
splitting coefficient offsets from our two sets of simulations and the
offsets from actual observations for intermediate-degree modes where there is
an overlap between ridge and mode fitting.  The theoretical differences agree
overall with the observations (\ie, magnitude and general trend), but not in
detail.  The effect of the distortion by the differential rotation is
significant, and affects the most the odd splitting coefficients. It explains
very well the discontinuity in the splittings coefficients seen as far back as
in \citet{korzennik90}, and more recently in \citet{rhodes99}.

\subsection{Influence of the Rotation Profile on the Splitting Corrections}

  The latitudinal dependence of the solar rotation rate, $\Omega(\theta)$,
must in principle be known to compute the perturbation of the leakage matrix
by the differential rotation (see Section~\ref{sec:PertRot}).
Equations~\ref{eq:Gll} and \ref{eq:delPhi} result from an explicit
parameterization of the near surface rotation with latitude given by
Equation~\ref{eq:omega}, and thus depend on the precise values adopted for
$B_2$ and $B_4$. Therefore the correction for the ridge rotational splittings
depends on a model of rotation, while the corrected splittings are meant to
allow us to derive that rotation profile.

  In order to estimate the effect of the rotation profile on the correction of
the splittings, we have synthesized theoretical ridges using three different
rotation profiles. Each set of ridges was then fitted and ridge frequency
splittings were computed, from which three sets of splitting coefficient
corrections were computed.

\begin{figure}[!p]
\epsscale{.85}
\plotone{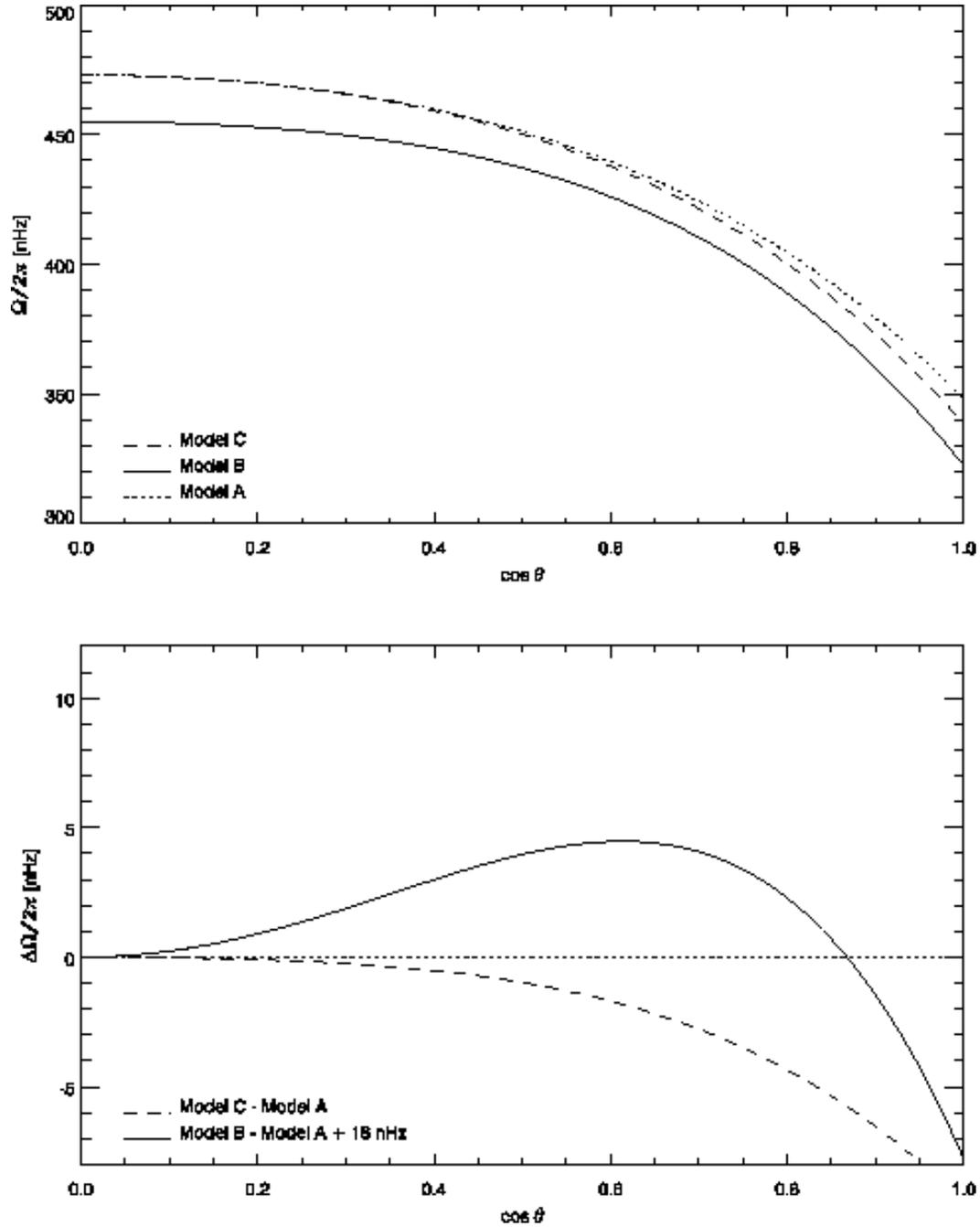}
\caption{Comparison of the latitudinal dependence of three different rotation
         profiles at the solar surface, used to estimate the effect of the
         differential rotation on the distortion of the eigenfunctions.
	       Model A corresponds to somewhat arbitrary values ($B_2=-75$ nHz,
	       $B_4=-50$ nHz); Model B to values derived from \citet{schouetal98}
	       ($B_2=-51$ nHz, $B_4=-81$ nHz); model C to values derived from
	       \citet{snodgrass90} ($B_2=-77$ nHz, $B_4=-57$ nHz).
         \label{fig:omega}}
\end{figure}

\begin{figure}[!p]
\epsscale{0.9}
\plotone{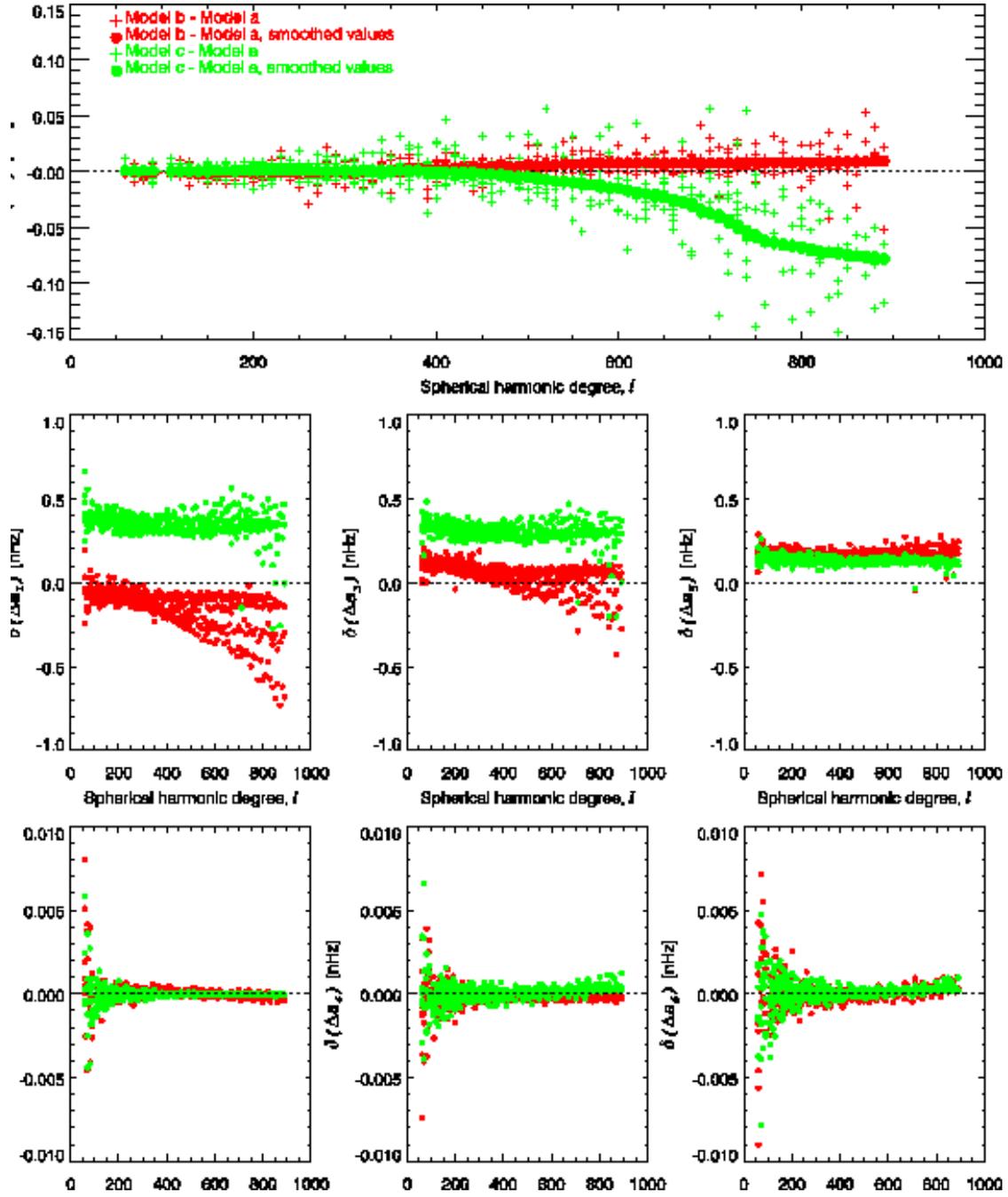}
\caption{Changes in the ridge to mode frequency and splittings coefficients 
         offsets caused by the distortion of the eigenfunctions by
         differential rotation, corresponding to the different rotation
         profiles illustrated in Figure~\ref{fig:omega}. 
         \label{fig:deltheo_omega}}
\end{figure}

  The first model, hereafter model A, used a somewhat arbitrary rotation
profile defined by $B_2=-75$ and $B_4=-50$ nHz, model B used a profile
inferred by \citet{schouetal98} using the MDI medium-$\ell$ splitting
coefficients, while model C used the profile given by \citet{snodgrass90}
based on the cross-correlation of Mt.\ Wilson magnetograms. The three profiles
are compared in Figure~\ref{fig:omega}.  Relative variations with respect to
model A are show in Table~\ref{tab:omegaSplit} and illustrated in
Figure~\ref{fig:deltheo_omega}. The first column of Table~\ref{tab:omegaSplit}
indicates the average relative change in the rotation profile, while the other
columns show the average and RMS of the respective parameter relative
corrections.

\begin{table*}[!t]
  \begin{center}
    \caption{Relative variation of ridge
             to mode corrections with respect to the rotation profile.
             \label{tab:omegaSplit}}\vspace{1em}
    \begin{tabular}{c|c|cccc}
      \hline
     & $\Delta\Omega$
                & $\Delta\nu$  & $\Delta a_1$ & $\Delta a_3$ & $\Delta a_5$ \\
     &    ~[\%] &         [\%] &         [\%] &         [\%] & [\%] 
      \\ \hline
 Model B $-$ A
     & $-4.4 $ & $+0.22 \pm 0.05 $ & $ -2.52 \pm 0.10$~\tablenotemark{(a)} & 
                 $+1.28 \pm 0.09 $~\tablenotemark{(b)} & $+10.55 \pm 0.08 $ \\
 Model C $-$ A
     & $+6.2 $ & $-1.65 \pm 0.52 $ & $ +6.47 \pm 0.06 $ & 
                 $+7.16 \pm 0.06 $ & $ +8.28 \pm 0.05 $ \\
      \hline
      \end{tabular}
\\
~$^{a}$ {computed for $n > 1$ only} \\
~$^{b}$ {computed for $n > 0$ only}   
  \end{center}   
\end{table*}

  This table shows that the frequency correction is affected by the choice of
the rotation profile, but at a relative level substantially smaller than the
relative change in the rotation profile. We should also note that the relative
change in the frequency correction is a function of degree, as it grows larger
at larger degrees. Table~\ref{tab:omegaSplit} also shows that the correction
in the $a_i$ coefficients is affected by the choice of $\Omega(\theta)$ by a
relative amount comparable to the relative change in the rotation
profile. Again we should point out that the change shows a variation with
degree and frequency. It is most pronounced for $a_1$ for the f-mode, but is
also substantial for $n=1$, while for $a_3$ only the f-modes show a
substantial effect. Of course some of this is the direct result of the
variation of $\Delta\Omega$ with latitude between the different models
used. Finally we should point out that the solar rotational profile is known
with a better precision than the numbers listed in Table~\ref{tab:omegaSplit}
suggest, resulting in smaller effects.

\section{Estimate of $\beta$ from Intermediate Degree Observations}

  \citet{korzennik99} presented a methodology to estimate $\beta$ from
intermediate degree observations and speculated, on the basis of a single
season of {\em Dynamics} observations, that Equation~\ref{eq:beta} might not
be satisfied. This methodology proceeds as follow: First, the leakage
asymmetry is quantified using a single factor, $S_x$, defined by:
\begin{equation}
S_x (n, \ell) = \frac{\sum_{\ell'=\ell-3}^{\ell+3} \; (\ell' - \ell) \;
  A_{n, \ell'}} {\sum_{\ell'=\ell-3}^{\ell+3} \; A_{n, \ell'}}
\end{equation}
using the values of $A_{n, \ell'}$ fitted to the sectoral limit spectra as
described in Section~\ref{sec:limitSpectra} and Equation~\ref{eq:7comp}. Next,
a correspondence between $S_x$ and $\beta$ is established using the complete
leakage matrix, namely:
\begin{equation}
S_x (\beta, \ell) = \frac{\sum_{\ell'=\ell-3}^{\ell+3} \; (\ell' - \ell) \;
  C_{\ell, \ell; \ell', \ell'}^2(\beta)} {\sum_{\ell'=\ell-3}^{\ell+3} \;
  C_{\ell, \ell; \ell', \ell'}^2(\beta)}
\end{equation}
Finally, using this correspondence \cite[shown to be nearly independent of
$\ell$ in Figure~4 of][]{korzennik99} the values of $S_x$ are converted to
observational estimates of $\beta$, hereafter refered as $\tilde{\beta}$.

\begin{figure}[!p]
\epsscale{1.5}
\plottwo{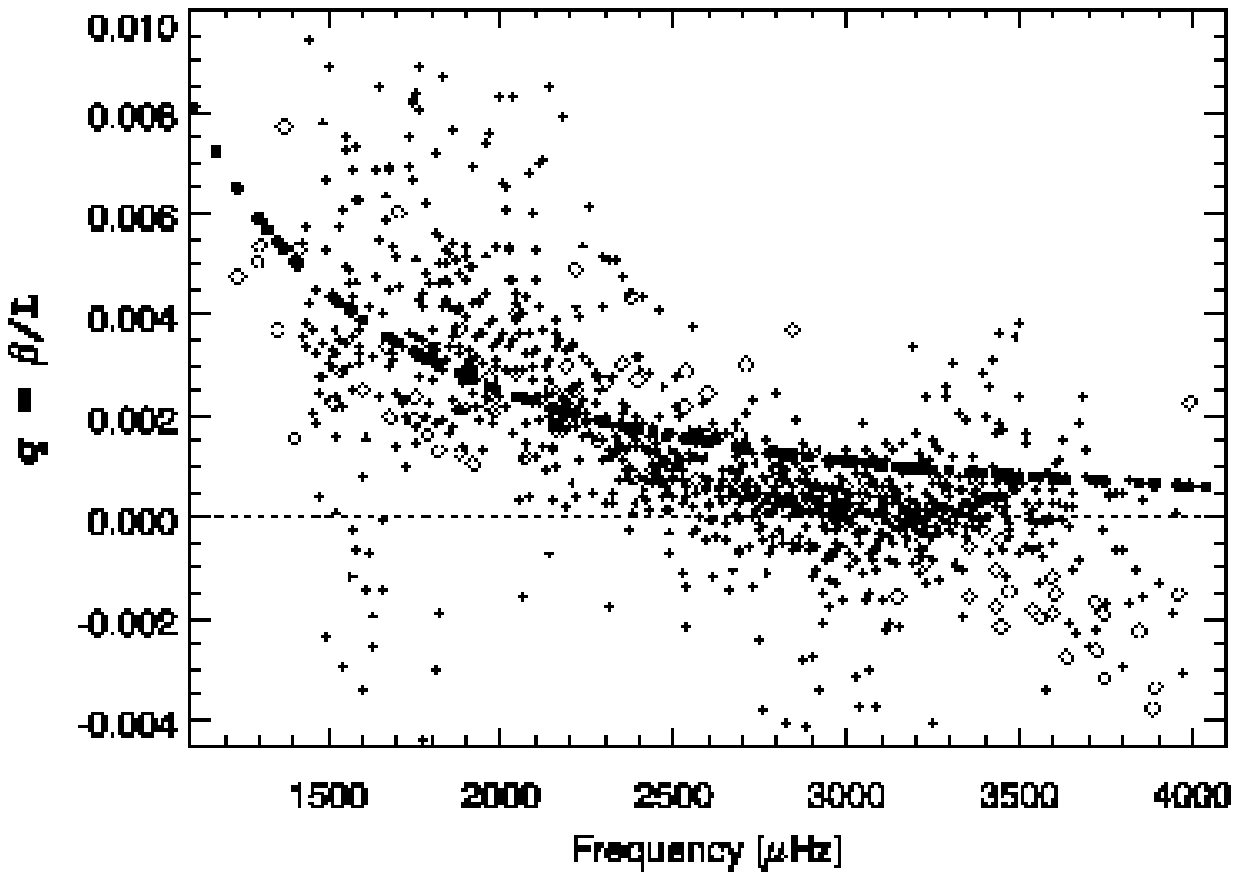}{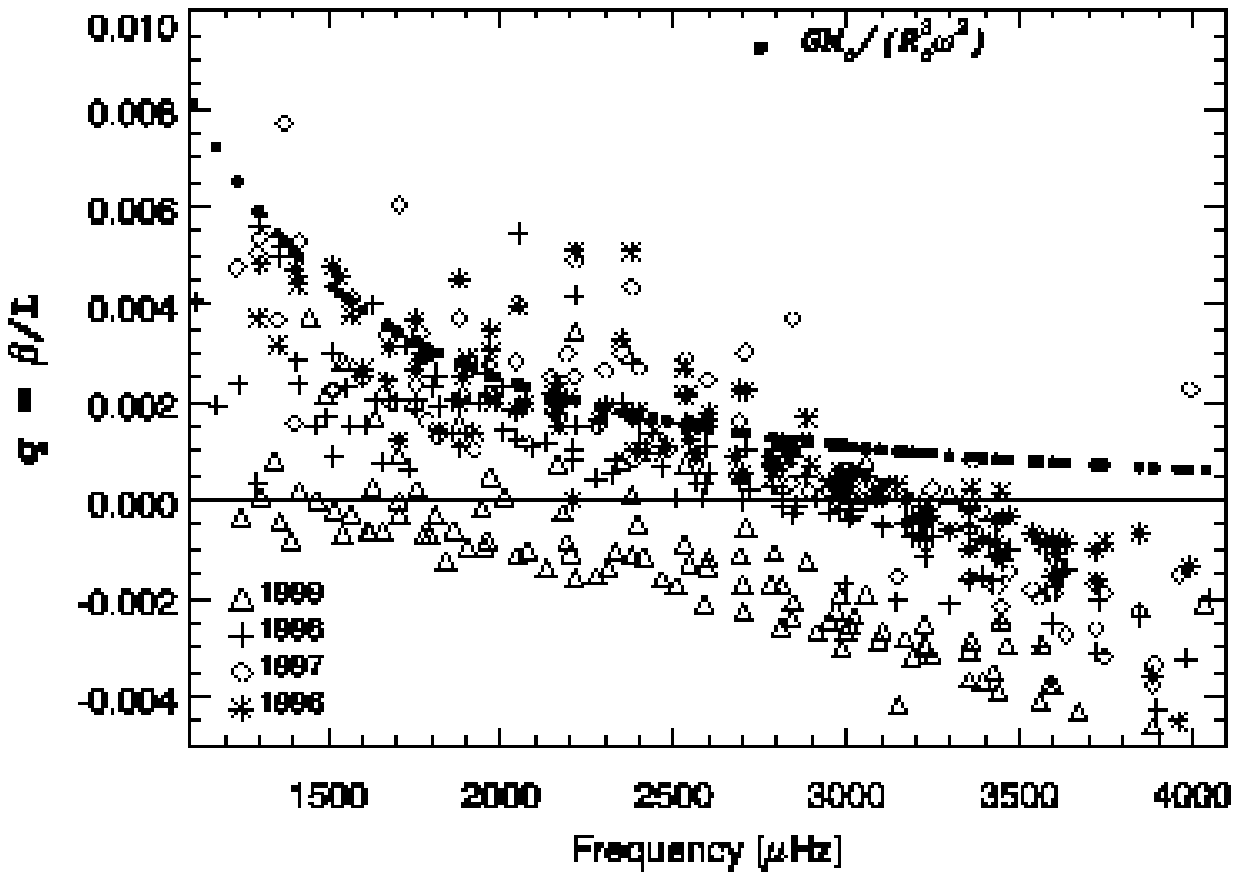}
\caption{Top panel: observed $q = \Frac{\tilde{\beta}}{L}$ estimated using the
         {\em Dynamics} 1997 data set using $S_x$ (diamonds) or using the
         observed ridge frequency (crosses), see text for detailed
         explanations. Bottom panel: observed $q$ for all four {\em Dynamics}
         data sets -- estimated using $S_x$. The theoretical value of
         $q=\Frac{G\;M_{\odot}}{R_{\odot}^3\; \omega^2}$ is also shown for
         comparison (filled dots). \label{fig:beta97}}
\end{figure}

  Alternatively, one can estimate $\tilde\beta$ using a forward approach,
where $\tilde\beta$ is adjusted as to produce theoretical ridge frequencies
--- by fitting synthetic spectra --- that matches the observed ridge
frequencies. Both methodologies agree, as shown in Figure~\ref{fig:beta97},
for a single year of {\em Dynamics} data.

  When this method was applied to successive years, using the same leakage
matrix, we found that $\tilde\beta$, while agreeing for 1996 and 1997, changed
noticeably in 1998 and substantially in 1999. Since it is highly unlikely that
$\beta$, the horizontal-to-vertical displacement ratio, would actually change
over the years, we came to the obvious conclusion that observed discrepencies
--- whether expressed in term of leakage asymmetry or frequency or splittings
differences --- resulted from some missing component in our modeling.

  Thus instead of trying to adjust the leakage matrix with some {\em ad hoc}
estimate of $\tilde\beta$, one must improve the leakage matrix estimate using
physically motivated contributions, as to produce an unbiased determination of
the high-degree mode frequencies.  In fact, we show in the next sections that
the differences of $\tilde\beta$ between 1998 and 1999 results from a time
variation of the plate scale error in the decomposition of the MDI images.

  We should also point out that \citet{schoubogart98} while investigating flow
and horizontal displacements from ring diagrams, using MDI {\em Dynamics}
observations, found the ratio $\beta$ to be in good agreement with the theory
\cite[see Figure 7 of][]{schoubogart98}. \citet{SchmidtEtal:99} have argued
that the observed velocity vectors in the photosphere appeared to be more
vertical than predicted, using MDI velocity power spectra to study the
centre-to-limb variation. But \cite{RhodesEtal:01} concluded that the ratio
$\beta$ is very close to the theoretical value, by studying $m$-averaged
intermediate degree GONG power spectra.

\section{Instrumental Effects}

  The MDI instrument has been very stable over the years. Nonetheless,
continuous exposure to solar radiation has increased the instrument front
window absorption resulting in a long term increase in its temperature
\citep{bush01}. Moreover, there is indirect evidence for a temperature
gradient from the center to the edge of the front window, associated with that
temperature increase. This gradient is believed to cause a small curvature
that converts the front window into a weak lens.

\begin{figure}[!p]
\epsscale{1.}
\plotone{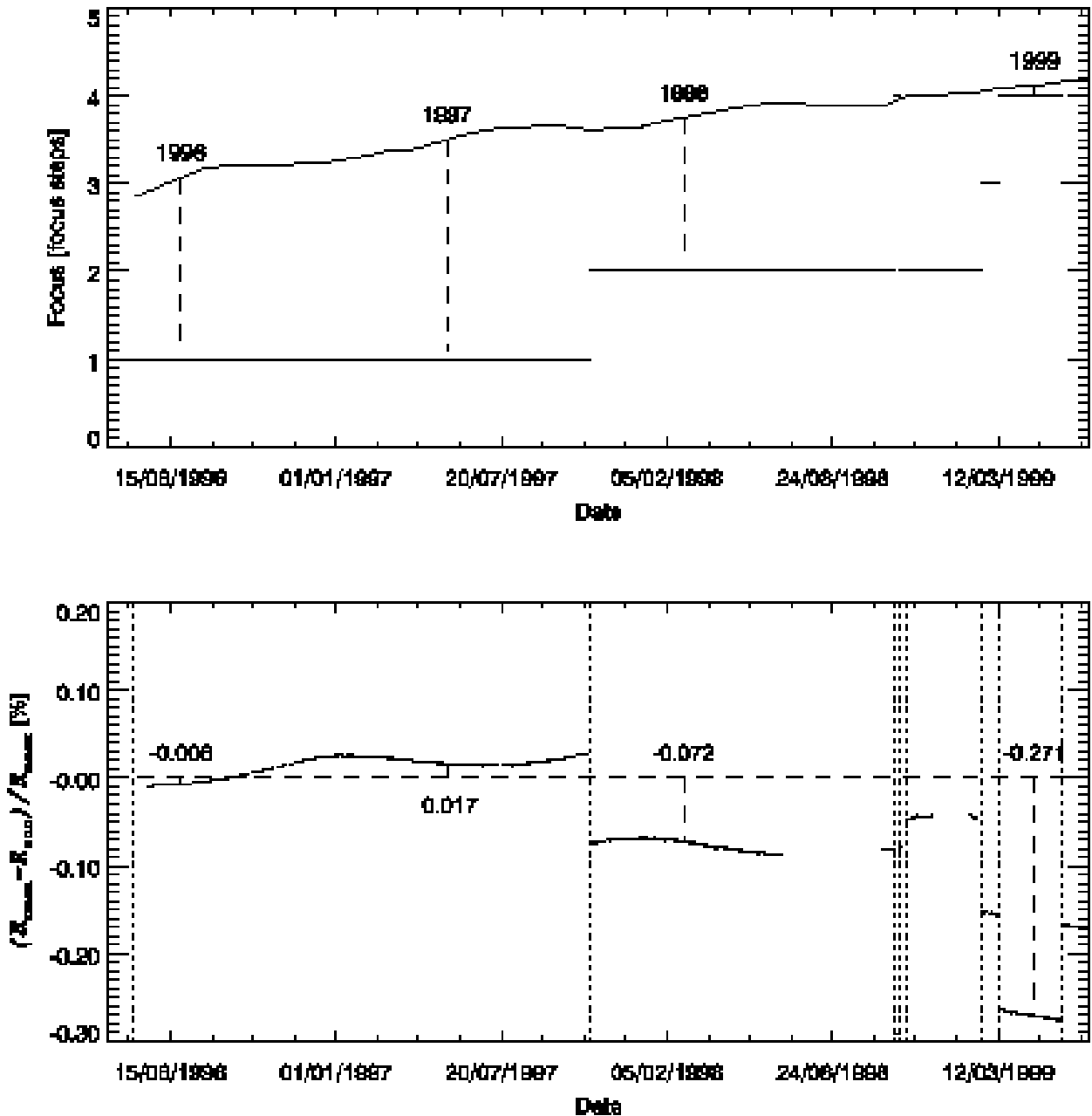}
\caption{Top panel: variation of the MDI mean focus as a function of time, as
         computed by \cite{bush01}. The horizontal lines indicate the focus
         position of the instrument at the time, and therefore the difference
         between the measured focus and the instrument focus position gives
         the amount of defocus for each {\em Dynamics} data set (vertical
         dashed lines).  Bottom panel: resulting plate scale error, \ie,
         relative difference between the observed mean solar image radius
         ($R_{\rm mean}$), calculated from each image using {\tt HGEOM}, and
         the radius determined using the initial plate scale ($R_{\rm sun}$,
         see text for a more detailed description).  \label{fig:focus}}
\end{figure}

   Evidence of this is a drift in the instrument focus (see top panel of
Figure~\ref{fig:focus}). The short term small variation in the focus is
correlated with the annual temperature change of the front window due to the
satellite orbit around the Sun \cite[see][for further details]{bush01}.  

  Image focusing on the MDI instrument is controled by inserting (or not)
blocks of glass of varying thickness into the light path
\cite[see][]{scherrer95}.  There are 9 possible focus positions, with one
focus step corresponding to approximately a third of a wave. Most of the time,
the image was slightly defocused on purpose to match the detector
resolution. The actual configuration in which the instrument was operating is
shown in the top panel of Figure~\ref{fig:focus}, with the corresponding
amount of defocus.

  A direct consequence of a change in focus is a variation of the image scale
at the detector.  The bottom panel of Figure~\ref{fig:focus} shows the
variation of the ratio of the radius calculated using the initial plate scale
relative to the mean image radius. The mean image radius is the instantaneous
average of the image semi-major and semi-minor axis (see below), while the
initial plate scale refers to a fixed estimate of the image radius computed in
early 1996.  The discontinuities are due to changes in the instrument focus
position and are indicated by vertical dashed lines.  The observed annual
variations as well as the small but systematic increase with time are
correlated with the aforementioned changes in the front window temperature.

  Besides the image scale, another relevant aspect of the optical
characteristics of the instrument is the distortion of the solar image.  The
shape of the solar limb on the detector can be approximated by an ellipse with
an eccentricity 100 times larger than the actual eccentricity of the sun.  The
difference beween the semi-major and the semi-minor axis of the MDI full-disk
image is $0.50 \pm 0.04$ pixels, while the mean radius is 488.45 pixels. The
orientation is such that the semi-major axis forms an angle of $34\Deg \pm
1\Deg$ with the equator in the clockwise direction.
Although the changes in the MDI focus affect the solar image ellipticity,
these variations are rather small, resulting in a distorted but relatively
stable image.

  Let us point out that the spherical harmonic decomposition of all the {\em
Dynamics} data has been carried out assuming a circular solar limb, with no
image distortion.  The image radius used for this decomposition was computed
using the value corresponding to the {\em initial} estimate of the MDI
instrument plate scale and rescaled according to the distance to the sun given
by the spacecraft orbit ephemeris.
The variations introduced by the change of focus position and the effect of
the front window were not all known or well characterized and thus were not
taken into account at the time the spatial decomposition was computed.
Since the computation of 500,000 coefficients for some 468,000 images -- each
1024$\times$1024 pixels -- is a rather intensive computing task, the spatial
decomposition has not yet been recomputed to take into account the known image
scale variations.

\subsection{Effects of Plate Scale Error}
\label{sec:PlatScalErr}

  Variations of the mean image radius relative to the initial plate scale,
even at the $10^{-3}$ level (as illustrated in the lower panel of
Figure~\ref{fig:focus}) will somewhat affect the amplitude of the spatial
leaks (see Equation~\ref{eq:klCr}) and must be taken into account in the
leakage matrix calculation.

\begin{figure}[!p]
\epsscale{2.}
\plottwo{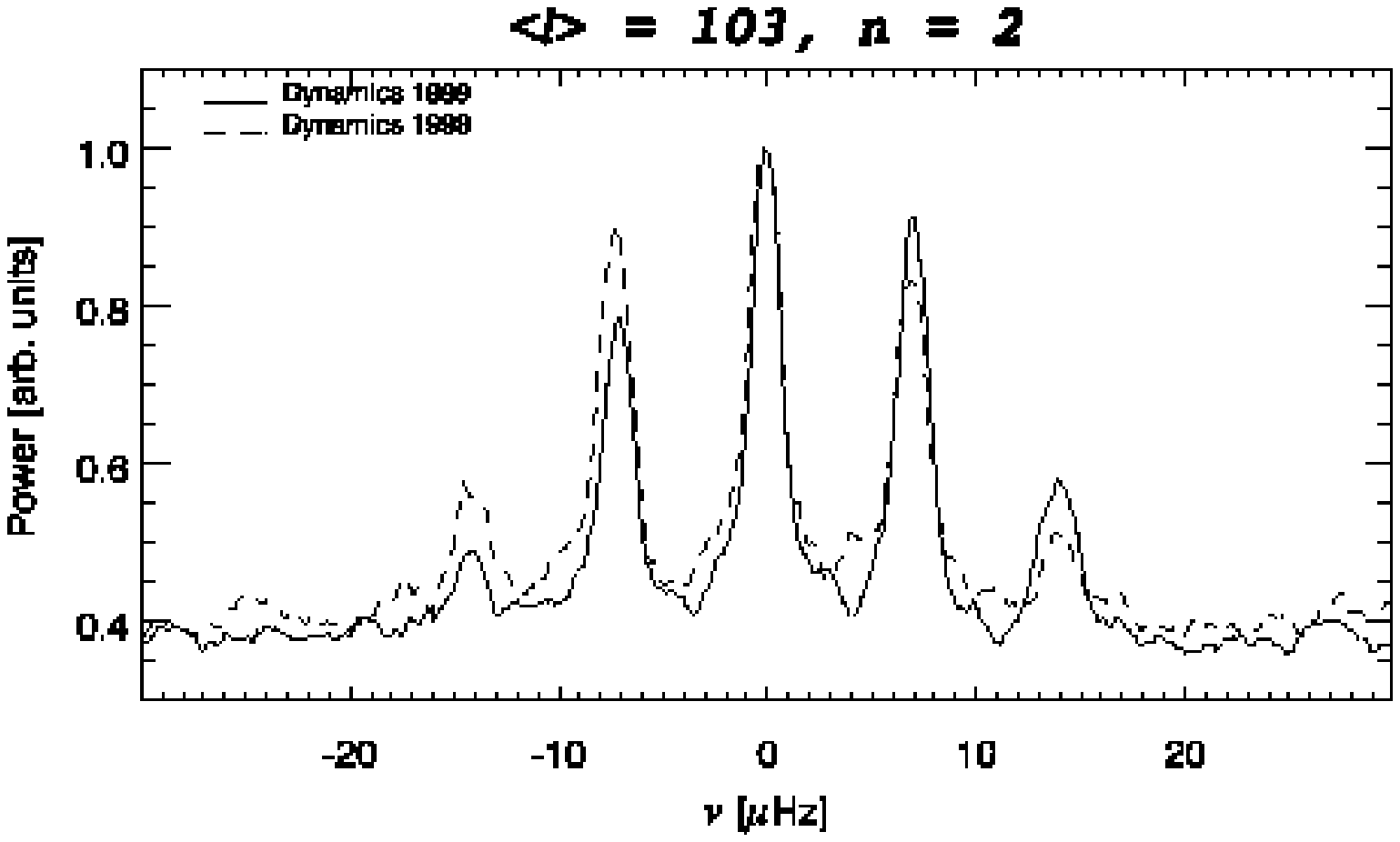}{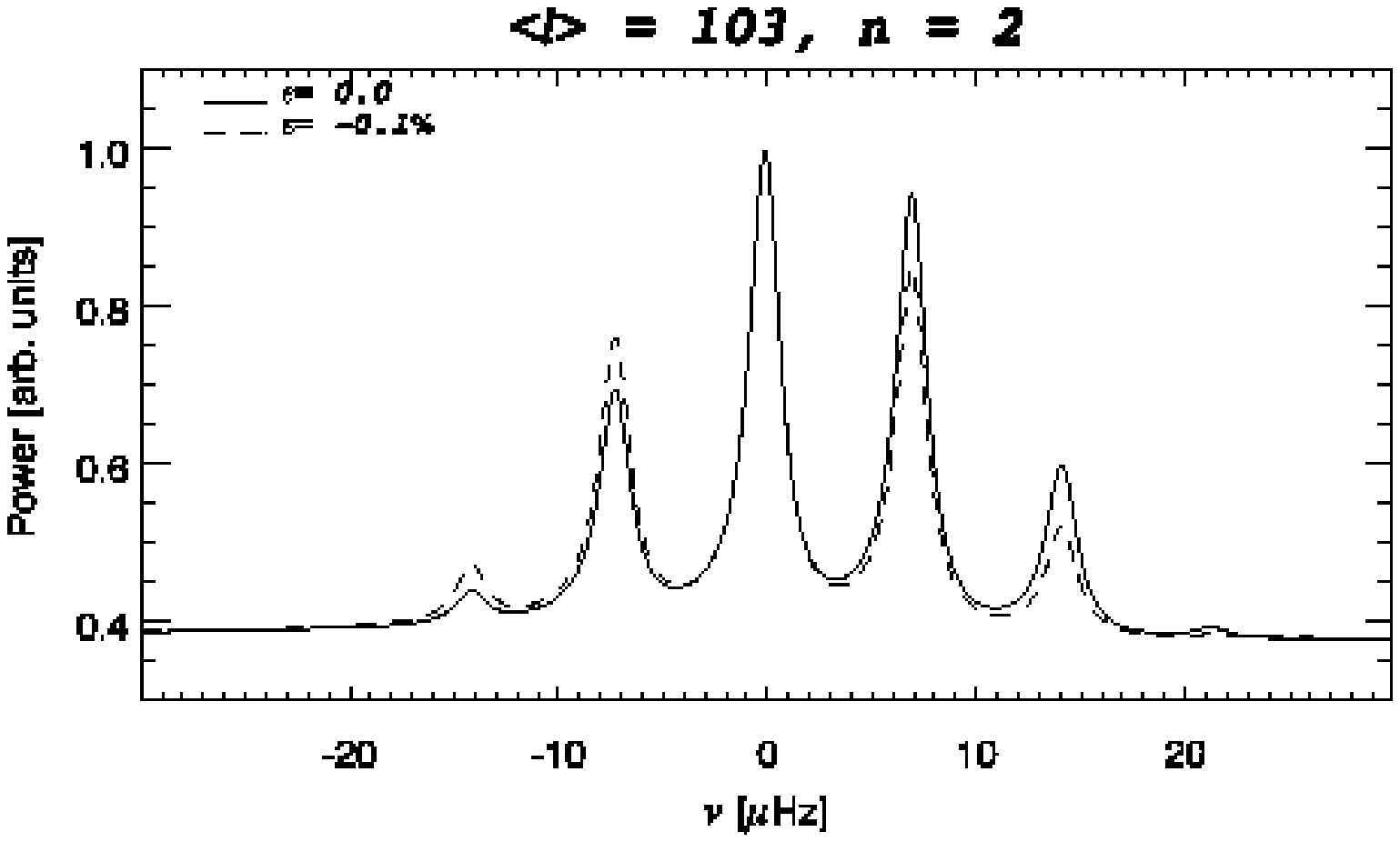}
\caption{Comparison of observed and simulated sectoral limit power spectra
         (top and bottom panels respectively). The top panel compares {\em
         Dynamics} 1998 to 1999 spectra while the bottom panel compares
         synthetic spectra computed using leakage matrix without and with the
         inclusion of a plate scale error of $-0.1\%$ (\ie, solar image radius
         is $0.1\%$ larger than the one assumed in the spatial decomposition).
	 \label{fig:spec_rad}}
\end{figure}

  In Figure~\ref{fig:spec_rad} we compare the computed limit spectrum (see
Section~\ref{sec:limitSpectra}) estimated for $n=2$ and $\ell=103$ for 1998
and 1999, in the upper panel. It clearly shows that the leakage asymmetry
changes sign between both epochs ($A_{\ell-1} < A_{\ell+1}$ in 1998, while
$A_{\ell-1} > A_{\ell+1}$ in 1999).  The lower panel of
Figure~\ref{fig:spec_rad} shows simulated power spectra computed using a
leakage matrix with and without a plate scale error of $-0.1\%$. As
expected\footnote{In this case, the leakage asymmetry is the result of two
factors: the contribution of the velocity horizontal component that leaks more
power at higher $\ell$ and a negative plate scale error that has the opposite
effect.}, the plate scale error causes a change in the leakage asymmetry
similar to the change observed between 1998 and 1999 observations.

  The plate scale errors for the different {\em Dynamics} epochs are given in
Table~\ref{tab:tableEpsilon}. Note also that it is the size of the equatorial
radius that will dictate the amplitude of the leaks for the sectoral
spectra. Since the shape of the solar image on the detector is nearly an
ellipse, the radius at the equator is slightly larger ($\approx .02\%$) than
the mean radius.


\begin{table*}[!t]
\begin{center}
\caption{Mean Plate Scale Errors
         \label{tab:tableEpsilon}}
\begin{tabular}{llc}
{Epoch}  & {Plate scale error, $\epsilon$} \\\hline\hline
      {1996} & $-0.0068 \pm 0.00013\%$  \\
      {1997} & $+0.0155 \pm 0.00005\%$  \\
      {1998} & $-0.0730 \pm 0.00018\%$  \\
      {1999} & $-0.2701 \pm 0.00022\%$  \\\hline
\end{tabular}
\end{center}
\end{table*}

\begin{figure}[!p]
\epsscale{0.8}
\plotone{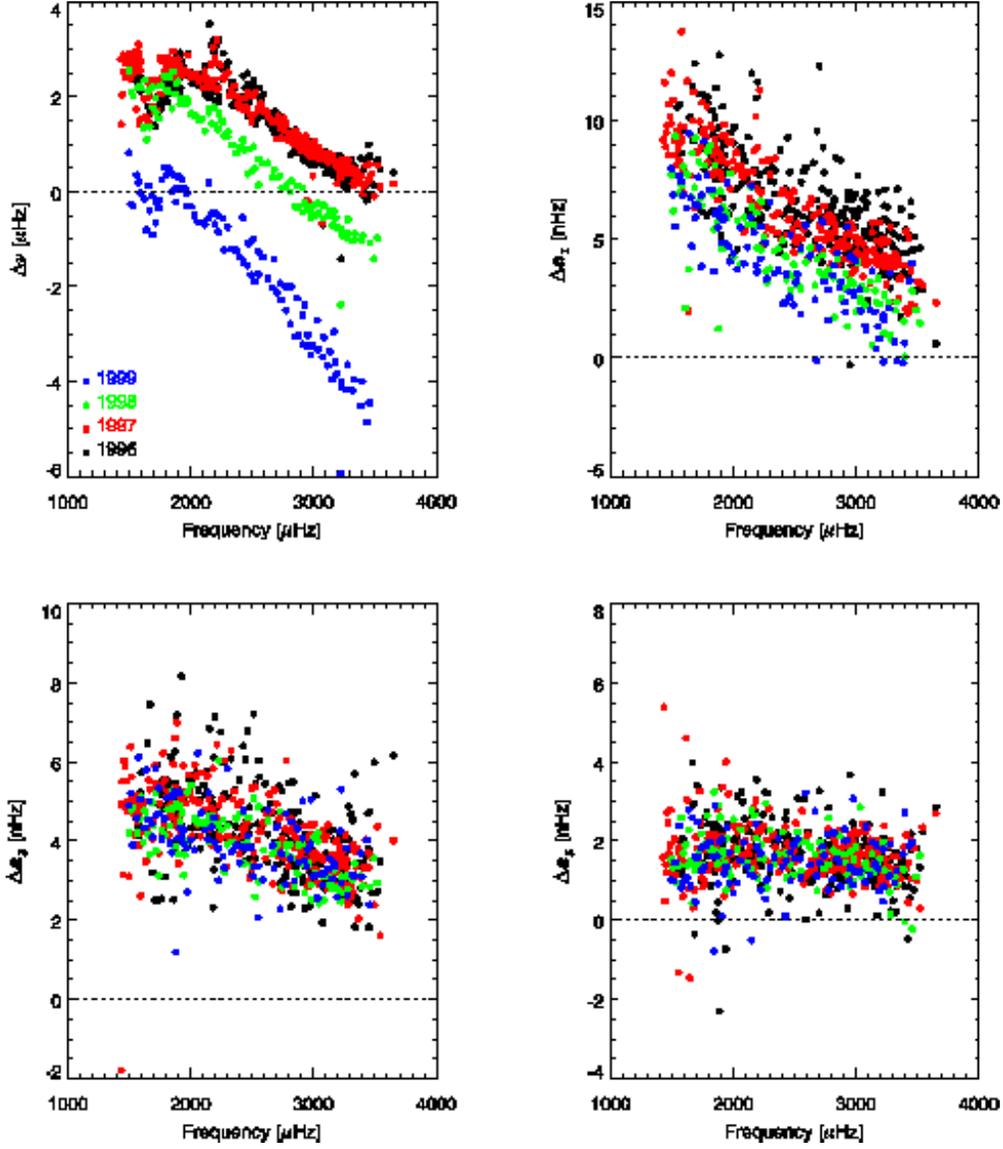}
\caption{Comparison of the differences between ridge and mode frequencies, and
         differences between ridge and mode odd splitting coefficients, as a
         function of frequency, for all the {\em Dynamics} data sets.
         \label{fig:plsclobs}}
\end{figure}

  The effect of the plate scale error can be clearly seen in the observed
ridge frequency offsets presented in Figure~\ref{fig:plsclobs}. The offsets
between ridge and mode frequencies for 1996 and 1997 are very similar, while
for 1998 and especially for 1999 they become substantial.  The values of
$\Delta a_1$ (\ie, the first {\em ridge} splitting coefficient offset)
also show some variation from year to year, although smaller.

\begin{figure}[!p]
\epsscale{0.8}
\plotone{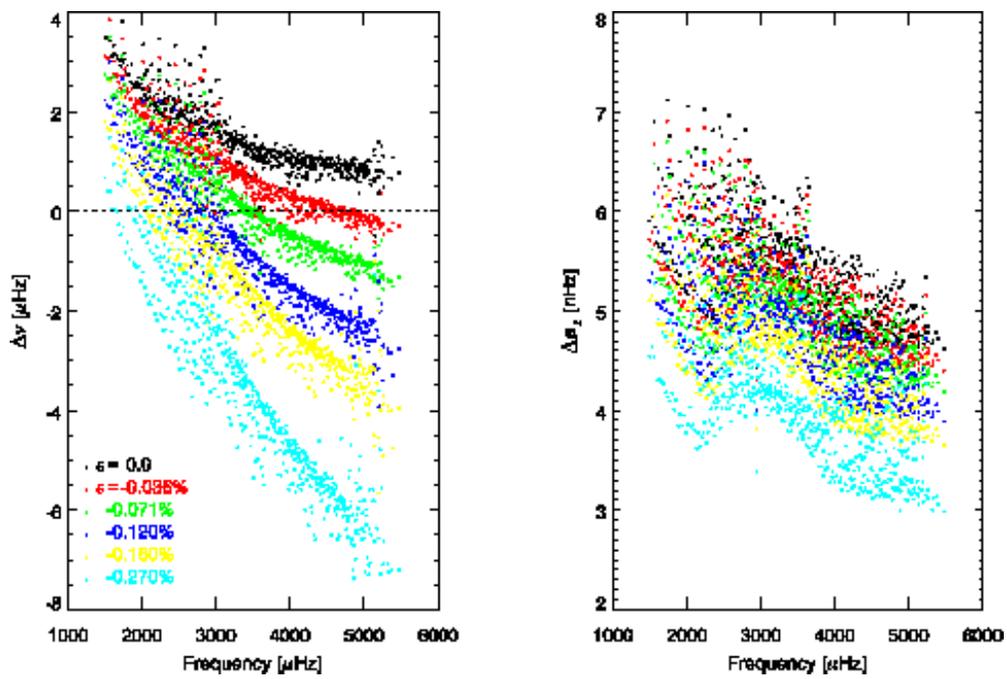}
\caption{Theoretical differences between ridge and mode frequencies (left
         panel) and between ridge and mode $a_1$ splitting coefficients (right
         panel), shown as a function of frequency, resulting from fitting
         synthetic spectra computed with leakage matrices with various plate
         scale errors ($\epsilon = 0, 0.036, 0.071, 0.120, 0.160$ and
         $0.270\%$).  \label{fig:frqpse}} 
\end{figure}

  We have modeled the effect of a plate scale error by computing the leakage
matrix resulting from decomposing images using the wrong radius, and including
the effect of the distortion of the eigenfunction by the differential
rotation.  These leaks were then used to generate synthetic power ridges that
were fitted using the same ridge fitting procedure described earlier.
Figure~\ref{fig:frqpse} shows the frequency offsets and rotational
splittings first coefficient offsets for different plate scale errors.
Such simulations reproduce rather well the variations observed in the data,
including the changes between epochs. Note that at this point we have not yet
included any image distortion.

  The effect of a small plate scale error on the ridge frequency can be
approximately modeled as a wavenumber scaling error in the spatial
decomposition, namely $\ell_{\epsilon} = \ell (1-\epsilon)$, causing an
apparent frequency shift of the ridge. Using a linear approximation, valid for
small $\epsilon$, this can be written as:
\begin{equation}
\Delta\nu   =   \nu(\ell_{\epsilon}) - \nu(\ell)
            =   \frac{\partial\nu}{\partial\ell} \Delta\ell
            =   - \frac{\partial\nu}{\partial\ell} \,\epsilon\,\ell
  \label{eq:dnuepsell}
\end{equation}

In practice the observed frequency differences result from the combination of
a plate scale error and some other terms. Thus, the variation of this
frequency difference with epoch, if the only change is a plate scale error,
can be measured, and corresponds to:
\begin{eqnarray}
  \zeta(P_1,P_2,P_3) & \stackrel{\mbox{def}}{=} &
  \frac{\Delta\nu(P_2) - \Delta\nu(P_1)}{\Delta\nu(P_3) - \Delta\nu(P_1)} \\
  & = &
- \frac{ \epsilon(P_2) -  \epsilon(P_1)}{ \epsilon(P_3) -  \epsilon(P_1)}
  \label{eq:dnupse}
\end{eqnarray}
where $\epsilon(P_i)$ is the plate scale error for the epoch $P_i$.

  Similarly, the variation of the rotational splittings first coefficient
differences, $\Delta a_1$, due to a plate scale error will be proportional to
first order to $\epsilon$. Our simulations have shown a residual dependence on
frequency and degree, namely:
\begin{equation}
  \Delta a_1(P_2) -  \Delta a_1(P_1) 
   \cong {\cal A} \;(\epsilon(P_2)-\epsilon(P_1)) 
\label{eq:da1pse}
\end{equation}
where
\begin{equation}
{\cal A} = {\cal A}_o
     + {\cal A}_\nu \; (\nu-\nu_o)
     + {\cal A}_\ell \; (\ell-\ell_o)
\label{eq:a1rel}
\end{equation}
and ${\cal A}_o = -580$ nHz, while ${\cal A}_\nu = -26$ nHz/mHz and ${\cal
A}_\ell = 0.13$ nHz, for $\nu_o = 3.333$ mHz and $\ell_o=150$, using our
normalization of the Clebsh-Gordon coefficients.

\begin{figure}[!p]
\epsscale{0.8}
\plotone{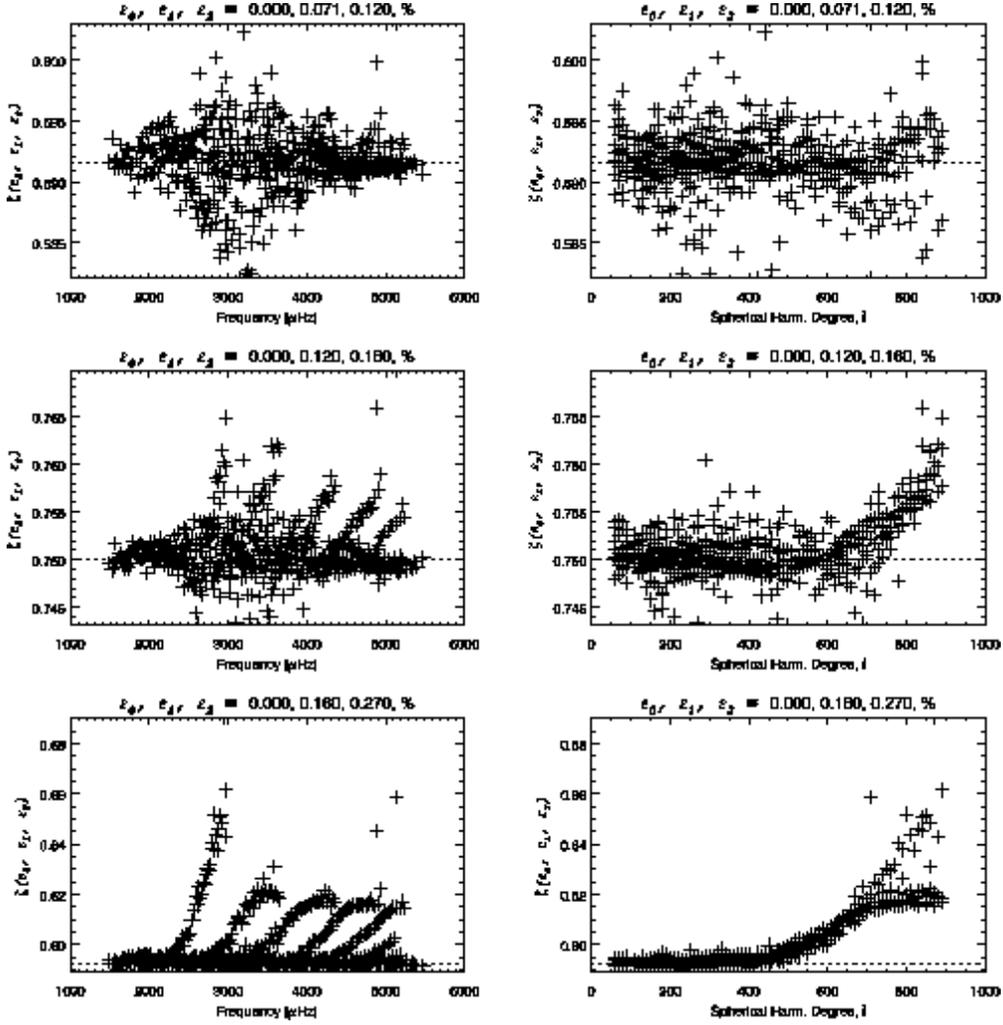}
\caption{Relation $\zeta(\epsilon_1, \epsilon_2, \epsilon_3)$
         (see Eq.~\ref{eq:dnupse}) that accounts for changes in the
         theoretical differences between ridge and mode frequencies resulting
         from different plate scale errors. The dotted lines indicate the
         value corresponding to the combination of plate scale errors given in
         Eq.~\ref{eq:dnupse}.  This relation is illustrated with three
         combination of plate scale errors to show that it breaks at large
         degrees for large plate scale error.
\label{fig:check_theo_plscale_frq}} 
\end{figure}

   We have verified, using our simulations, that for plate scale errors of a
fraction of a percent Equations~\ref{eq:dnupse} and \ref{eq:a1rel} are
satisfied. These simulations show that the plate scale induced frequency error
scales with $\epsilon$ according to Equation~\ref{eq:dnupse} to better than
0.01\% on average, with a scatter of 0.3\%, for plate scale errors up to
$0.12\%$. Simulations with $\epsilon=0.16\%$ show a departure from
Equation~\ref{eq:dnupse} at the 2\% level for $\ell \ge 700$, while
simulations for $\epsilon=0.27\%$ show a departure from
Equation~\ref{eq:dnupse} at the 5\% level for $\ell \ge 450$, as illustrated
in Fig.~\ref{fig:check_theo_plscale_frq}.

\begin{figure}[!p]
\epsscale{0.8}
\plotone{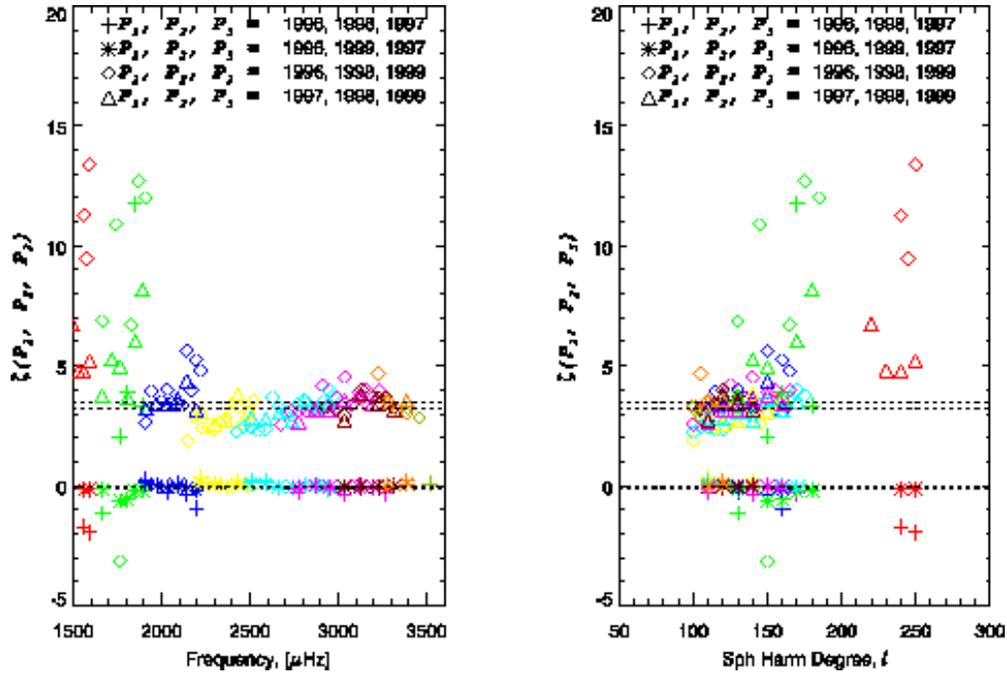}
\caption{Relation $\zeta(P_1, P_2, P_3)$ (Eq.~\ref{eq:dnupse}) for the four
         possible combination of {\em Dynamics} observations, as a function of
         frequency (left panel) and degree (right panel). The horizontal dash
         lines correspond to the expected values (see right-hand-side
         of Eq.~\ref{eq:dnupse}). This confirms that the changes seen in
         Fig.~\ref{fig:plsclobs} are primarily due to the change in the plate
         scale error.  \label{fig:check_obs}}
\end{figure}

  Figure~\ref{fig:check_obs} illustrates how well Equation~\ref{eq:dnupse} is
satisfied for the observed frequency differences. It shows that for the modes
whose frequency is above 1.9 mHz (\ie, $n \ge 2$) the relation in
Equation~\ref{eq:dnupse} is well satisfied, hence the change in frequency
differences between epochs can be explained by the change in plate scale.

\begin{figure}[!p]
\epsscale{0.8}
\plotone{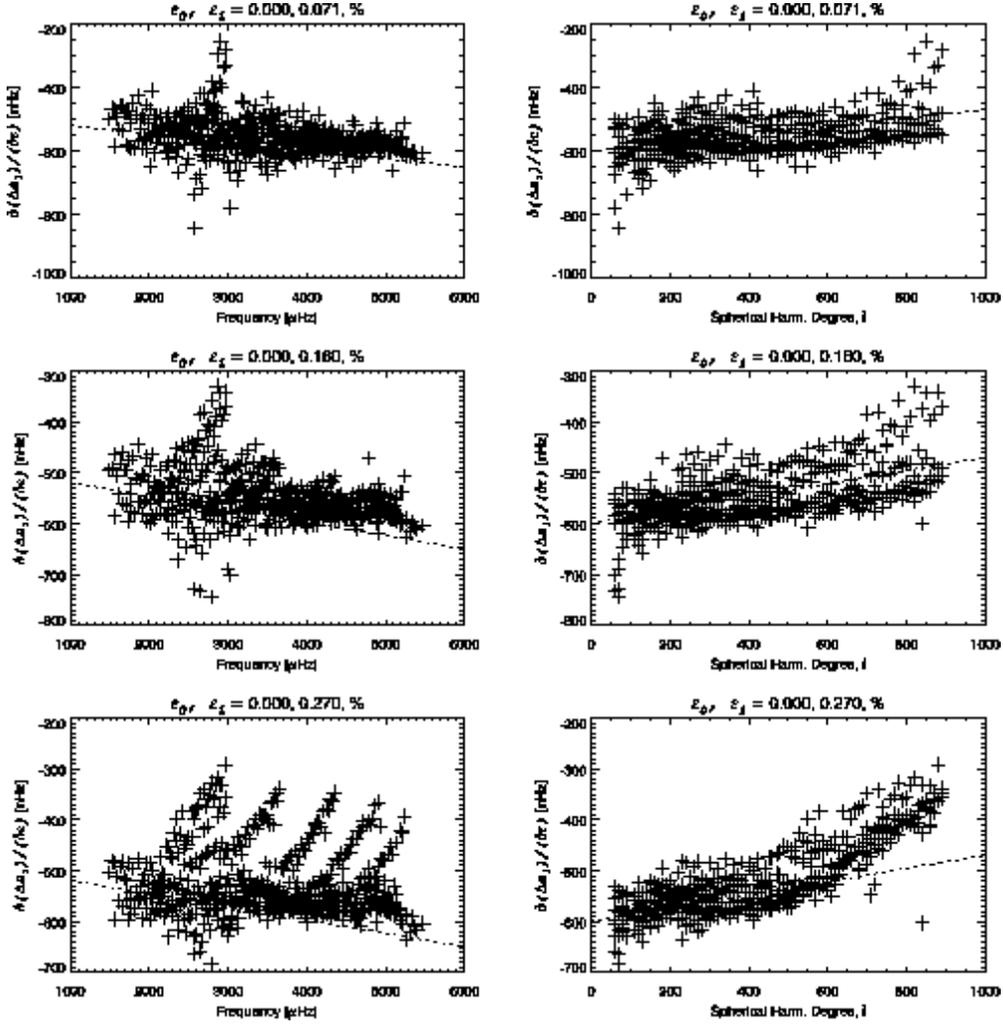}
\caption{Relation between $\Delta a_1$ and $\epsilon_1, \epsilon_2$
         (see Eq.~\ref{eq:da1pse}) that expresses the changes in the
         theoretical differences between ridge and mode $a_1$ splitting
         coefficients resulting from different plate scale errors. The dotted
         lines correspond to the value predicted by Eq.~\ref{eq:a1rel}. This
         relation is illustrated with three combination of plate scale errors
         to show that it breaks at large degrees for large plate scale error.
         \label{fig:check_theo_plscale_a}}
\end{figure}

\begin{figure}[!p]
\epsscale{0.8}
\plotone{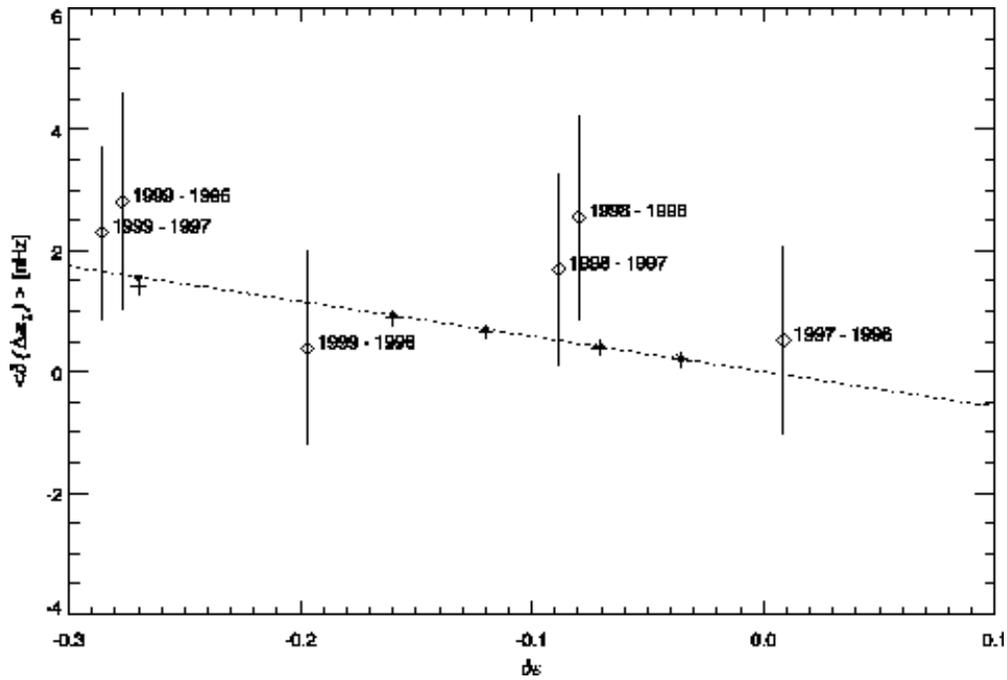}
\caption{Average change in the observed differences between ridge and mode
         $a_1$ splitting coefficients (diamonds) as a function of change in
         plate scale errors, for all possible {\em Dynamics} observation
         combinations. The error bars correspond to the standard deviation
         about the mean. The crosses and stars symbols correspond to our
         simulations (all the modes or only the modes with degree up to 250
         respectively) and the dotted line illustrates the relation expressed
         by Eqs~\ref{eq:da1pse} and \ref{eq:a1rel}.  \label{fig:check_obs_a1}}
\end{figure}

 On the other hand our simulations show that Equation~\ref{eq:a1rel} is only
satisfied to 2\% on average, but the plate scale induced changes in $a_1$ are
themselves much smaller than the corresponding changes in frequency. Our
simulations show that the relation breaks down at large degree for large
plate scale errors, as illustrated in Fig.~\ref{fig:check_theo_plscale_a}.
Finally, Figure~\ref{fig:check_obs_a1} depicts the correlation of the change
in $a_1$ with epoch, and shows that in practice the scatter in the data makes
it hard to establish whether Equation~\ref{eq:a1rel} is actually satisfied by
the measurements.

\subsection{Effect of Image Distortion}

  So far we have ignored the distortion of the observed solar image.  Further
progress requires the inclusion of the image distortion and the instrument
MTF, adding one more level of complexity, but an element that is expected to
be somewhat stable with time.

  A full ray tracing of the optical design of the MDI instrument, in the {\em
Dynamics Program} configuration, predicts distortions (Scherrer, private
communication) that can be modeled as:
\begin{equation}
  \epsilon_r \equiv \frac{\Delta r}{r} = a_r ((\frac{r}{r_{\rm m}})^2 - 1)
\label{eq:raddist}
\end{equation}
where $a_r=1.1 \times 10^{-3}$, $r$ is the distance from the detector center,
$\Delta r$ the image distortion, and $r_m$ is the observed image mean radius.

  However, this expression does not account for the elliptical shape of the
MDI image, an additional distortion that is also on the order of 0.1\%. A
possible explanation is that the CCD detector is tilted with respect to the
focal plane. The shape of the solar image resulting from such a tilt is indeed
elliptical, as shown in Appendix A, and a tilt of $\sim2\Deg$ can account
for the observed 0.1\% effect.

\begin{figure}[!p]
\epsscale{1.8}
\plottwo{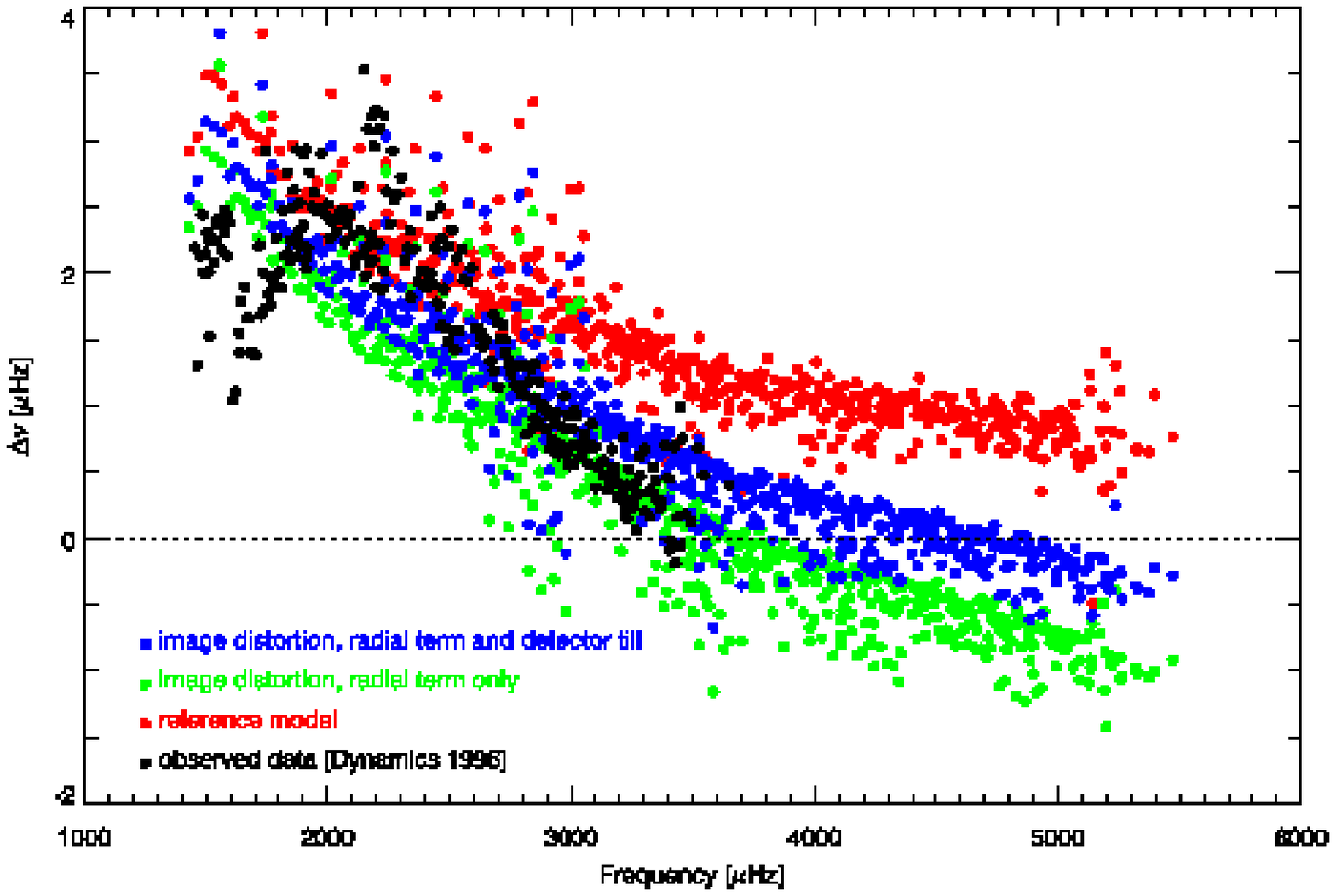}{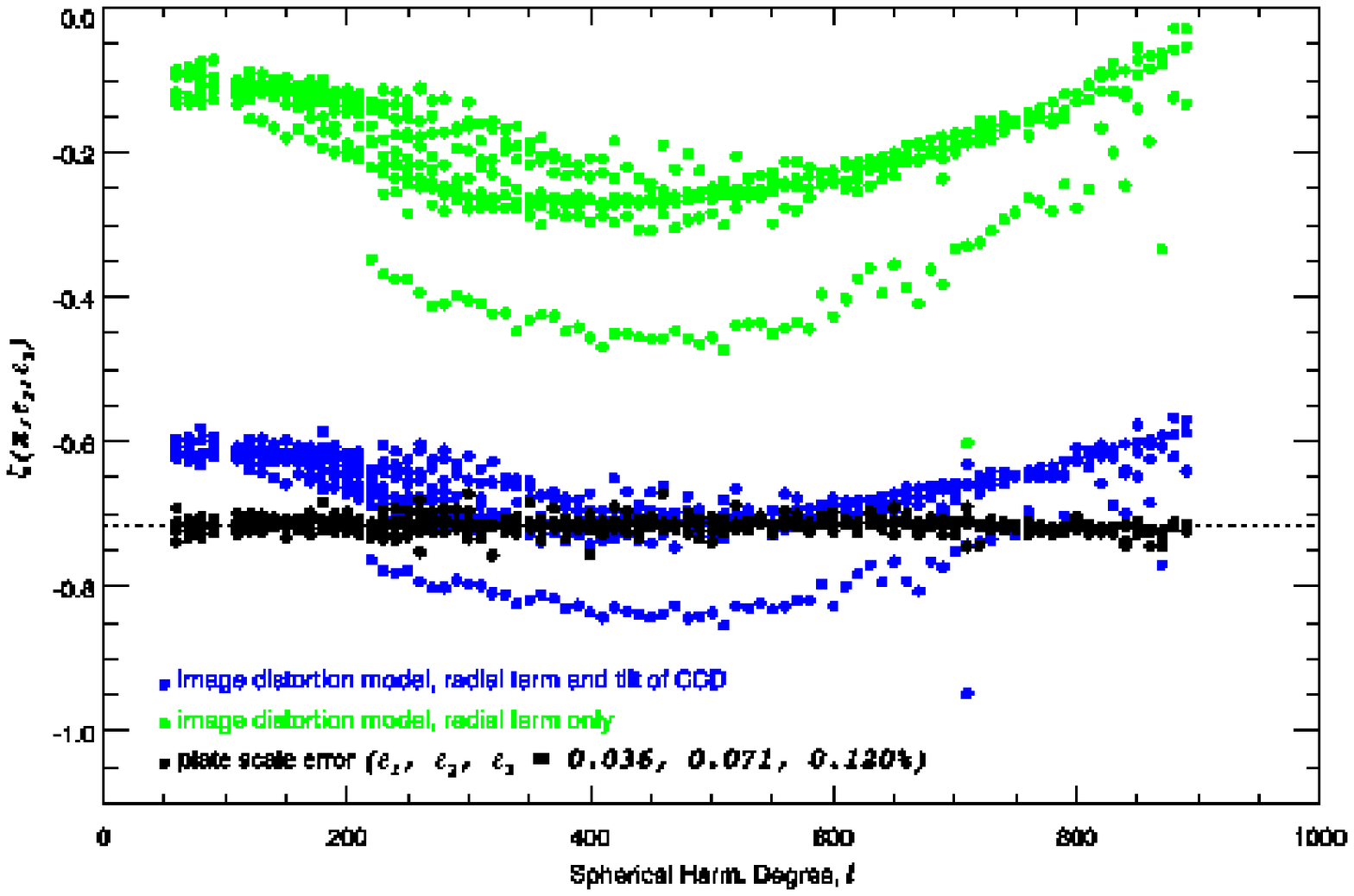}
\caption{
%
%
         Top panel: Comparison of observed and theoretical differences between
         ridge and mode frequencies. The effect of including an image
         distortion (in green and blue) in the leakage matrix is compared to
         our reference model (in red) and to the observations (in black).
         The bottom panel shows that the effect of the image distortion is
         different from an image scale error: the relation expressed by
         Eq~\ref{eq:dnupse} is not satisfied.
         \label{fig:delobsXdist}}
\end{figure}

  We have computed leakage matrices that include some image distortion. These
were used to generate simulated power ridges that were then fitted as
described earlier. In one set of leakage matrix computation we only included
an image distortion as described in Equation~\ref{eq:raddist} (hereafter Model
1). In the second set we included the elliptical distortion resulting from a
tilt of the detector as well the radial distortion (hereafter Model 2). The
resulting frequency differences, as well as the observed ones for 1996, are
shown in Figure~\ref{fig:delobsXdist}.  The distortion of the eigenvalues by
differential rotation was included in these calculations.

  The effect of an image distortion on the frequency differences, although
similar, is different from the effect of a plate scale error. Indeed, the
relation expressed in Equation~\ref{eq:dnupse} is not satisfied for the
simulations that include an image distortion, as shown in the lower panel of
Figure~\ref{fig:delobsXdist}.

\begin{figure}[!p]
\epsscale{.9}
\plotone{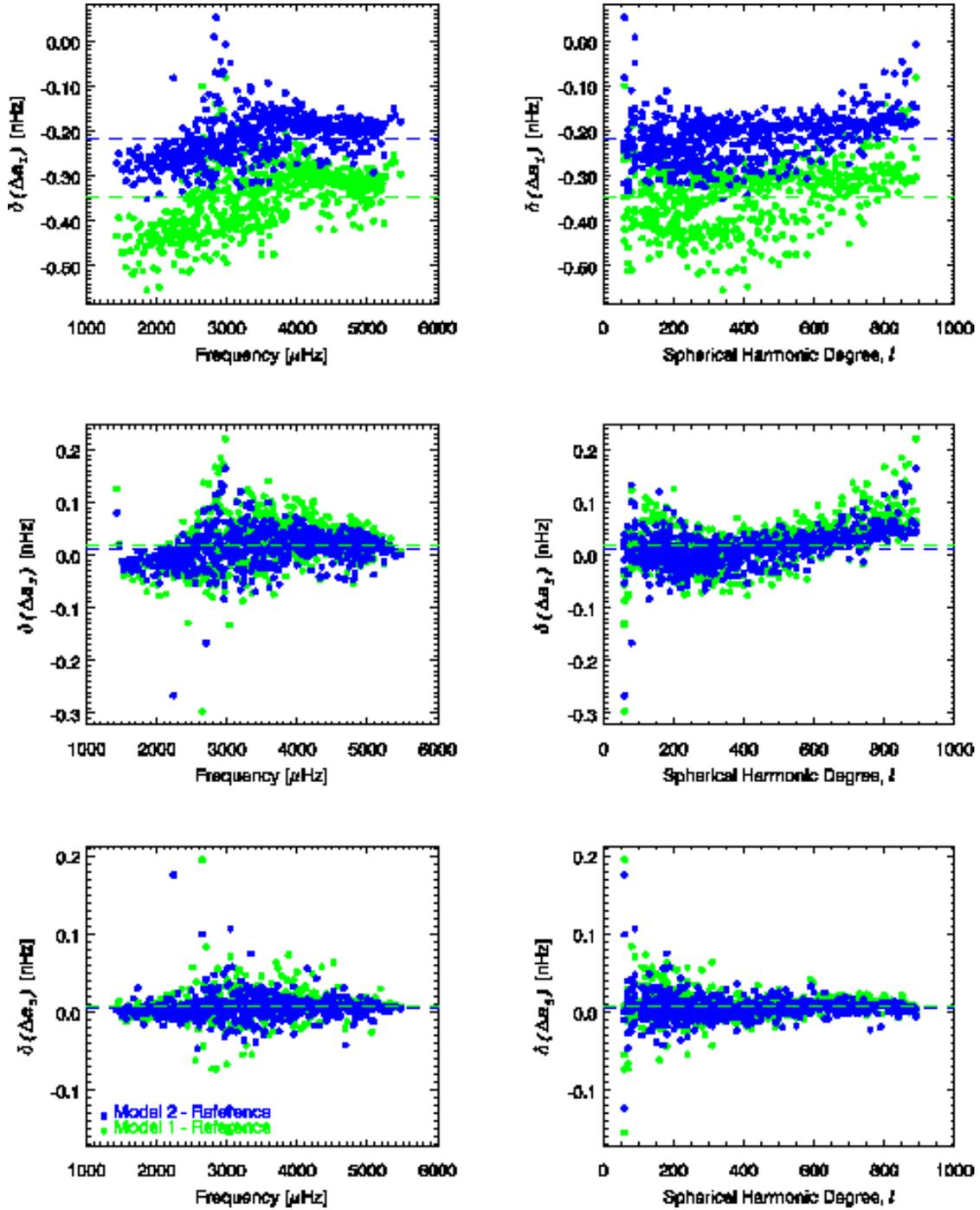}
\caption{
%
%
         Changes in odd splitting coefficients offsets for two image
         distortion models with respect to our reference model. Model 1
         includes only a radial image distortion (green points), Model 2
         includes both the radial term and the tilt of the CCD (blue points).
         Horizontal lines are the respective means.
         \label{fig:delaiXdistor}}
\end{figure}

  The effect of an image distortion on the odd splitting coefficients is
comparable in amplitude to the effect of a similar plate scale error, as
illustrated in Figure~\ref{fig:delaiXdistor}. The effect on the even
coefficients is negligible. The mean change in the splitting coefficient
offsets is listed in Table~\ref{tab:imgDistor}.

%

\begin{table*}[!t]
\begin{center}
\caption{Mean Change in Splitting Coefficient Offsets Due to Image Distortion 
             \label{tab:imgDistor}}
\begin{tabular}{rrr}\hline\hline
~ & Model 1 $-$ Reference  & Model 2 $-$ Reference \\\hline
$\delta(\Delta a_1)$ & $-0.3457 \pm 0.0686$ &  $-0.2162 \pm 0.0459$ \\
$\delta(\Delta a_2)$ & $~0.0017 \pm 0.0019$ &  $~0.0010 \pm 0.0012$ \\
$\delta(\Delta a_3)$ & $~0.0205 \pm 0.0401$ &  $~0.0127 \pm 0.0290$ \\
$\delta(\Delta a_4)$ & $-0.0017 \pm 0.0018$ &  $-0.0010 \pm 0.0011$ \\
$\delta(\Delta a_5)$ & $~0.0075 \pm 0.0119$ &  $~0.0051 \pm 0.0107$ \\
$\delta(\Delta a_6)$ & $~0.0001 \pm 0.0005$ &  $~0.0001 \pm 0.0004$ \\
\hline
\end{tabular} \\
Model 1: radial term only \\
Model 2: radial term and tilt of CCD
\end{center}
\end{table*}


\subsection{Effect of the Instrumental Point Spread Function}

  Another point to consider in the leakage matrix calculation is the effect of
the instrumental point spread function (PSF). Not only will the smearing of the
image by the PSF affect the leakage matrix (especially at high degrees), but
it will change with time as the focus changes.

  To estimate the PSF of the MDI instrument and its variation with epoch, we
ran the {\tt HGEOM} procedure on MDI full-disk images. {\tt HGEOM}, a
procedure part of the GONG reduction and analysis software package ({\tt
GRASP}), returns an estimate of the azimuthally averaged modulation transfer
function (MTF, \ie, the Fourier transform of the PSF), following the
methodology described in \cite{toner93}. This method reduces the 2D solar
image to a radial profile (by computing the mean over concentric annuli) and
computes the Hankel transform of this profile. By exploiting the zero-crossing
properties of the Hankel transform one recovers the true image dimensions and
estimates the MTF.

\begin{figure}[!p]
\epsscale{0.85}
\plotone{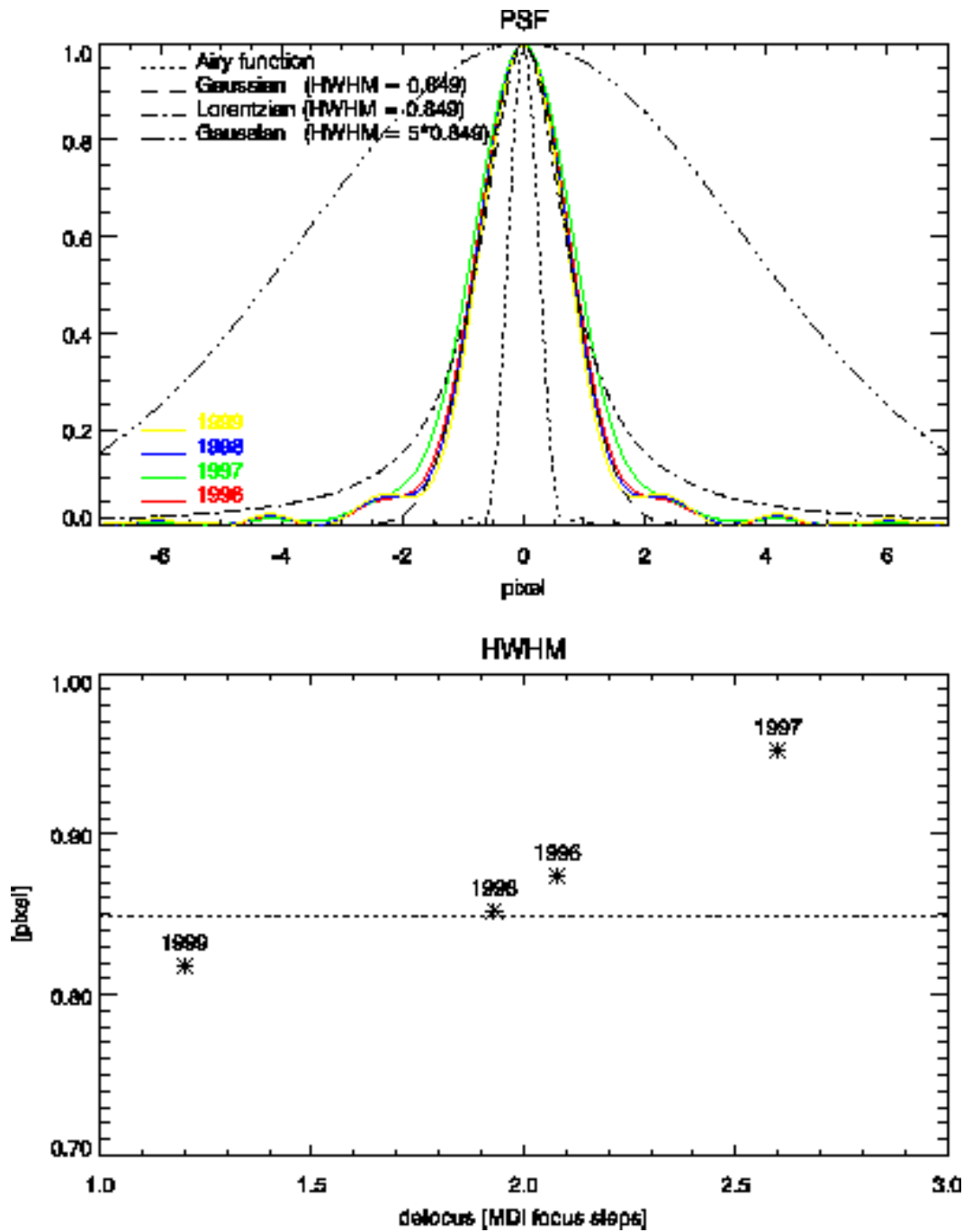}
\caption{Top panel: estimates of MDI azimuthally averaged point spread
         function (PSF), computed from four images corresponding to each
         consecutive year of the {\em Dynamics} program. These are compared
         to the Airy function corresponding to the instrument aperture
         diffraction limit and to a Gaussian and a Lorentzian with a matching
         half width at half maximum (HWHM).
         Bottom panel: HWHM of the azimuthally averaged point spread functions
         as a function of image defocus.
         \label{fig:psf}}
\end{figure}

  Figure~\ref{fig:psf} (top panel) shows the PSF for five different full-disk
images taken by MDI on five consecutive years.  Although their shapes are very
similar, the width of the PSF is increasing with the amount of defocus, as
expected (Figure~\ref{fig:psf}, lower panel).  While the PSF of MDI displays a
substantial tail beyond 2 pixels, its core can be approximated by a Gaussian,
with a half width at half maximum of $0.8 \pm 0.1$ pixels. 
The image sizes estimated by {\tt HGEOM} are consistent with the the plate
scale based on the MDI standard reduction pipe-line to within .04 pixels or
.008\%. This is, of course, to be expected, since the {\em level 1.8} MDI data
processing uses the same algorithm for the limb definition as the GONG
procedure.

\begin{figure}[!p]
\epsscale{1.0}
\plotone{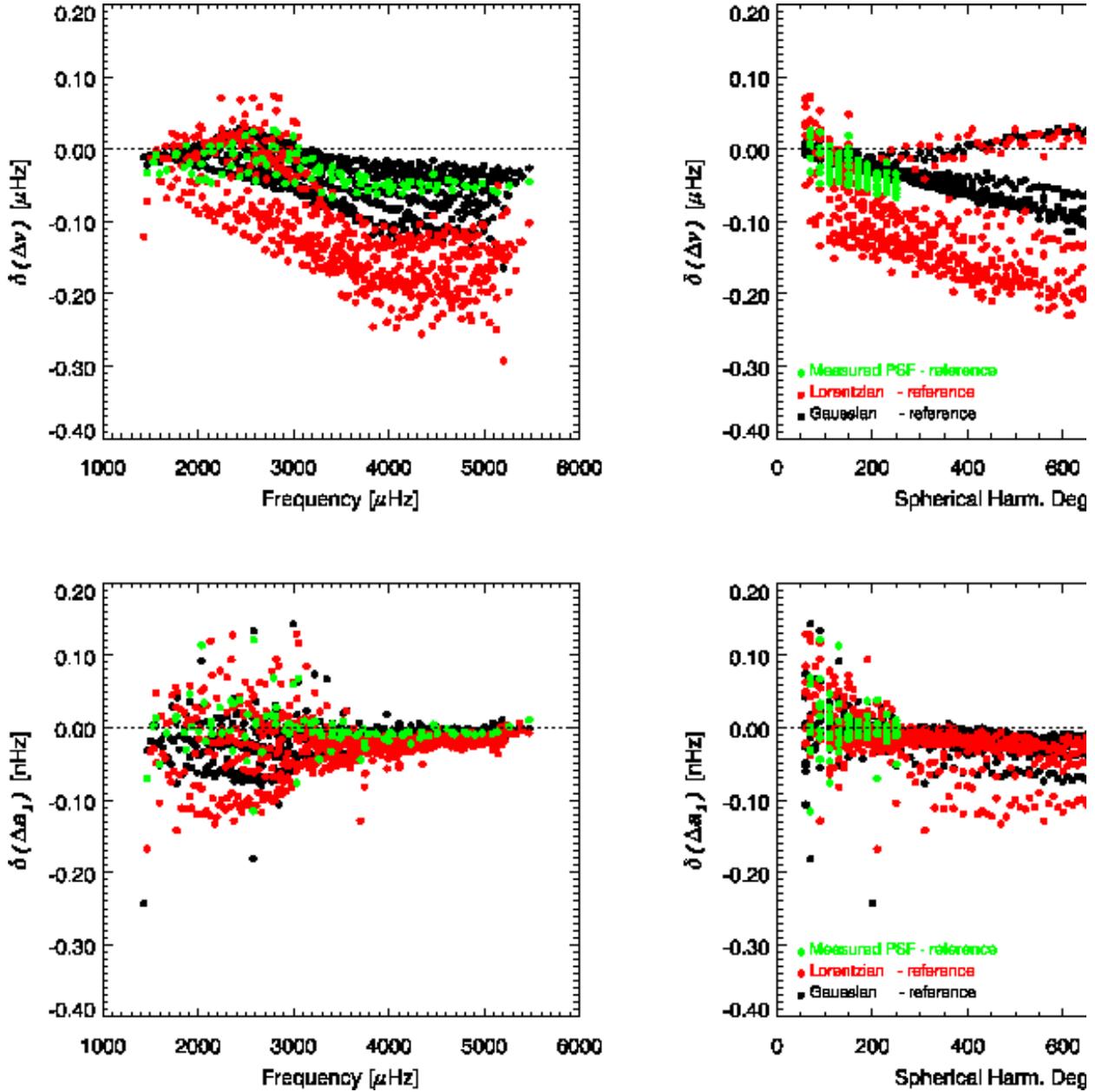}
\caption{
%
%
%
         Change in the theoretical differences between ridge and mode
         frequency (top panels) and and between ridge and mode $a_1$ splitting
         coefficient (bottom panels), with respect to a reference case, as a
         function of frequency (left panels) or degree (right panels). The
         reference case uses a leakage matrix computation that does not
         include the effect of the instrumental PSF. This is compared to the
         cases where the leakage matrix computation includes a) a Gaussian PSF
         (HWHM = 0.71 pixel, black points), b) a Lorentzian PSF (HWHM = 1.0
         pixel, red points), or c) the PSF estimated by {\tt HGEOM} using a
         {\em Dynamics} 1996 image (green points).
         These changes are on the order of 10\% for the frequency differences
         and 2\% for the $a_1$ coefficient differences.
\label{fig:fig_comppsf}}
\end{figure}

  Simulated {\rm ridge} frequencies resulting from including the PSF in the
computation of the leakage matrix were computed. We used a Gaussian PSF, a
Lorentzian PSF as well as the PSF estimated by {\tt HGEOM}.  The resulting
changes in the frequency offsets and splitting coefficient offsets are shown
in Figure~\ref{fig:fig_comppsf} as a function of frequency and degree. The
changes in frequency offsets are on the order of 0.1 \uHz\ (\ie, a 5 \%
effect), while changes in $\Delta a_1$ are on the order of 0.1 nHz (or a 2\%
effect).

\begin{figure}[!p]
\epsscale{1.}
\plotone{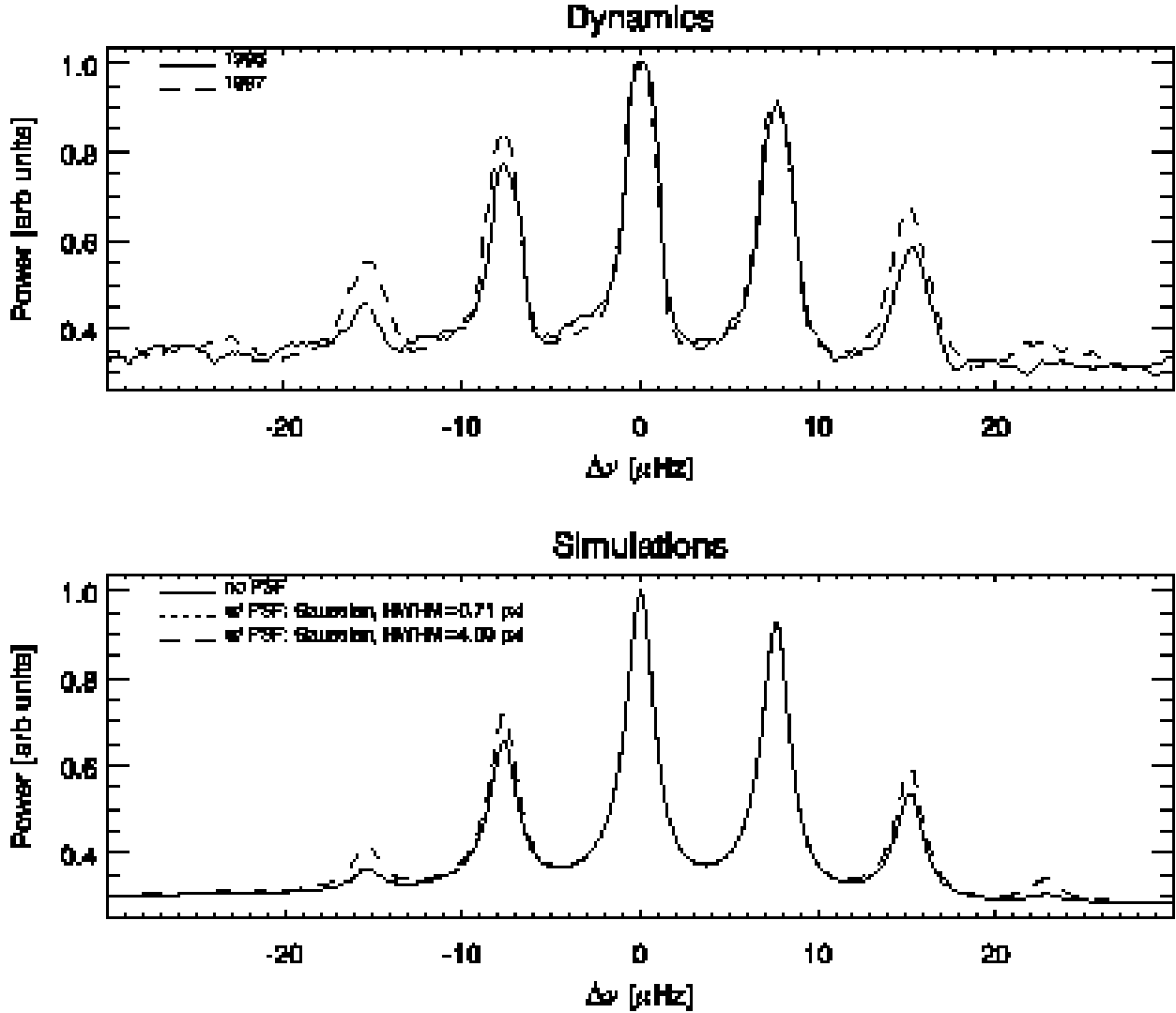}
\caption{Observed (top panel) and synthetic (middle panel) sectoral limit
         power spectra for $n=2$ and $\ell=92$. The observed spectra
         correspond to the two consecutive years with the largest change in
         PSF width (see Fig~\ref{fig:psf}), that might account for the observed
         change in leaks amplitude.
         The three synthetic spectra correspond to leakage matrices computed
         using no PSF, a Gaussian PSF with a HWHM of 0.71 pixel (a good
         approximation of the measured azimuthally averaged PSF) and a
         Gaussian PSF with a HWHM of 4.0 pixel (a substantially wider
         PSF). Only when very wide PSF is included does our simulations
         produce a noticeable change in the amplitude of the leaks.
         \label{fig:spec_psf}}
\end{figure}

  Variations of the width of the PSF will affect predominantly the amplitude
of the leaks with a small effect on the leakage asymmetry \cite[as shown
in][]{rabello-soares01}.
 The change of the limit power spectrum computed for 1996 and 1997 (as shown
in Fig.~\ref{fig:spec_psf} -- top panel) is qualitatively similar to the
variation of simulated power spectra synthesized using leakage matrix
computations that includes the effect of the PSF for two different widths
(Fig.~\ref{fig:spec_psf} -- lower panel).
 Our simulations seem to reproduce the observed changes in amplitude, but for
a change in PSF width nearly 5 times larger than the change observed in the
corresponding azimuthal averages.  This might result from the fact that the
true PSF is not azimuthally symmetric and that the azimuthal averages we used
so far are in fact poor approximations.

\begin{figure}[!p]
\epsscale{1.}
\plotone{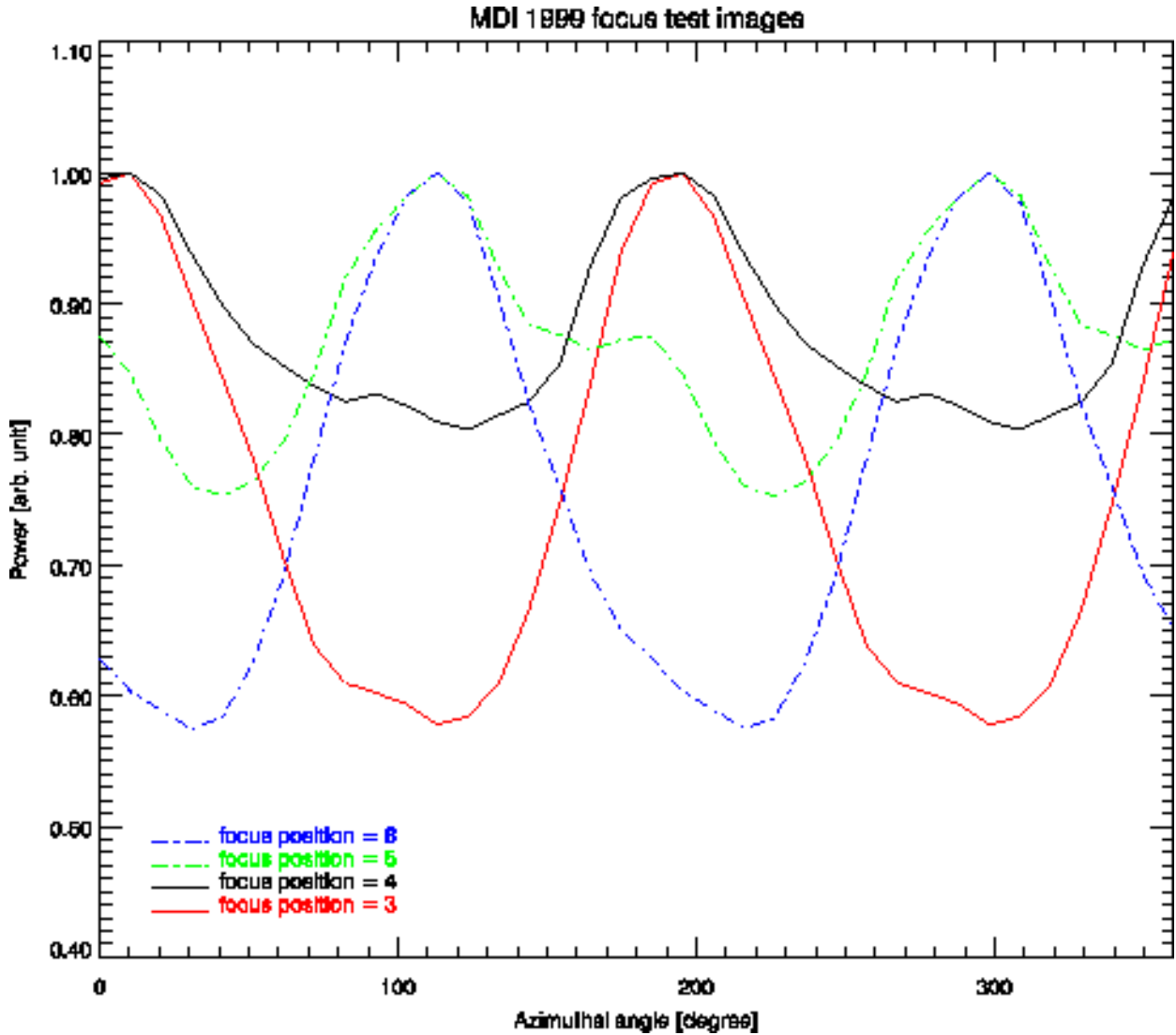}
\caption{
         Observed power spectra as a function of azimuthal angle for a
         horizontal wave number, $k_h$, of 230 pixel$^{-1}$, computed from
         1999 MDI images taken at four different focus positions, \ie\ focus
         steps 3, 4, 5 and 6. The best focus position at the time was
         determined to be at focus step 4.1 (see Fig~\ref{fig:focus}). The
         angular dependence of these power spectra indicates that the
         instrumental PSF is not uniform with respect to the azimuthal
         orientation on the detector. Note how the non-uniformity increases
         with increasing defocus and the phase shift of the azimuthal
         dependence between either side of the best focus position.
         \label{fig:spec_psf_az}}
\end{figure}

  \citet{schoubogart98} also noticed that the power spectrum of MDI images
varies with azimuth (see Figure 1 of their paper).
Figure~\ref{fig:spec_psf_az} shows power as a function of the azimuthal angle
for horizontal wavenumber $k_h = 230$ pixel$^{-1}$ for test images taken in
1999 at four focus settings. This figure shows not only that the power is a
function of azimuth, but, as first noticed by Duvall (private communication),
the phase of this angular dependence changes with focus position, changing by
nearly $90\Deg$ between focus positions 4 and 5, corresponding to a change
from positive to negative defocus (the best focus at disk center in 1999 was
at $\approx 4.1$, as shown in the upper panel of Figure~\ref{fig:focus}).
Such a change can be explained by the combination of an astigmatism with a
positive or negative defocus. Such configuration would produce a PSF that is a
function of the azimuthal angle (Bush, private communication).

  \citet{tarbell97}, using phase-diversity analysis, estimated the PSF of the
MDI instrument in focus and in its high resolution configuration (\ie, when the
image is magnified by a factor 3.2). This PSF shows a clear angular
dependence, with a secondary peak with an amplitude of 0.15, attributed to the
residual optical aberrations of the instrument components.

 Finally, we should also point out that the best focus changes across the CCD
by nearly one focus step. This is caused by some residual field curvature and
will produce a radial variation of the PSF. The optical characteristics of the
MDI instrument are such that the instrumental PSF has to be a function of both
radius and azimuth.

\subsection{Effects of Image Orientation, $P$ and $B_{\rm o}$ Angles}

  The spherical harmonic decomposition is {\em in principle} computed with
respect to an orientation aligned with the solar rotation axis. The location
of this axis is determined by two angles, the position angle, $P$, and the
heliographic latitude of the central point of the disk, $B_{\rm o}$. The roll
angle of the spacecraft is maintained to keep MDI $P_{\rm eff}$ to be zero
(where $P_{\rm eff}$ is the position angle of the northern extremity of the
rotation axis with respect to MDI CCD detector column orientation).  However,
Cliff Toner (private communication) estimated the value of $P_{\rm eff}$ by
inter-comparing MDI and GONG images, drift scans, and observations taken
during the November 1999 transit of Mercury in front of the Sun.
He found that $P_{\rm eff} = 0.19\Deg \pm 0.04\Deg$ using May 2000 observation
and $P_{\rm eff} = 0.20\Deg \pm 0.05\Deg$ based on the transit of Mercury.

  The precision to which $B_o$ is known is directly related to the precision
of the Carrington elements, $i_C$ and $\Omega_C$ (\ie, the inclination of the
solar equator to the ecliptic and the longitude of the ascending node of the
solar equator on the ecliptic plane respectively). \citet{Giles:99} showed
that these are not perfect and that $\Delta i_C = -0.091\Deg \pm 0.012\Deg$
and $\Delta \Omega_C = -0.1\Deg \pm 0.1\Deg$. The error on $B_o$ is therefore
no larger than $0.1\Deg$.

\begin{figure}[!p]
\epsscale{0.8}
\plotone{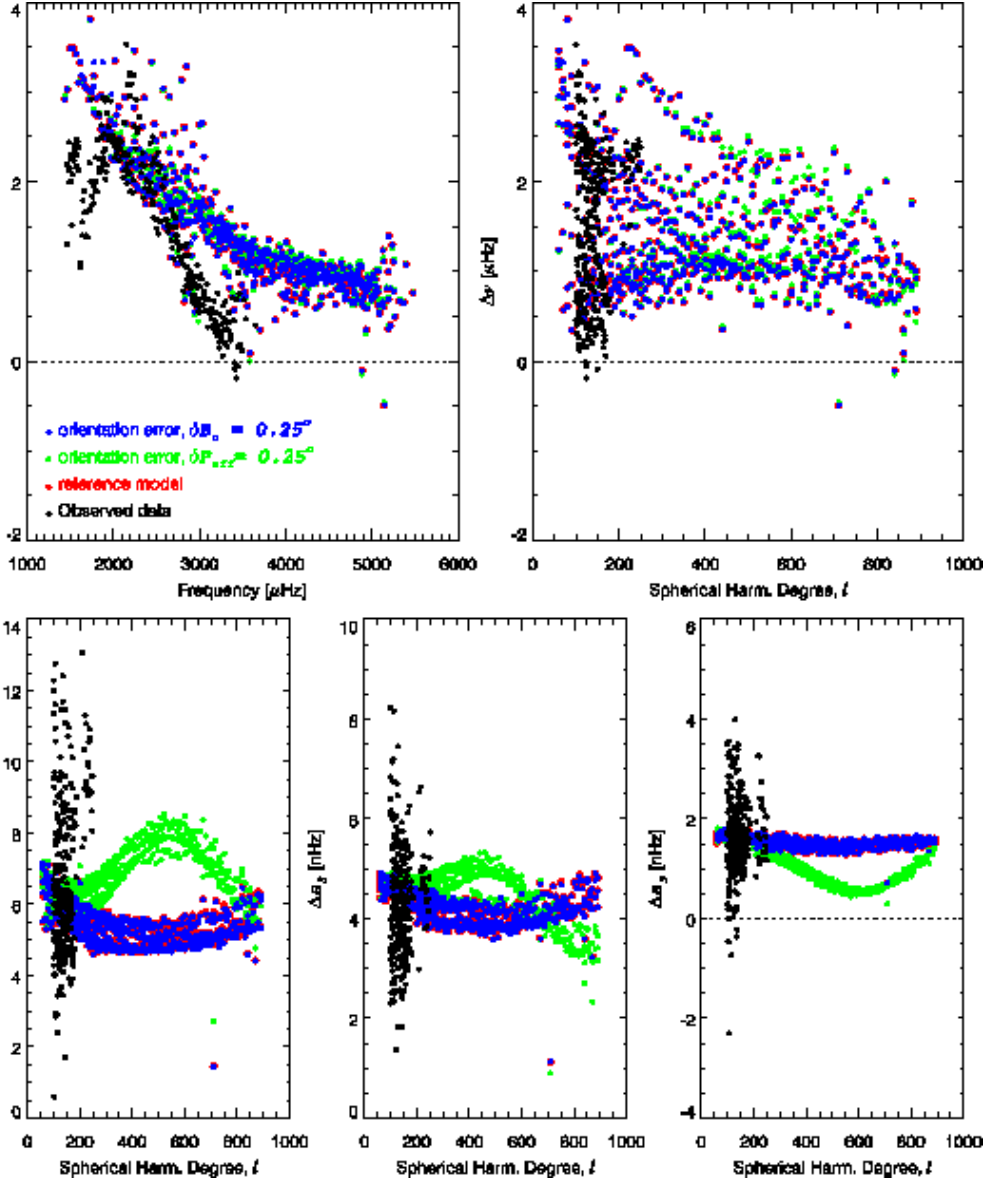}
\caption{
%
%
%
         Comparison of observed and theoretical differences between ridge and
         mode frequencies and between ridge and mode odd splitting
         coefficients.  The effect of including an error in the solar rotation
         axis orientation in the computation of the leakage matrix ($\delta
         P_{\rm eff} = 0.25\Deg$, in green, and $\delta B_{\rm o} = 0.25\Deg$,
         in blue) is compared to our reference model (in red) and to the
         observations ({\em Dynamics} 1996, in black).
         \label{fig:p0b0}}
\end{figure}

  To assess the effect of an error in the image orientation used for the
spherical harmonic decomposition, we computed simulated frequency differences
by fitting simulated ridges based on a leakage matrix computed with either an
error in $P_{\rm eff}$ or $B_{\rm o}$ of $0.25\Deg$.  Figure~\ref{fig:p0b0}
compares the observed frequency differences for 1996 to the theoretical
differences resulting from including such orientation error.  The effect of a
quarter of a degree error is unnoticeable for the frequency differences (less
than 1\% of relative change of the offsets, for either angle), while an error
in $P_{\rm eff}$ affects noticeably the odd splitting coefficients. This
should not come as a surprise: an error in the orientation of the image with
respect to the rotation axis will affect the relative mode power as a
functions of $m/\ell$ -- in a way conceptually similar to the distortion of
the eigenfunctions by the differential rotation, see Section~\ref{sec:PertRot}
-- and will therefore affect the ridge rotational splittings.  The smooth
dependence of $\Delta a_i$ with $\ell$ for a conservative $0.25\Deg$ error in
$P_{\rm eff}$ seen in Figure~\ref{fig:p0b0} is not readily seen in the MDI
measurements. But this effect is small and is most likely masked by the random
scatter and other systematics present in the measurements.

\section{Results}

\subsection{Error Budget}

  Table~\ref{tab:cmpErrCorr} compares the range of uncertainties resulting
from ridge fitting, \ie\ the formal error bars of the ridge fitting, to the
magnitude of the ridge to mode corrections, for both the frequency and the
rotational splitting coefficients. While the uncertainties resulting from the
fitting scale with the ridge width, Table~\ref{tab:cmpErrCorr} shows that the
required corrections are about an order of magnitude larger than the
uncertainties.  On the other hand, to produce useful high-degree mode
parameters we need to be able to compute these corrections with a precision
substantially smaller than the uncertainties.

\begin{table*}[!t]
\begin{center}
\caption{Comparison between Uncertainty and Ridge to Mode Correction
         \label{tab:cmpErrCorr}}\vspace{1em}
\begin{tabular}{l|c|ccc}\hline
                  &  frequency & 
                    \multicolumn{3}{c}{frequency splittings [nHz]}           \\
\cline{3-5}
                  &  [\uHz]          &  $a_1$     &   $a_3$    &      $a_5$  \\
\hline
uncertainty$^{(a)}$ &  0.1 -- 0.6    & 0.3 -- 1.0 & 0.2 -- 0.8 & 0.15 -- 0.7 \\

largest correction$^{(b)}$ &        4.0       &      10.0  &       5.0  &       2.0   \\
\hline
\end{tabular}\\
~$^{(a)}$ given range corresponds to a ridge width range of 10  -- 60  \uHz\\
~$^{(b)}$ for the $100 \le \ell \le 900$ range
\end{center}
\end{table*}

  Table~\ref{tab:errBudget} presents an error budget for the ridge to mode
correction, breaking it up as a function of the different elements used to
compute the correction. The listed typical values are based on our best
estimates of these different elements resulting from the current
analysis. 

  Now that these effects are better understood and that we better understand
the MDI instrument imaging imperfections, we can -- and will in the near
future -- reprocess the {\em Dynamics} data, and include the correct image
scale and the known image distortion in the spatial
decomposition. Table~\ref{tab:optErrBudget} presents an error budget projected
for the best achievable values of these elements.  The sobering conclusion of
this exercise is that our estimate of the achievable precision of the
frequency and splitting coefficients corrections are still comparable to the
fitting uncertainties.

\begin{table*}[!t]
\begin{center}
\caption{Error Budget for the Ridge to Mode Correction
         \label{tab:errBudget}}\vspace{1em}
\begin{tabular}{l|c|c|c}\hline
                              & typical  &  \multicolumn{2}{c}{corresponding
change in} \\
\cline{3-4}
parameter                     & values   &  frequency [\uHz] & $a_1$ [nHz] \\
\hline
plate scale                   & 0.01 -- 0.27 \%&  0.2 -- 6.0  & 0.06 -- 1.6  \\
differential rotation         & 4.4  --  6.2 \%& 0.01 -- 0.8  & 0.4  -- 0.7  \\
image distortion              & 0.1\%          &  1.0 -- 1.5  & 0.2  -- 0.4  \\
instrument PSF                & (*)            & 0.05 -- 0.15 & 0.05 -- 0.1  \\
image orientation, $P_{\rm eff}$           
                                & $0.25\Deg$   & 0.03         & 0.01 -- 2.0  \\
image orientation, $B_{\rm o}$  & $0.1\Deg$    & 0.004        & 0.012        \\
\hline
\end{tabular}\\
(*) Changes resulting from using a Gaussian versus a Lorentzian PSF \\
Note also that the values tabulated are unsigned magnitude \\
\end{center}
\end{table*}

\begin{table*}[!t]
\begin{center}
\caption{Optimal Error Budget for the Ridge to Mode Correction
         \label{tab:optErrBudget}}\vspace{1em}
\begin{tabular}{l|c|c|c}\hline
                                & achievable & \multicolumn{2}{c}{change in} \\
\cline{3-4} 
parameter                       & values     &  frequency [\uHz] & $a_1$ [nHz] \\
\hline
plate scale                     & 0.01       \% & 0.220      & 0.06  \\
differential rotation (*)       & 0.03       \% & 0.054      & 0.045 \\
image distortion                & 0.01       \% & 0.150      & 0.04  \\
instrument PSF                  & ($\dag$)      & 0.015      & 0.01  \\
image orientation, $P_{\rm eff}$& $0.05\Deg$    & 0.006      & 0.40  \\
image orientation, $B_{\rm o}$  & $0.02\Deg$    & 0.001      & 0.002 \\
\hline\hline
correction estimated precision &               & 0.272      & 0.409 \\
fitting uncertainties          &               & 0.1 -- 0.6 & 0.3 -- 1.0 \\
\hline
\end{tabular}\\
(*) values scaled from our worst case, see discussion in text \\
($\dag$) 1/10th of corresponding values in table~\ref{tab:errBudget}
\end{center}
\end{table*}

  For the frequency corrections, the dominant contributions come from the
plate scale error and the image distortion.  If both could be known to an
accuracy of $10^{-5}$, the precision of the correction would be further
reduced to $0.06$ \uHz, a {\em useful} value since it would then become
substantially smaller than the uncertainties.

  The precision of the fit of an ellipse to the limb of the image is likely to
be $10^{-5}$ (\ie, fitting 5 parameters to some 3000 limb positions). But it
would be presumptuous to assert that the algorithmic definition of the limb
can be accurate to that level. Characterizing the instrument image distortion
to the $10^{-5}$ level is also very challenging. The image distortion in the
MDI instrument is only at the $10^{-3}$ level (corresponding to displacements
smaller than 0.95 pixels). We thus need to know the distortion with a relative
precision of 1\% (\ie, 1/100 of a pixel) to reach the $10^{-5}$ level.

  For the splitting coefficient $a_1$ corrections, the dominant contributions
come from the $P$ angle, and might prove to be a fundamental limitation for
the use of high degree rotational splittings. If we could reduce the error in
the position angle by an additional order of magnitude (\ie, to $0.005\Deg$),
the precision of the correction in $a_1$ would be further reduced to less than
0.1 nHz, a value now smaller than the fitting uncertainties. At that level,
contributions from the precision of the plate scale, differential rotation and
image distortion to this correction are comparable to that from the image
orientation.

 The determination of the shape of the differential rotation just below the
surface can be improved by including high degree splittings, but that shape
must be known to correct these splittings, as they are the result of ridge
fitting. It is likely that a bootstrapping method --- or some other optimized
methodology --- can be devised to disentangle this problem. This still leaves
the uncertainty of the image orientation as a substantial contribution.

  On the other hand, if we focus on the precision of relative changes, \ie\
differences between epochs, we can be more optimistic. Indeed, the residual
systematic errors introduced by the imperfect corrections are likely to be
nearly constant with time. 

\subsection{Changes with Solar Cycle}

  The four {\em Dynamics} data epochs analyzed for this study were taken
during the rising part of solar cycle 23. The first set, observed in 1996
corresponds to a period when the sun's magnetic activity was nearly at its
minimum, while for the forth data set, taken in 1999, the sun was near maximum
activity.  The average relative sunspot index for the four epochs is shown in
Table~\ref{tab:sunspotIndex}, \ie\ the ratio of the sunspot number to the
maximum value of cycle 23, reached in 2000.  We can thus study the changes
of high-degree p-mode parameters with the solar cycle by comparing results
from our four data sets

\begin{table*}[!t]
\begin{center}
\caption{Relative Average Sunspot Index
         \label{tab:sunspotIndex}}
\begin{tabular}{ccc}
     &                & Sunspot number change \\
Year & Relative Index & with respect to 1996  \\
\hline
 1996 &  7 \% &  ~   \\
 1997 & 16 \% & 11 \\
 1998 & 46 \% & 47 \\
 1999 & 72 \% & 79 \\
\hline
\end{tabular}
\end{center}
\end{table*}

\subsubsection{Frequency}

  To estimate high-degree mode frequency changes with the solar cycle, we have
corrected the ridge frequencies for the plate scale error (as described in
Section~\ref{sec:PlatScalErr}, Eq.~\ref{eq:dnuepsell}). We thus assume that
all other instrumental effects are either constant with time or that the
impact of their variation is small, effects such as the image distortion, the
instrumental PSF, and the error in the $P$ and $B_{\rm o}$ angles.

\begin{figure}[!p]
\epsscale{0.9}
\plotone{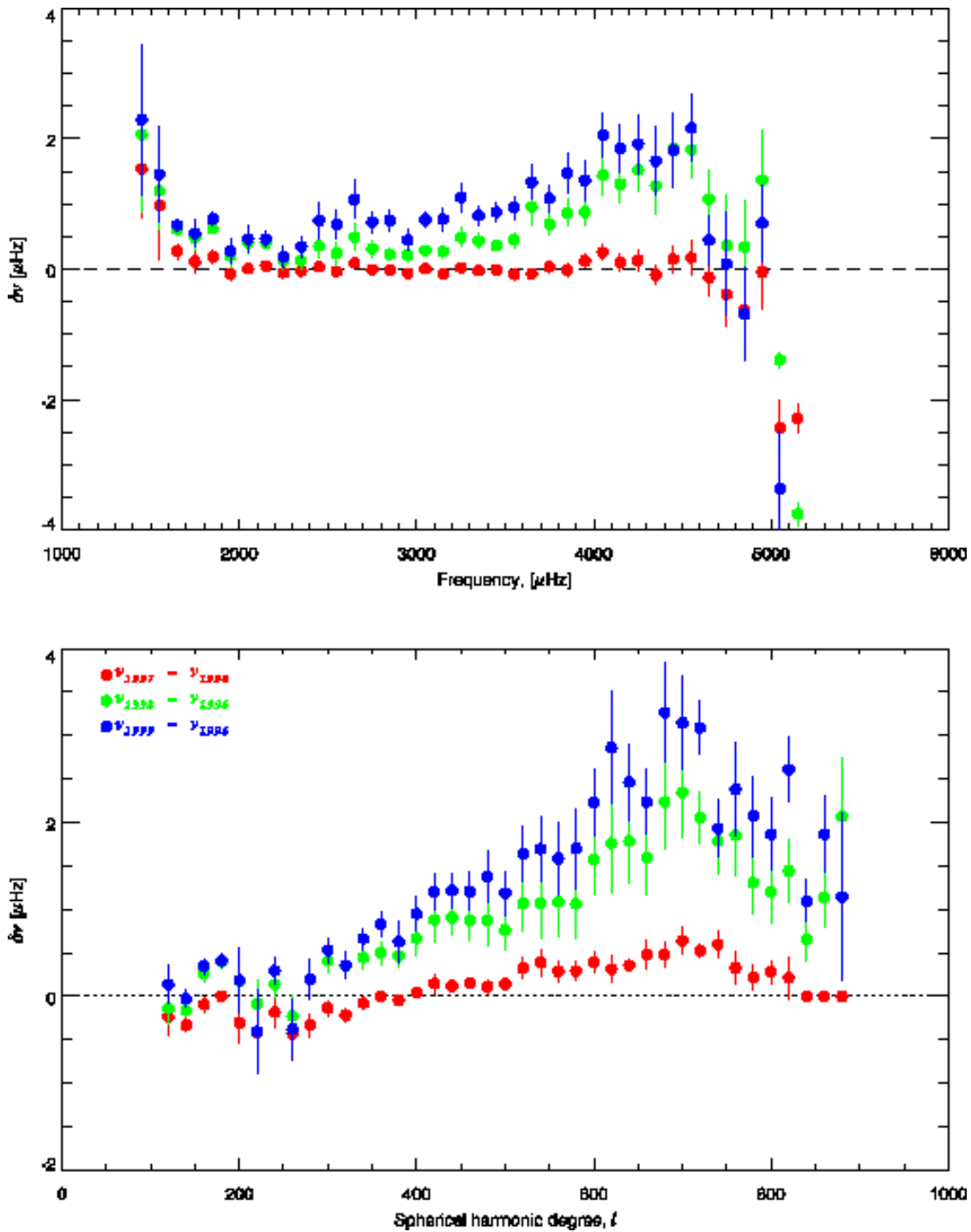}
\caption{Binned frequency changes with respect to values for 1996, after
         correction for respective plate scale errors. The differences were
         averaged over bins 100 $\mu$Hz wide in frequency (top panel) and 20
         degree wide (bottom panel); the error bars represent the standard
         deviation of the mean.
         \label{fig:freqXcycle2}}
\end{figure}

\begin{figure}[!p]
\epsscale{0.9}
\plotone{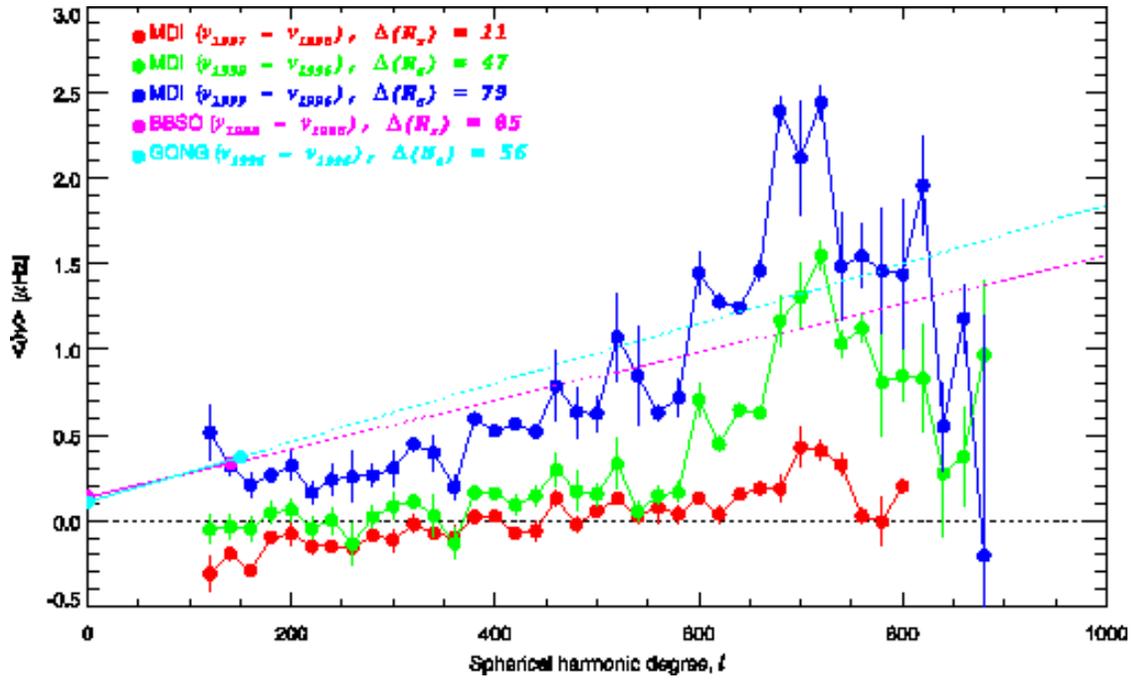}
\caption{Mean frequency changes around 3 mHz (\ie, the frequency change
         averaged over the 2.6 to 3.4 mHz frequency range) -- with respect to
         1996 values -- as a function of degree. These values are compared to
         linear extrapolations of comparable determinations based on Big
         Bear Solar Observatory observations taken between 1986 and 1988
         \citep{libbrecht90} and on GONG observations taken between March 1996
         and July 1998 \citep{Howe+Komm+Hill:1999}.
         \label{fig:lvalXcycle}}
\end{figure}

  Figure~\ref{fig:freqXcycle2} shows frequency changes with respect to 1996,
as a function of frequency and degree. These plots show that these differences
increase with frequency, degree and sunspot number. To compare these changes
to other investigators measurements, we show in Figure~\ref{fig:lvalXcycle}
mean frequency changes around 3 mHz (\ie, the frequency change averaged over
the 2.6 to 3.4 mHz frequency range), with respect to 1996 values, as a
function of degree. These are compared to estimates based on
intermediate-degree observations ($\ell \le 150$) taken during the raising
part of cycle 22 by \citet{libbrecht90}, a period that corresponds to a change
in the sunspot number index of 85, and the raising part of cycle 23 by
\citet{Howe+Komm+Hill:1999}, a period that corresponds to a change in the
sunspot number index of 56. A linear extrapolation with respect to degree of
the shifts based on intermediate-degree observations agrees rather well up to
$\ell = 600$. This confirms that these shifts scale with the inverse of the
mode mass, indicating that the source of the perturbation responsible for
these shifts is located very close to the surface.
  There are also claims of a sharp drop of the frequency shift at high
frequencies. In \citet{libbrecht90}, the drop is seen around 4 mHz, while in
our measurements it seems to occur at higher frequency (\ie, around 4.5 mHz;
see top panel of Figure~\ref{fig:freqXcycle2}).
  One also notices that besides a linear increase of the frequency shift with
degree, there is a {\em bump} centered on $\ell$=700. 

\begin{figure}[!p]
\epsscale{0.9}
\plotone{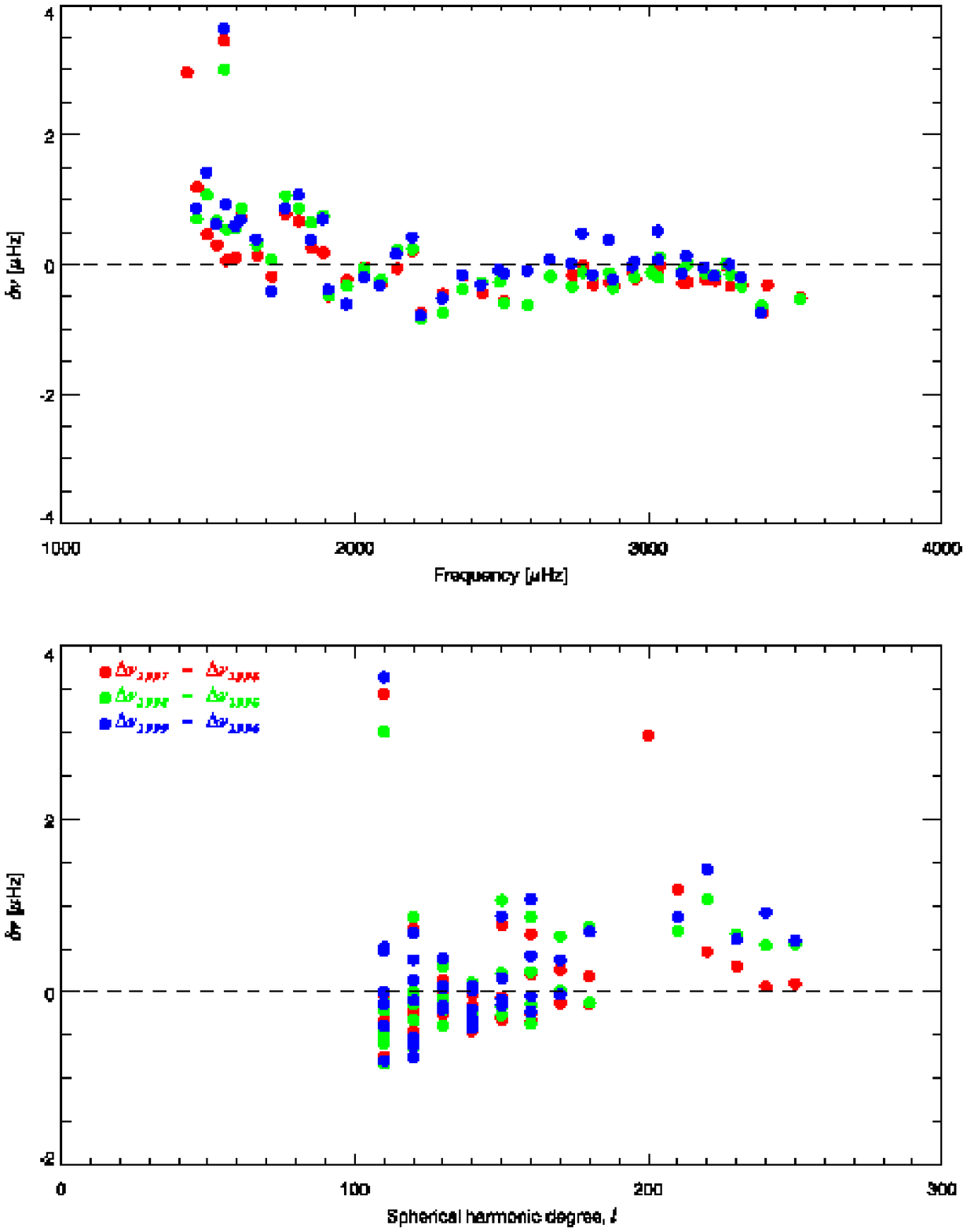}
\caption{Residual frequency offsets with respect to values for 1996, as a
         function of frequency and degree. 
         \label{fig:freqXcycle}}
\end{figure}

  We have checked our assumption that all other instrumental effects can be
neglected for our estimate of the frequency shifts shown in
Figure~\ref{fig:freqXcycle2}. Figure~\ref{fig:freqXcycle} presents residual
frequency offsets with respect to 1996, namely the the quantity
$\Delta\nu(P_i) - \Delta\nu(1996)$, where $\Delta\nu = \tilde\nu' - \nu$, $
\tilde\nu'$ is the ridge frequency corrected for the plate scale error, and
$\nu$ the mode frequency computed from resolved modes at low and intermediate
degrees.  This figure shows that while there remains a residual discrepancy,
there are no systematic changes with epoch, except for the values at low
frequencies ($\nu < 1.9$ mHz).  We can thus be confident that the variations
seen Figure~\ref{fig:freqXcycle2} are dominated by solar cycle effects, except
for the increase in the frequency shift at low frequency, that is most likely
an artifact resulting from some residual systematic errors.

\subsubsection{Splitting Coefficients}

\begin{figure}[!p]
\epsscale{0.8}
\plotone{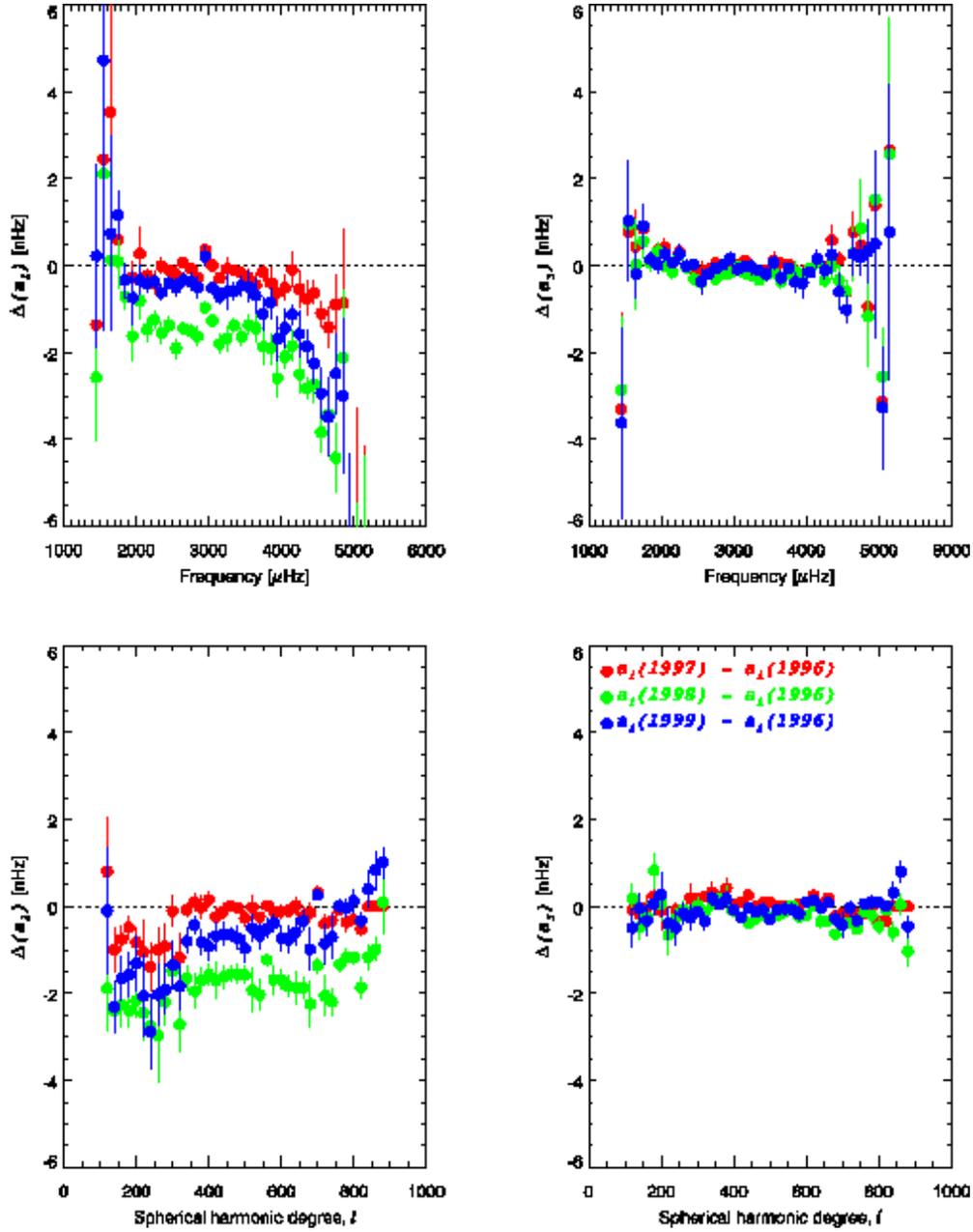}
\caption{Changes in $a_1$ and $a_3$ splitting coefficients -- relative to
         values for 1996 -- for different epochs, as a function of frequency
         and degree, after correction for respective plate scale errors. The
         changes were averaged over bins 100 $\mu$Hz wide in frequency (top
         panel) and 20 degree wide (bottom panel); the error bars represent
         the standard deviation of the mean. \label{fig:a1xcycle2}}
\end{figure}

 The splitting coefficients were corrected for the plate scale error and their
change with solar cycle estimated. Figure~\ref{fig:a1xcycle2} shows the
changes of the odd coefficients ($a_1$ and $a_3$) with respect to 1996, for
different epochs as a function of frequency and degree.  The plots show that
the changes of the $a_1$ coefficient increase with frequency, while displaying
only a weak dependence on degree. Although the shift increases with activity,
it is larger in 1998 than in 1999. The changes for $a_3$ are very small, and
the scatter large enough to be consistent with zero.

  To check if instrumental effects, other than the plate scale error, can be
neglected and are not causing the changes seen in $a_1$, we again computed the
residual $a_1$ offsets with respect to 1996, namely the the quantity $\Delta
a_1(P_i) - \Delta a_1(1996)$, where $\Delta a_1 = \tilde a_1' - a_1$, $ \tilde
a_1'$ is the ridge fitting value corrected for the plate scale error, and
$a_1$ the mode fitting value.  While there remains a residual discrepancy,
there appears to be no systematic variations with frequency, degree, or epoch.

  Similarly we have looked at the changes in the even splitting coefficients
and in the residual offsets with respect to 1996, also as a function of
frequency and degree. The changes in $a_2$ appear systematic with frequency,
degree and epoch, while the scatter in the residual $a_2$ offsets are large
and inconclusive.

\begin{figure}[!p]
\epsscale{0.9}
\plotone{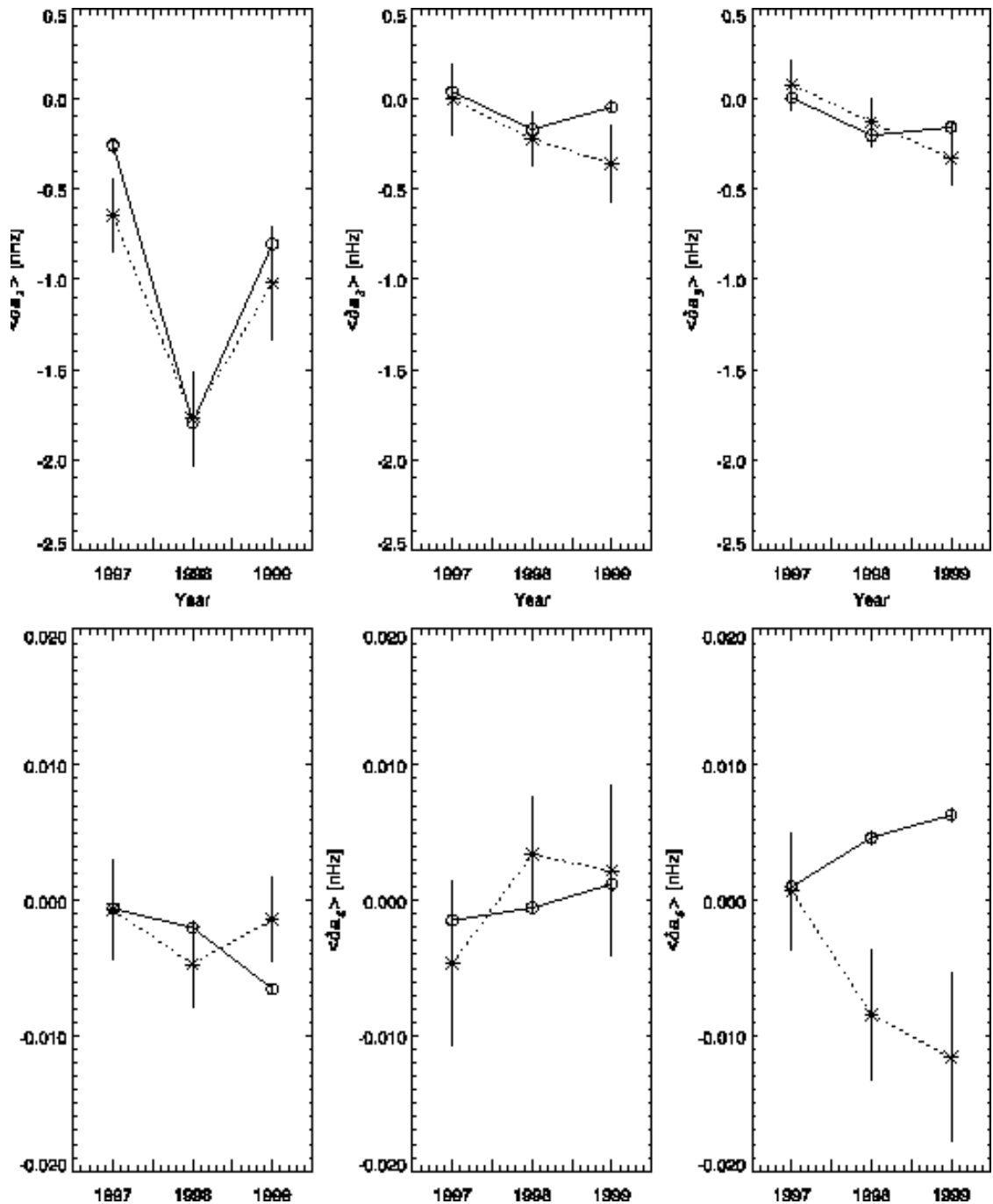}
\caption{Mean changes in odd and even splitting coefficients with epoch
         (circles) compared to mean changes in residual offsets (stars, see
         text). Since both sets track each other (except for $a_6$), we are
         led to conclude that the changes are due to systematic errors.
         \label{fig:delaiX}}
\end{figure}

  In Figure~\ref{fig:delaiX}, the mean splitting coefficient change and the
mean residual offset, corresponding to the average over all available modes,
are plotted as a function of epoch.  Since the mean splitting coefficient
changes track the mean residual offsets (except for $a_6$), we must conclude
that the changes with epoch result from residual systematic errors, and are
not an indication of changes due to solar activity.

\subsubsection{Asymmetry Parameter}

\begin{figure}[!p]
\epsscale{0.89}
\plotone{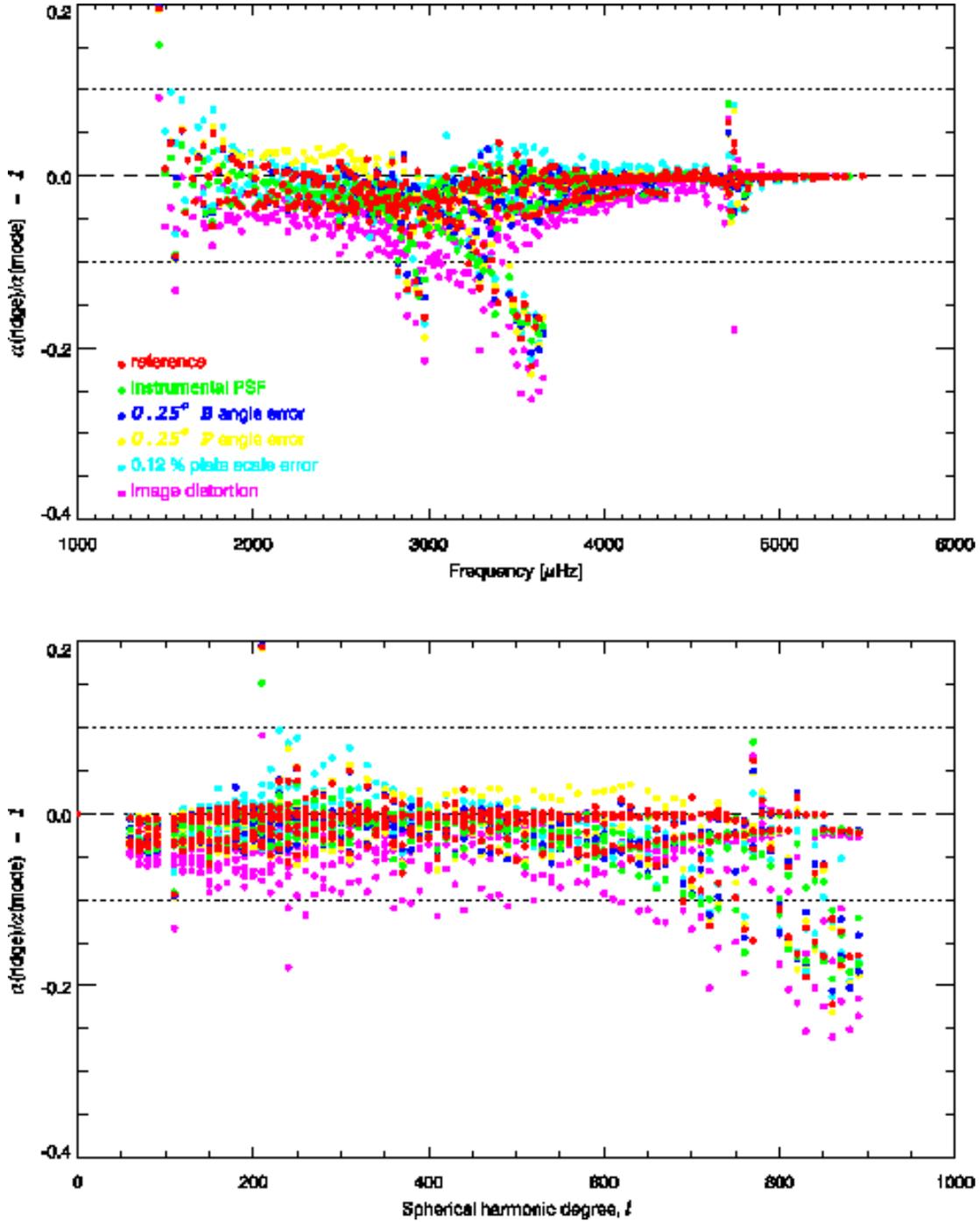}
\caption{
%
%
         Comparison of the ridge asymmetry resulting from fitting synthesized
         ridges with the mode asymmetry used as input. The different colors
         represent results for different simulations. Up to $\ell = 700$ the
         ridge asymmetry is within 10\% of the mode value. The large
         excursions are for low order modes (\ie, for $n=0$ and 1). Also, the
         feature at $\nu = 4.7$ mHz occurs where the asymmetry goes through
         zero.
\label{fig:comp_asym}}
\end{figure}

  Unlike the other mode parameters, the estimate of the asymmetry parameter is
not much affected by the fact that we observe and fit ridges instead of
individual modes. \citet{korzennik99} pointed out that the asymmetry of the
ridge seems to be the same as the asymmetry of the modes that are blended into
that ridge. We found from our simulations that these differences are smaller
than 10\% up to $\ell=700$ when including the various instrumental effects
described in Section 5, as illustrated in Figure~\ref{fig:comp_asym}.

\begin{figure}[!p]
\epsscale{0.9}
\plotone{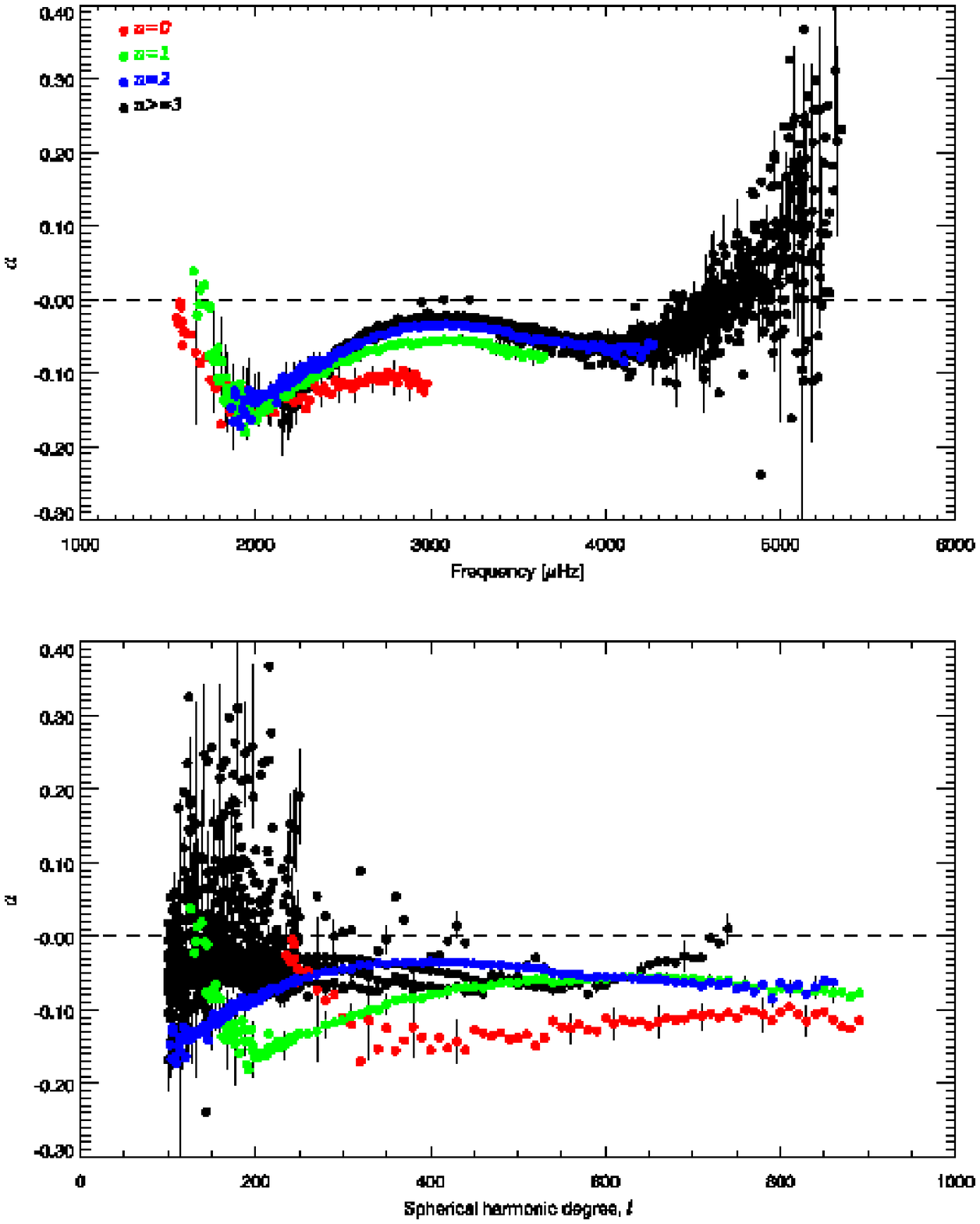}
\caption{
%
%
         Ridge asymmetry for the 1996 {\em Dynamics} epoch, as a function of
         frequency and degree. Error bars are shown only for every 5th data
         point, the low order modes are color coded.
\label{fig:obsalpha}}
\end{figure}

\begin{figure}[!p]
\epsscale{0.9}
\plotone{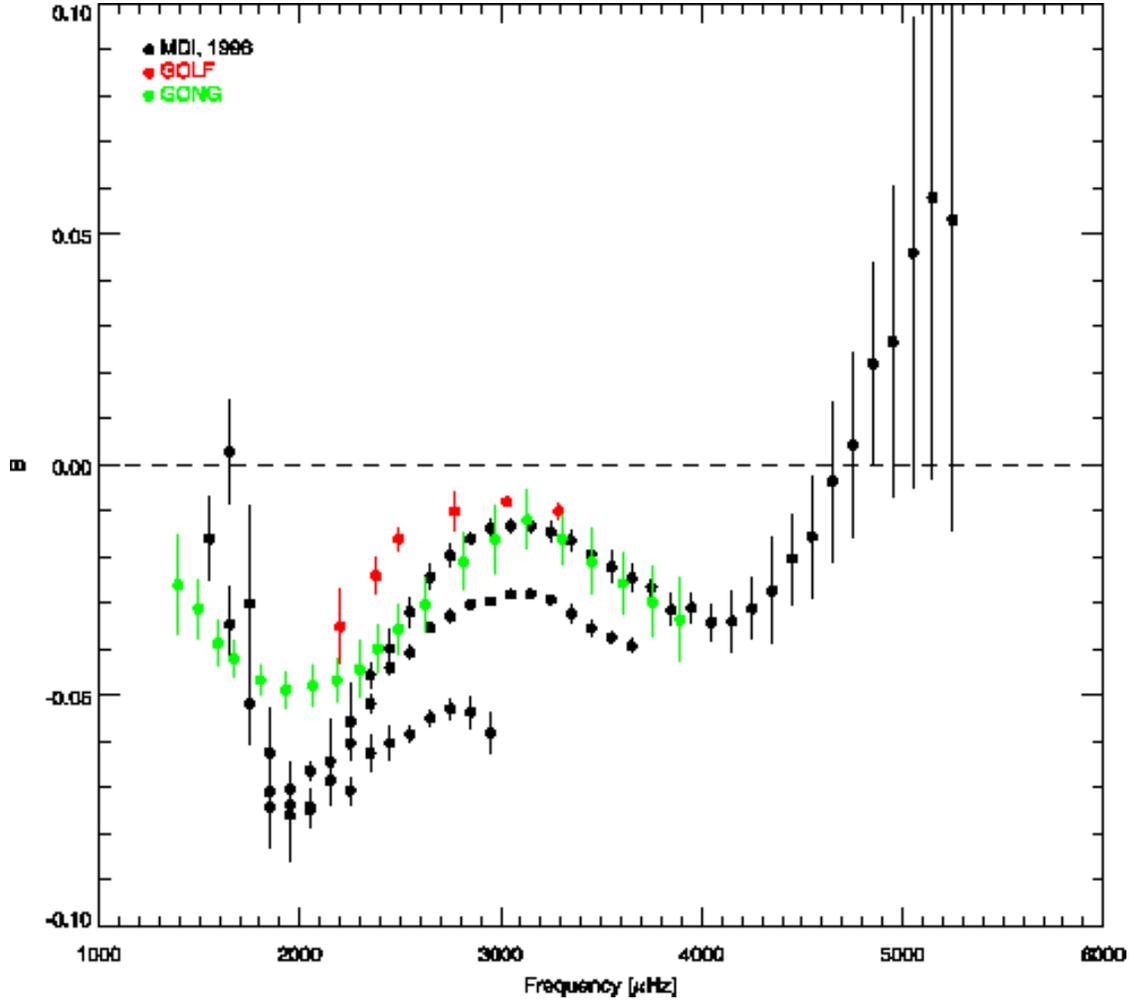}
\caption{Comparison of the profile asymmetry estimates from 1996 MDI {\em
         Dynamics} data (this work) with published values from GOLF and GONG
         observations. The MDI data were binned in frequency for $n=0, 1$ and
         $n\ge 2$ separately. The error bars for the MDI and GONG values
         represent the scatter of the data.
         \label{fig:obsb}}
\end{figure}

  Figure~\ref{fig:obsalpha} shows the measured asymmetry coefficients for the
1996 {\em Dynamics} observations.  The asymmetry is mainly a function of
frequency, except for the low order modes ($n=0$ and $n=1$) whose asymmetry is
larger. 
  In Figure~\ref{fig:obsb} we compare our determinations of the mode
asymmetry\footnote{we plotted the asymmetry parameter $B$, as defined in
\citet{Nigam+Kosovichev:1998}, where $B = \frac{\alpha/2}{(1 - \alpha^2/2)}$}
with values obtained for low-degrees ($\ell < 3$) using GOLF data
\citep{thiery00} and for intermediate degrees using GONG observations
\citep{basu00}. Our measurements agree very well with the GONG determinations
in the 2.4 to 3.8 mHz range.

\begin{figure}[!p]
\epsscale{0.9}
\plotone{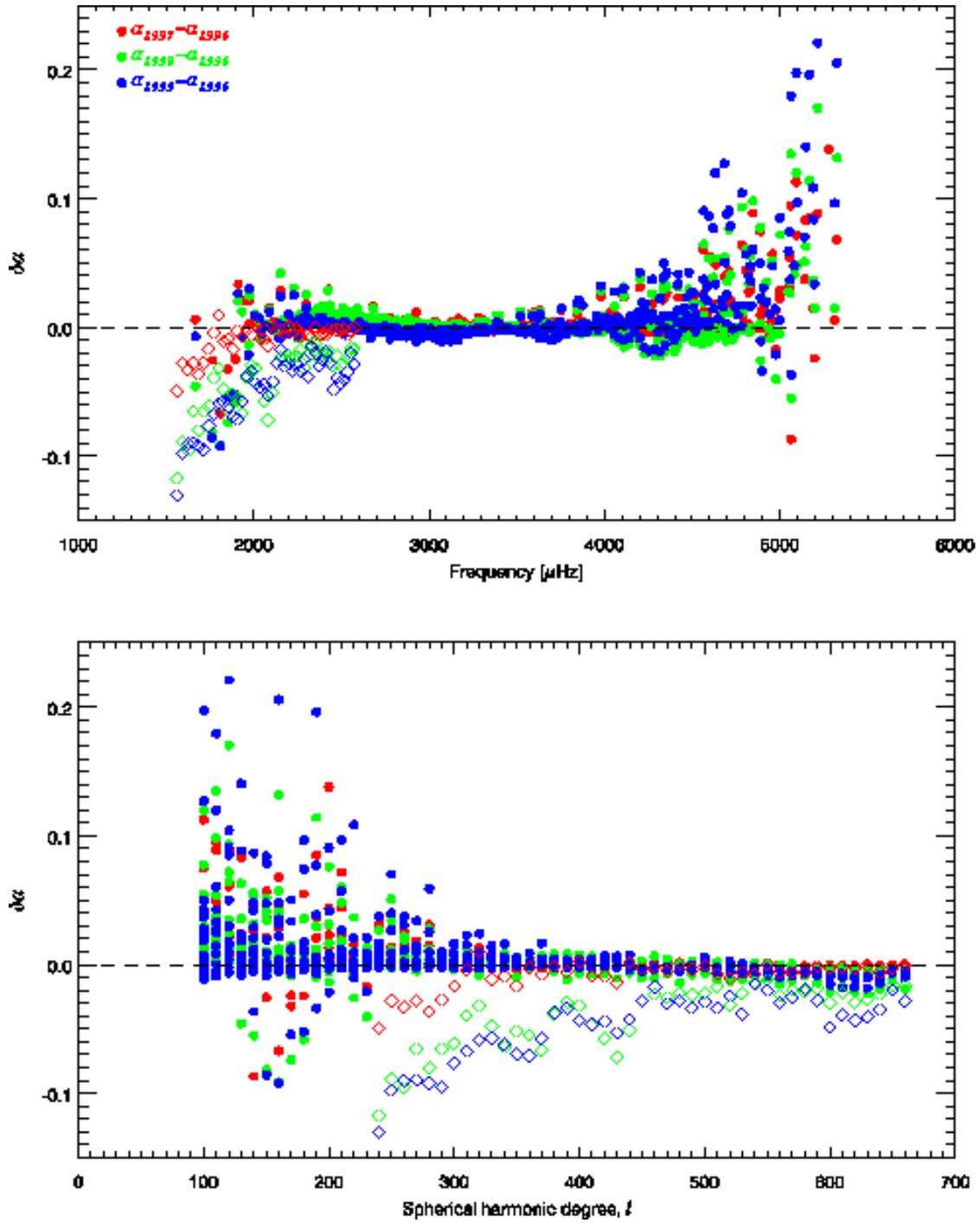}
\caption{
%
%
         Changes of the asymmetry parameter -- with respect to 1996 -- for
         different epochs, as a function of frequency and
         degree. Changes for the f-modes are indicated by diamonds.
         \label{fig:obsasym}}
\end{figure}

  In Figure~\ref{fig:obsasym} we compare the asymmetry for different epochs,
shown as a function of frequency and degree. Except for the $f$-mode, there
seems to be no change with epoch, while the asymmetry of the $f$-mode appears
to have changed systematically, suggestive of a solar activity effect.

\subsubsection{Linewidth}

  Measurements of the mode linewidths provide clues to the damping mechanisms
of the solar oscillations and are directly related to the lifetimes of the
modes. Previous measurements have shown that the linewidths increase both with
frequency and degree \citep{hill96}.

\citet{korzennik90} proposed a simple formula to estimate the mode linewidth
from the the ridge linewidth, namely:
\begin{equation} 
  \Gamma_{\rm ridge}^2 = \Gamma_{\rm mode}^2 + \Gamma_{\rm WF}^2 
                                  + (s\, \frac{d\nu}{d\ell})^2
\label{eq:ridgeCorr}
\end{equation}
where $\Gamma_{\rm WF}$ is the width of the window function, $s\,d\nu/d\ell$
corresponds to the effective leakage matrix width in frequency space, where
the empirical coefficient $s \approx 0.5$.

\begin{figure}[!p]
\epsscale{0.89}
\plotone{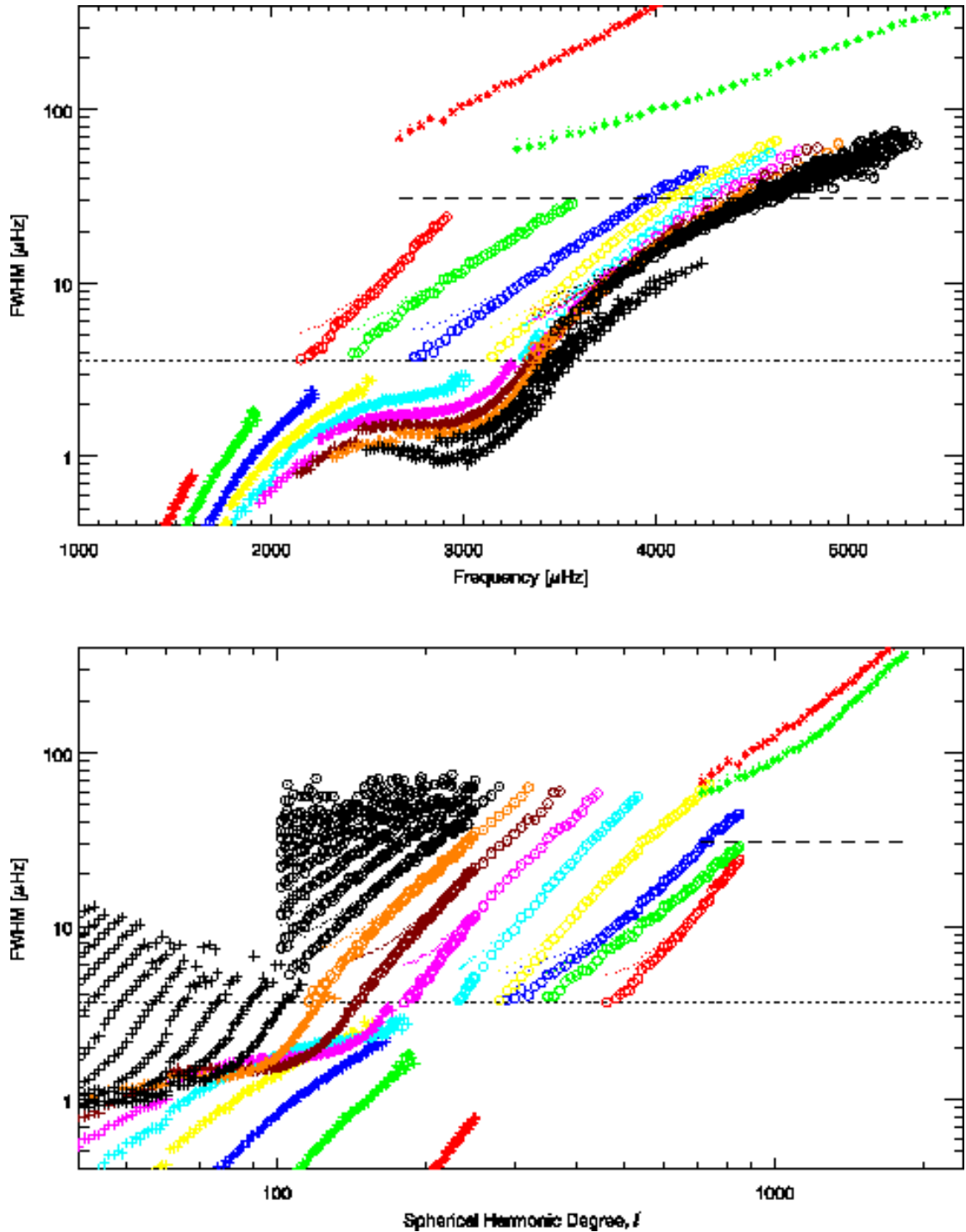}
\caption{Corrected MDI linewidths (circles, this work) compared to MDI low and
         intermediate degree values derived from 360-day-long {\em Structure}
         data (crosses) and to very high degree corrected linewidth estimates
         from MDI in high-resolution mode (stars) as published by
         \citet{duvall98}. Uncorrected linewidths are shown as dots. The
         horizontal lines correspond to the FWHM of the window functions for
         the MDI observations.
\label{fig:wdtfrq}}
\end{figure}

  Figure~\ref{fig:wdtfrq} shows the estimated mode linewidth for the 1996 {\em
Dynamics} observations, as well as linewidths estimated for low and
intermediate degrees ($0 < \ell < 250$) based on 360-day long time-series of
MDI {\em Structure Program} \citep{schou99}. 
Preliminary estimates of linewidths for high-$\ell$ modes ($700 < \ell <
1800$) based on high-resolution MDI observations \citep{duvall98} are also
included. Note that modes whose corrected linewidth are smaller than the width
of the window function are not shown in Figure~\ref{fig:wdtfrq}, since the
correction scheme of Eq.~\ref{eq:ridgeCorr} is not valid.

  Our simple minded linewidth correction scheme connects reasonably well the
MDI {\em Structure} linewidths (\ie, low and intermediate mode fitting) with
the corrected MDI {\em Dynamics} linewidths, but we could not reconcile the
\cite{duvall98} determinations based on high-resolution MDI observations.  We
do see in our measurements the trend seen in \citet{duvall98} at higher
degrees, namely that the linewidth of the $f$-modes increases with degree
faster than the linewidth of the $p_1$ modes. Our measurements suggest a cross
over around $\ell=1000$, while the \citet{duvall98} estimates imply a cross
over below $\ell=700$.

\begin{figure}[!p]
\epsscale{1.0}
\plotone{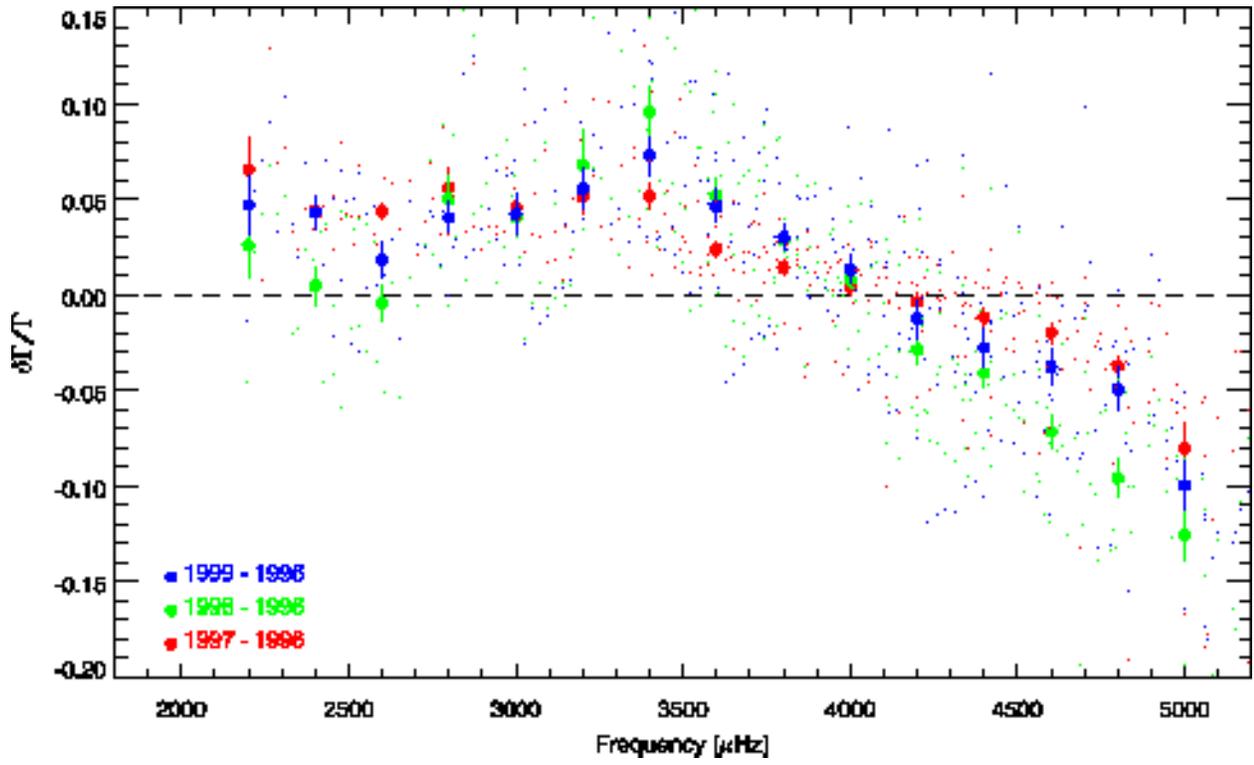}
\caption{
%
%
         Raw and binned linewidth relative changes (dots and circles
         respectively), with respect to 1996, as a function of frequency, and
         restricted to modes common to all epochs. The ridge linewidths were
         first corrected according to Equation~\ref{eq:ridgeCorr}.
\label{fig:wdtcycle}}
\end{figure}

  Figure~\ref{fig:wdtcycle} shows the relative change of the corrected
linewidths with epoch -- using 1996 as reference -- as a function of
frequency. The magnitude of these changes -- on the order of 5 to 10\% -- do
not correlate with solar activity, nor do they scale with the plate scale
error. The fact that the linewidths of all three later epochs have changed in a
systematically similar way with respect to 1996 is intriguing and suggestive
of residual systematic errors at the 5 to 10\% level.

\section{Conclusion}

  We believe that we have shown that one can construct a physically motivated
model (rather than some {\em ad hoc} correction scheme) of the ridge power
distribution that results in a methodology that can produce an unbiased
determination of high-degree modes, once the instrumental characteristics are
well understood and precisely measured.

  We have gained substantial insight in the understanding and modeling of the
instrumental effects of the Michelson Doppler Imager, including plate scale
error, image distortion, point spread function and image orientation.  Now
that we better understand the MDI instrument imaging imperfections, we can --
and will in the near future -- reprocess the {\em Dynamics} data, and include
the correct image scale and the known image distortion in the spatial
decomposition.

  We have produced an error budget for the characterization of high degree
mode parameters. Beyond MDI {\em Dynamics} observations, this has direct
applications on the characterization of the GONG+ instruments and the
reduction of the GONG+ observations as well as on the characteristics of the
Helioseismic/Magnetic Imager (HMI) instrument planned for the Solar Dynamics
Observatory (SDO) mission, if one wishes to make effective use of the high
degree modes accessible to these experiments.

  Last but not least, we also see changes with solar activity in the high
degree frequencies and the asymmetry of the $f$-mode. Our attempt to see
changes in splitting coefficients and mode linewidths remained inconclusive.

\acknowledgments
\section*{Acknowledgments}

  We wish to thank Rock Bush providing us with the MDI best focus estimates,
Phil Scherrer for the MDI image distortion model, J\o{}rgen
Christensen-Dalsgaard for estimates of the horizontal-to-vertical displacement
ratio from his solar model, Ted Tarbell for the PSF of MDI in high resolution
mode, and Cliff Toner for his estimates of MDI roll angle.  Sunspot index data
provided by the SIDC, RWC Belgium, World Data Center for the Sunspot Index,
Royal Observatory of Belgium.

 The Solar Oscillations Investigation - Michelson Doppler Imager project on
SOHO is supported by NASA grant NAG5--8878 and NAG5--10483 at Stanford
University.  SOHO is a project of international cooperation between ESA and
NASA.
 SGK was supported by Stanford contract PR--6333 and NASA grant NAG5--9819.

\appendix

\section{Elliptical Image Distortion}

  The observed elliptical shape of the solar image is likely to be the result
of an unwanted tilt of the detector plane with respect to the focal plane.
This appendix shows how the combination of the distortion resulting from this
tilt with the optical package distortion (or cubic distortion) can be
estimated.  First the cubic distortion is applied to the undistorted image
$(x_0,y_0)$
\begin{eqnarray}
x & = & x_0 + \epsilon_r x_0 \\
y & = & y_0 + \epsilon_r y_0
\end{eqnarray}
where
$\epsilon_r = \epsilon_r(r_0)$ is given by Equation~\ref{eq:raddist} and $r_0
= \sqrt{x_0^2 + y_0^2}$, while $(x,y)$ are the reference coordinates in
the focal plane that would correspond to the coordinates for a perfectly
aligned detector.

  To model the image distortion due to a tilt of the CCD, the reference
coordinate system $(x,y)$ is first rotated around the optical axis (the
$z$-axis) by an angle $\beta$, which defines the direction around which the
detector is tilted (the $x'$-axis):
\begin{eqnarray}
x' & = & \phantom{-} x \cos\beta + y \sin\beta \\
y' & = & -x \sin\beta + y \cos\beta .
\end{eqnarray}
The effect of a tilt by an angle $\alpha$ around the $x'$-axis is then
computed using a simple pinhole camera model. It is easy to show that such
a model leads to the following equations:
\begin{eqnarray}
\label{eq:ccdtilt1}
x'' & = & x' (1 + \frac{y'}{f} \, \alpha) - x' \frac{\alpha^2}{4} \\
y'' & = & y' (1 + \frac{\alpha^2}{2} + \frac{y'}{f} \, \alpha) - 
y' \frac{\alpha^2}{4} - \frac{r_{\mbox{m}}^2}{f} \, \alpha  
\label{eq:ccdtilt2}
\end{eqnarray}
where $f$ is the effective focal length and $r_{\rm m}$ the observed image
mean radius, both of which are measured in the same units as $x'$ and $y'$.
Note that the second term in both equations assures that the distorted and
undistorted images have the same mean radius, while the third term in the
second equation keeps the center of both images unchanged.

Finally, the coordinate system is rotated back, around the $z$-axis, to be
co-aligned with the original reference coordinates system:
\begin{eqnarray}
x''' & = & x'' \cos\beta - y'' \sin\beta \\
y''' & = & x'' \sin\beta + y'' \cos\beta 
\end{eqnarray}
and the image distortion is given by:
\begin{eqnarray}
\delta_x = x''' - x_0 \label{eq:distxtotal} \\
\delta_y = y''' - y_0 \label{eq:distytotal}
\end{eqnarray}
where $(x''', y''')$ are the observed (\ie\ actual) coordinates and $(x_0,
y_0)$ are the undistorted ones.

Due to the tele-photo nature of the MDI optical package \citep{scherrer95},
the effective focal length $f$ defined in Equations \ref{eq:ccdtilt1} and
\ref{eq:ccdtilt2} is not the one that reproduces the image size.
A complete ray trace
of the MDI optical system (Scherrer, private communication) produces an
estimate for $f$ of 13000 pixels.

\begin{figure}[!p]
\epsscale{0.8}
\plotone{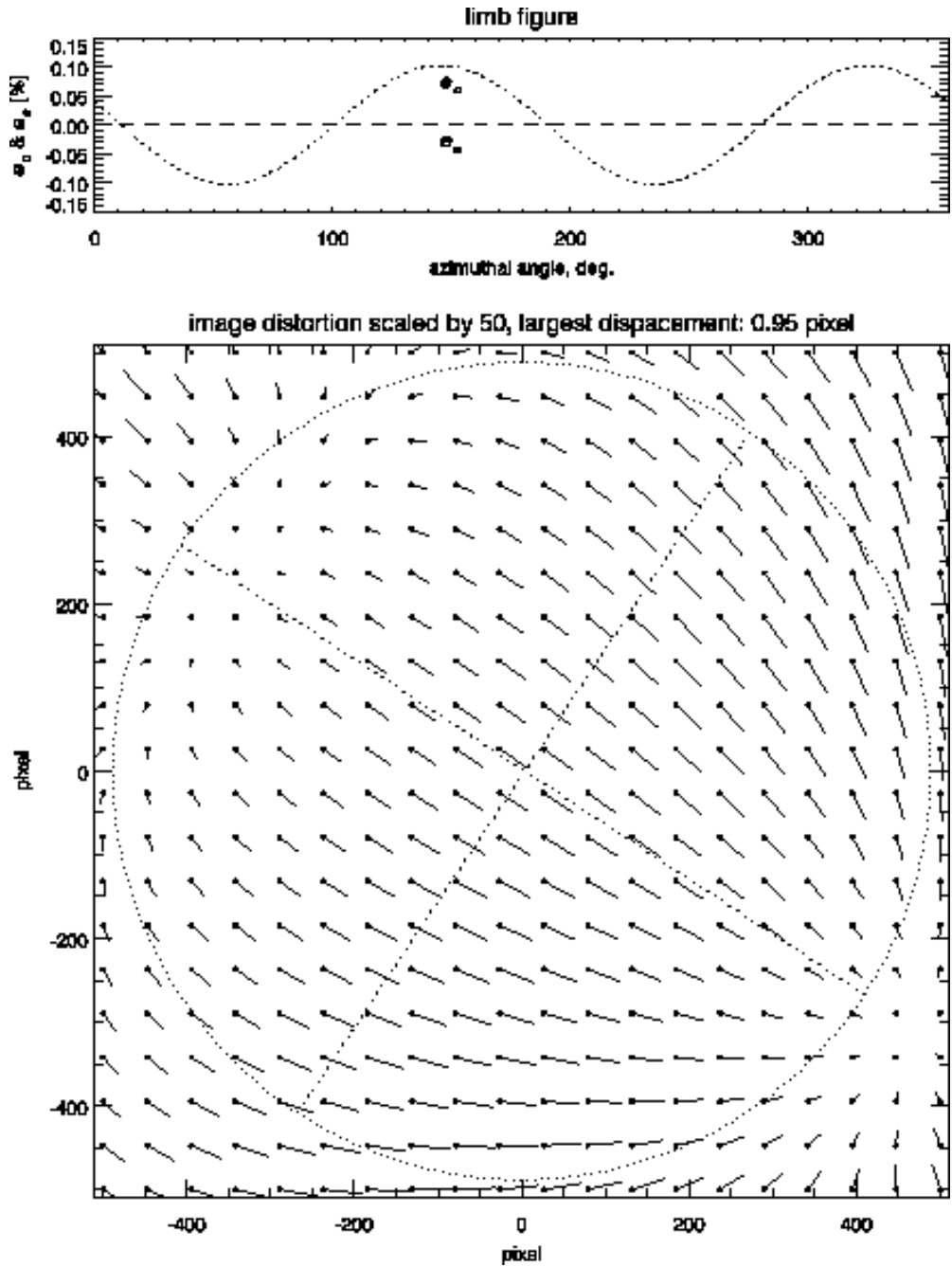}
\caption{Map of the image distortion in the plane of the CCD. The distortion
         has been multiplied by a factor 50 to make it visible. The largest
         displacement inside the solar limb is 0.95 pixel. The top panel
         illustrates how the distorted departs from a circle (\ie, $e_c =
         (x/r)^2 + (y/r)^2 - 1$) and how well it fits an ellipse (\ie, $e_e =
         (x/a)^2 + (y/b)^2 - 1$).
\label{fig:distmap}}
\end{figure}

  Various observational methods have been used to determine the parameters of
the distortion, some of which are described below. Results are summarized in
Table \ref{tab:distor}.
  With $\beta = 56\Deg$ and $\alpha = 2.59\Deg$, the model reproduces rather
well the nearly elliptical images of MDI.  The resulting shape of the limb
matches an ellipse to $10^{-6}$ (to be compared to the 0.1\% departure from a
circle), and the resulting image distortion pattern is shown in
Figure~\ref{fig:distmap}.

  An extensive series of ground tests were completed prior to launch to verify
the performance of the MDI instrument \citep{zayer94}. In one of them, a grid
target was imaged to evaluate, amongst other things, the image distortion. The
distortion was determined by measuring the change in shape of the grid target
as it was imaged onto different positions within the field of view.  This
distortion was fitted to the parameterized representation expressed by
Equations~\ref{eq:distxtotal} and \ref{eq:distytotal}, leaving $\alpha$, $f$
and $a_r$ to be freely adjustable parameters. The fit lead to $a_r$ some 11\%
higher and $\alpha$ some 5\% lower than expected. This fit also lead to a
value of $f$ some 50\% higher than predicted.  The higher effective focal
length is surprising since ray tracing is expected to produce a reliable value
for this parameter.  While these are pre-flight measurements, and thus cannot
account for changes that could have occurred during deployment of the
satellite, the nature of the manufacturing and assembly of the optical package
make such post-launch changes highly improbable.  On the other hand it is
possible that some subtle aspect of the ground test has not been accounted
for.

  In flight off-center continuum intensity images were acquired in February
2001 --- with MDI operating in focus position 3 --- to further study the image
distortion.  When the solar images are centered in different parts of the
detector, the distortion will affect them in a different way and the shape of
the limb will change with each image position.  Using again the image
distortion parameterized by Equations~\ref{eq:distxtotal} and
\ref{eq:distytotal}, the predicted shape of the limb can be computed as a
function of image position, and compared to the observed limb profiles.  The
best match between predicted and observed limb shapes was achieved for $\alpha
= 2.6\Deg$, $\beta = 56\Deg$ and $f = 12940$.

  Finally, images obtained during rolling maneuvers of the spacecraft in March
1997 and November 2001 were also used to characterize the distortion.  These
images were averaged for each roll angle and rotated back to $P_{\rm eff } =
0$. The resulting images, which are dominated by the supergranulation, are
then used to estimate the differential distortion by correlating each image
with an image in the center of the time range.  In this case $f$ was kept
fixed while the same distortion model was again fitted to the data.  Two
estimates were made using these images.  The first (hereafter {\em method 1})
is made using the offset of the center of the images, while the other
(hereafter {\em method 2}) is made using the stretching of the images.
The results are inconsistent as summarized in Table~\ref{tab:distor}.  In
particular {\em method 2} always produces higher values of $\alpha$ than {\em
method 2}.  Again, a larger value of $f$ would be required to reconcile
results from both methods.

\begin{table}[!ht]
\begin{center}
\caption{Summary of the Characterization of the Model of MDI Image Distortion}
\label{tab:distor}
\begin{tabular}{l|c|c|c|c}
Method                 &  $\alpha$ & $\beta$ & $f$  & $a_r$ \\\hline

Match limb shape       &   2.59$\Deg$ &    56$\Deg$ & Fixed    & Fixed \\
Pre-flight target grid &   2.45$\Deg$ &    ~        & 19604.4  & $1.2\times 10^{-3}$ \\
Image offset           &    2.6$\Deg$ &    56$\Deg$ & 12940.0  & Fixed  \\\hline
1997 roll, method 1    &   1.80$\Deg$ &    52$\Deg$ & Fixed    & Fixed  \\
1997 roll, method 2    &   1.97$\Deg$ &    59$\Deg$ & Fixed    & Fixed  \\
2001 roll, method 1    &   1.71$\Deg$ &    53$\Deg$ & Fixed    & Fixed  \\
2001 roll, method 2    &   2.43$\Deg$ &    50$\Deg$ & Fixed    & Fixed  \\\hline
\end{tabular}
\end{center}
\end{table}

  Although the parameters resulting from the different methods used to
characterize the image distortion of MDI do not agree as well expected, the
distortion pattern resulting from a tilt of the detector is a rather good
model of the observed distortion.  The fact that the model parameters are so
different when using different methods might be attributed to the complicated
instrumental PSF of the MDI optical package that is not fully known and
therefore cannot be taken into account. In particular the different
observational methods described above use features of the image at different
spatial scales (or spatial wavenumbers) to estimate the image distortion.  The
PSF may also explain why the ellipticity of the images changes significantly
when the images are strongly defocused.  Another possible explanation may, of
course, be that other types of distortions, such as non square pixels, are
present.

\clearpage
\onecolumn

%
%
%

\clearpage

\clearpage

\clearpage

\clearpage

\begin{thebibliography}{}
%
\bibitem[Bachmann et~al.(1995)]{bachmann95} Bachmann, K.T., Duvall, T.L.Jr.,
Harvey, J.W., and Hill, F. 1995, \apj, 443, 837
%
\bibitem[Basu \& Antia(2000)]{basu00} 
Basu, S., and Antia, H.M. 2000, \apj, 531, 1088
%
\bibitem[Bush, Chu, \& Kuhn(2001)]{bush01} 
Bush, R.~I., Chu, K., \& Kuhn, J.~R.\ 2001, American Geophysical Union, Fall
Meeting 2001, abstract \#SH11B-0715, 715
%
\bibitem[Christensen-Dalsgaard(1997)]{cd97}
Christensen-Dalsgaard J., 1997,
Lecture Notes, Aarhus Universitet, Denmark
%
\bibitem[Christensen-Dalsgaard et~al.(1996)]{cd96}
Christensen-Dalsgaard J., et~al., 1996, Science  272, 1286
%
\bibitem[Duvall et~al.(1998)]{duvall98} 
Duvall, T.L., Kosovichev, A.G., Murawski, K. 1998, \apj, 505, L55
%
\bibitem[Giles(1999)]{Giles:99}
Giles, P, 1999, Ph.D. Dissertation, Stanford University.
%
\bibitem[Gough et~al.(1980)]{GoughEtal:80}
Gough, D.O., 1980, in {\em Nonradial and Nonlinear Stellar Pulsation},
ed.\ H.A.\ Hill and W.A.\ Dziembowski (Springer-Verlag, Berlin), p. 273

\bibitem[Hill et~al.(1996)]{hill96}
Hill, F.~et al.\ 1996, Science, 272, 1292 
%
\bibitem[Howe, Komm \& Hill(1999)]{Howe+Komm+Hill:1999} 
Howe, R., Komm, R. and Hill, F. 1999, \apj, 524, 1084
%
%
%
\bibitem[Korzennik(1999)]{korzennik99}
Korzennik S.G., 1999, ESA SP 418: Structure and Dynamics of the Interior
of the Sun and Sun-like Stars, eds. S.G. Korzennik, A. Wilson, ESA 
Publications Division, Noordwijk, The Netherlands, p. 933 
%
\bibitem[Korzennik(1990)]{korzennik90}
Korzennik S.G., 1990, Ph.D. thesis, Univ. of California, Los Angeles
%
\bibitem[Libbrecht \& Woodard(1990)]{libbrecht90}
Libbrecht, K.G, and Woodard, M.F. 1990, Nature, 345, 779
%
\bibitem[Libbrecht \& Kaufman(1988)]{libbrecht88} 
Libbrecht, K.G., and Kaufman, J.M. 1988, \apj, 324, 1172
%
\bibitem[Nigam \& Kosovichev(1998)]{Nigam+Kosovichev:1998}
Nigam, R.~\& 
Kosovichev, A.~G.\ 1998, \apjl, 505, L51 
%
\bibitem[Rabello-Soares et~al.(2000)]{rabello-soares00}
Rabello-Soares, M.~C., Basu, S., 
Christensen-Dalsgaard, J., \& Di Mauro, M.~P.\ 2000, \solphys, 193, 345 
%
\bibitem[Rabello-Soares et~al.(2001)]{rabello-soares01}
Rabello-Soares, M.~C., Korzennik, S.~G., \& Schou, J.\ 2001, Proceedings of 
the SOHO 10/GONG 2000 Workshop: Helio- and asteroseismology at the dawn of 
the millennium, 2-6 October 2000, Santa Cruz de Tenerife, Tenerife, 
Spain.~Edited by A.~Wilson, Scientific coordination by P.~L.~Pall{\' 
e}.~ESA SP-464, Noordwijk: ESA Publications Division, ISBN 92-9092-697-X, 
2001, p.~129 - 136, 464, 129 
%
\bibitem[Rhodes et~al.(2001)]{RhodesEtal:01}
Rhodes, E.~J., Reiter, J., Schou, J., Kosovichev, A.~G., \& Scherrer, P.~H.\
2001, \apj, 561, 1127
%
\bibitem[Rhodes et~al.(1999)]{rhodes99}
Rhodes E.J., Reiter J., Kosovichev A.G., et~al., 1999, ESA SP 418:
Structure and Dynamics of the Interior
of the Sun and Sun-like Stars, eds. S.G. Korzennik, A. Wilson, ESA 
Publications Division, Noordwijk, The Netherlands, p. 73
%
\bibitem[Ritzwoller \& Lavely(1991)]{Ritzwoller+Lavely:91}
Ritzwoller, M.~H.~\& Lavely, E.~M.\ 1991, \apj, 369, 557 
%
\bibitem[Scherrer et~al.(1995)]{scherrer95}
Scherrer P.H., Bogart R.S., Bush R.I., et~al., 1995, \solphys. 162, 129
%
\bibitem[Schou(1999)]{schou99}
Schou, J.\ 1999, \apjl, 523, L181 
%
\bibitem[Schou et~al.(1998)]{schouetal98}
Schou et al., 1998, \apj, 505, 390
%
\bibitem[Schou \& Bogart(1998)]{schoubogart98}
Schou J., Bogart R.S., 1998, \apj, 504, L131 
%
\bibitem[Schmidt, Stix, \& W{\" o}hl(1999)]{SchmidtEtal:99}
Schmidt, W., Stix, M., \& W{\" o}hl, H.\ 1999, \aap, 346, 633 
%
\bibitem[Snodgrass \& Ulrich(1990)]{snodgrass90}
Snodgrass, H.~B.~\& Ulrich, R.~K.\ 1990, \apj, 351, 309 
%
\bibitem[Tarbell et~al.(1997)]{tarbell97}
Tarbell T.D., Acton D.S., Frank Z.A., 1997,
Adaptive Optics 96, OSA Technical Digests Vol. 13
%
\bibitem[Thiery et~al.(2000)]{thiery00} 
Thiery, S., Boumier, P., Gabriel, A.H., et~al. 2000, \aap,
355, 743 
%
\bibitem[Toner \& Jefferies(1993)]{toner93}
Toner, C.~G.~\& Jefferies, S.~M.\ 1993, \apj, 415, 852 
%
\bibitem[Woodard(2000)]{Woodard:00}
Woodard, M.~F.\ 2000, \solphys, 197, 11
%
\bibitem[Woodard(1989)]{woodard89}
Woodard, M.~F.\ 1989, \apj, 347, 1176
%
\bibitem[Zayer et~al.(1993)]{zayer94}
Zayer, I., Bogart, R.~S., Hoeksema, J.~T., \& Tarbell, T.\ 1993, AAS/Solar 
Physics Division Meeting, 25, 1192 
%
\end{thebibliography}
\end{document}